\documentclass[prd,onecolumn,showpacs,floatfix,superscriptaddress,nofootinbib]{revtex4-2}

\usepackage{graphicx}
\usepackage{epsfig}
\usepackage{bm}
\usepackage{amssymb}
\usepackage{float}
\usepackage{amsmath}
\usepackage{dcolumn}
\usepackage{cancel}

\usepackage{array}
\usepackage{mathrsfs}
\usepackage{dcolumn}
\usepackage{graphicx}
\usepackage{amsmath}
\usepackage{amsfonts}
\usepackage{amssymb}
\usepackage{microtype}
\usepackage{subfigure}
\usepackage{makeidx}
\usepackage{bm}
\usepackage{epsf}
\usepackage{color}
\usepackage{multirow,dcolumn}
\usepackage{graphicx}
\usepackage{mathrsfs}
\graphicspath{{Images/}}

\def\doi{http://doi.org}

\def\be{\begin{equation*}}
\def\ee{\end{equation*}}

\begin{document}

\title{Motion of particles around a magnetically charged Euler-Heisenberg black hole with  scalar hair and the Event Horizon Telescope}

\author{Dionysios P. Theodosopoulos}
	\email{dionysistheodosopoulos@gmail.com}
	\affiliation{Physics Department, National Technical University of Athens, 15780 Zografou Campus, Athens, Greece}

\author{Thanasis Karakasis}
\email{thanasiskarakasis@mail.ntua.gr}
\affiliation{Physics Department, National Technical University of Athens, 15780 Zografou Campus, Athens, Greece}
	
\author{George Koutsoumbas}
	\email{kutsubas@central.ntua.gr}
	\affiliation{Physics Department, National Technical University of Athens, 15780 Zografou Campus, Athens, Greece}

\author{Eleftherios Papantonopoulos}
\email{lpapa@central.ntua.gr} \affiliation{Physics Department, National Technical University of Athens, 15780 Zografou Campus, Athens, Greece}

\vspace{4.5cm}

\begin{abstract}
We study the motion of uncharged particles and photons in the background of a magnetically charged Euler-Heisenberg (EH) black hole (BH) with scalar hair. The spacetime can be asymptotically (A)dS or flat. After investigating particle motions around the BH and the behavior of the effective potential of the particle radial motion, we determine the contribution of the BH parameters to the geodesics. Photons follow null geodesics of an effective geometry induced by the corrections of the EH non-linear electrodynamics. Thus, after determining the effective geometry, we calculate the shadow of the BH. Upon comparing the theoretically calculated BH shadow with the images of the shadows of M87* and Sgr A* obtained by the Event Horizon Telescope collaboration, we impose constraints on the BH parameters, namely the scalar hair ($\nu$), the magnetic charge ($Q_{m}$) and the EH parameter ($\alpha$).
\end{abstract}

\maketitle

\flushbottom

\tableofcontents

\section{Introduction}

Einstein's theory of General Relativity (GR) \cite{Einstein:1916vd,Schwarzschild:1916uq,Penrose:1964wq} predicts the existence of black holes (BHs), which are remarkable regions of spacetime characterized by their event horizon, a boundary from which even light cannot escape. According to the theory of GR, BHs are the result of matter undergoing gravitational collapse. Moreover, BHs can potentially bridge the gap between gravity and quantum mechanics \cite{Hawking:1976ra}, which is a longstanding aspiration in physics. Modern observational techniques, such as gravitational wave observations and very-long-baseline interferometry (VLBI), provide an opportunity to test some quantum properties of BHs, as proposed in \cite{Giddings:2017jts, Giddings:2019jwy}. Exploring BHs is undoubtedly very important in gaining deeper insights into gravity at energy scales beyond our current reach on Earth.

Magnetically charged BHs have been extensively studied mainly in connection to their stability. In addition to the Maxwell-Einstein theory, string theory predicts generalized magnetically charged Reissner-Nordstr\"om BH solutions as a generalization of the magnetically charged Garfinkle-Horowitz-Strominger BH solutions  \cite{Gibbons:1987ps,Garfinkle:1990qj}. Magnetic monopoles are closely connected to magnetically charged BHs. In \cite{Lee:1991qs}, it was shown that a  magnetic monopole may be generated as a classical instability in a magnetically charged Reissner-Nordstr\"om solution. The magnetic monopoles are hypothetical particles predicted in string theories \cite{Wen:1985qj} and Grand Unified Theories (GUT) \cite{tHooft:1974kcl,Polyakov:1974ek}. It has also been proven in \cite{Mavromatos:2016mnj,Mavromatos:2018drr} that gravitational monopoles \cite{Barriola:1989hx} may yield self-gravitating magnetic monopoles. Nevertheless, magnetic monopoles have not been observed in nature. Dirac has shown that the existence of a magnetic monopole in the Universe implies the quantization of the electric charge \cite{Dirac:1931kp}. Dyonic black holes have also been investigated \cite{Priyadarshinee:2021rch,Mahapatra:2020wym}.

In the theory of GR, the process of continuous gravitational collapse is believed to result in the presence of singularities, which are considered undesirable \cite{Penrose:1964wq,Hawking:1970zqf,Senovilla:1998oua}. Although the cosmic censorship conjecture states that singularities should always be hidden behind the event horizons of BHs \cite{Penrose:1969pc, Wald:1997wa}, it is still preferable to discover solutions that entirely avoid singularities. As a result, significant efforts have been dedicated to the search for regular BH solutions. This endeavor started with the pioneering work of Bardeen \cite{J.Bardeen}, where regular BHs were proposed in connection to magnetic charges. Additionally, Ayón-Beato and García \cite{Ayon-Beato:2000mjt} have proven that the Bardeen spacetime is a regular solution of the Einstein gravitational equations coupled to non-linear electrodynamics (NLED).

One of the first developed NLED theories is the Born-Infeld (BI) electrodynamics \cite{Born:1934gh}, which was initially introduced as a classical solution to address the electron self-energy problem. While NLED models are useful in the search for regular BH solutions, they also hold significance when considering loop corrections to quantum electrodynamics (QED). These corrections become necessary when attempting to describe the strong-field regime of the electromagnetic field, such as when addressing the self-energy problem of a point charged particle. Notably, Euler-Heisenberg (EH) NLED serves as a particularly relevant example \cite{Heisenberg:1936nmg}. Its action arises from the effective action of QED, accounting for one-loop corrections, and introduces two relativistic invariants constructed from the electromagnetic field-strength tensor. Another noteworthy model is the Bronnikov NLED \cite{Bronnikov:2000vy}, which only incorporates one of the aforementioned relativistic invariants. In this model, regular BH solutions exist as long as they carry only magnetic charge and not electric charge. Moreover, it has been established that NLED models frequently emerge as low-energy approximations in different string theories or supersymmetric theories. For instance, the BI NLED is recognized as the effective description of world-volume gauge fields on D-branes in the low-energy limit \cite{Fradkin:1985qd,Tseytlin:1986ti}. Similarly, the EH NLED serves as an effective description in the low-energy limit of the BI NLED, accurately approximating the supersymmetric action of particles with spin-1/2 and spin-0 that are minimally coupled \cite{Bern:1993tz,Dunne:2004nc,Jacobson:2018kso}. A way to detect the effect of the EH theory has been proposed in \cite{Brodin:2001zz}, while some other fetures of EH NLED has been assessed in \cite{Kruglov:2007bh, Kruglov:2017ymn}.

The coupling of the EH Lagrangian to the Ricci scalar via the volume element allowed the finding of BHs. One of the BH solutions to the EH theory has been derived in \cite{Yajima:2000kw}, where analytical solutions were obtained for the magnetically and electrically charged cases and also for dyons.  Electrically charged BHs were found in \cite{Ruffini:2013hia} and \cite{Amaro:2020xro}, and geodesics around the EH BH have been studied in \cite{Amaro:2020xro}. Also, in \cite{Chen:2022tbb}, motions of charged particles around the EH AdS BH were studied. A study of the thermodynamics of these BHs was performed  in \cite{Magos:2020ykt, Dai:2022mko}, while the stability of these BHs, calculating the  quasinormal modes, was investigated in \cite{Breton:2021mju}. Rotating BHs were found in \cite{Breton:2019arv, Breton:2022fch}, while the EH theory in modified gravity theories was studied and BH solutions were analyzed in \cite{Stefanov:2007bn, Guerrero:2020uhn, Nashed:2021ctg}.

A self-interacting scalar field minimally coupled to gravity was introduced in the Lagrangian of the EH theory in \cite{Karakasis:2022xzm} and hairy singular BHs were found, which can be considered as generalization of the EH BHs of \cite{Yajima:2000kw} and the hairy BHs of \cite{Gonzalez:2013aca}. BHs with scalar hair in the scenario of NLED have also been discussed in \cite{Barrientos:2016ubi}. The hairy BH solution of \cite{Karakasis:2022xzm} is characterized by five parameters, the BH mass, the EH parameter, the magnetic charge of the BH, the scalar charge of the scalar field and the cosmological constant. A magnetically charged hairy BH is obtained when the EH parameter vanishes, while when the scalar charge vanishes we get the BH solution of \cite{Yajima:2000kw}. The hairy BH solution of \cite{Gonzalez:2013aca} is recovered when both the  EH parameter and the magnetic charge vanish. It was found that a scalar field dresses the BH of \cite{Karakasis:2022xzm} with secondary scalar hair, and the size of the BH decreases as the magnitude of the scalar field increases, while it increases as the gravitational mass increases. The thermodynamical properties of the solution have also been discussed, as well as the energy conditions. The thermodynamical properties of the found solution are interesting, as the scalar field gains entropy for the BH by the addition of a linear term in the entropy, and hence, the hairy BHs are thermodynamically preferred.

The properties of charged BHs can be revealed through the study of geodesics around these objects. By exploring the geodesics and solving the geodesic equations, we can get information about the structure of a BH. This study allows us to determine whether uncharged test particles outside the event horizon of a BH follow stable circular orbits. The motion of charged particles in the Reissner-Nordstr\"om spacetime has been discussed in \cite{Olivares:2011xb}. Geodesics of the magnetically charged Garfinkle-Horowitz-Strominger stringy BH \cite{Garfinkle:1990qj} have been analyzed in \cite{Kuniyal:2015uta, Soroushfar:2016yea}, revealing the absence of stable circular orbits outside the event horizon for massless test particles. In \cite{Gonzalez:2017kxt}, the motion of massive particles with electric and magnetic charges in the background of a magnetically charged Garfinkle-Horowitz-Strominger stringy BH was investigated. Bound and unbound orbits were discovered for critical values of the BH magnetic charge and the magnetic charge of the test particle. Two observables, the periastron shift and the Lense-Thirring effect, were also studied in \cite{Gonzalez:2017kxt}. Additionally, all the trajectories depended on the electric and magnetic charges of the test particles. Furthermore, the trajectories of the hairy BHs \cite{Gonzalez:2013aca} have been discussed in \cite{Gonzalez:2015jna}, where it has been found that particles complete an oscillation in an angle less than $2\pi$. Also, the geodesics of test particles around rotating BHs were calculated in \cite{Gonzalez:2020kbv, Gonzalez:2019xfr, Gonzalez:2018lfs}.

All the possible trajectories around the EH BHs  were studied in \cite{Amaro:2020xro}. The geodesic equations were analytically integrated and the corresponding effective potentials were analysed. It was found that the stable and unstable circular orbits of massive test particles are barely modified due to the EH NLED contribution. For the photon trajectories it was found that the vacuum polarization effect is significant due to the non-linear EH electromagnetic field. The influence of the angular momentum and charge of the particle around the EH AdS BH on the Lyapunov exponent has been studied in \cite{Chen:2022tbb}. For the specific parameters of the BH, the spatial regions, where the chaos bound is violated, were found by fixing the particle charge and changing its angular momentum.

From an observational perspective, BHs are observed in a wide range of astrophysical environments, and there is a considerable amount of direct and indirect evidence supporting the existence of supermassive black holes (SMBHs) with masses reaching up to $10^{10}M_{\odot}$ (solar mass). It is widely accepted that SMBHs are located at the centers of most sufficiently massive galaxies, including our own \cite{Lynden-Bell:1969gsv,Kormendy:1995er}, and that they play a crucial role in powering active galactic nuclei – intensely luminous central regions of galaxies that often surpass the brightness of the rest of the galaxy itself. For a comprehensive review on astrophysical BHs, refer to \cite{Bambi:2019xzp}.

The photon sphere, where photons follow unstable circular paths, and the gravitational bending of light result in the formation of a ``shadow" around an accreting BH. This shadow is accompanied by a bright emission ring \cite{Luminet:1979nyg,Lu:2014zja,Cunha:2018acu,Gralla:2019xty,Narayan:2019imo}. The BH shadow corresponds to the interior of a critical curve, which is a curve that describes the path of light rays that asymptotically approach a bound photon orbit when traced back from a distant observer to the BH. Essentially, the BH shadow is a closed curve on the sky that separates capture orbits from scattering orbits. For further study on BH shadows, see \cite{Chen:2022scf,Wang:2022kvg,Perlick:2021aok,Dokuchaev:2019jqq}, while remarks on photon surfaces might be found in \cite{Claudel:2000yi, Adler:2022qtb}. VLBI surveys, which involve collecting signals from different radio sources using multiple radio telescopes on Earth, effectively create a virtual large telescope with a size determined by the maximum separation between the individual telescopes. These surveys are expected to be capable of detecting the shadows cast by SMBHs \cite{Falcke:1999pj}.

VLBI has made significant progress in studying BH shadows through the collaborative efforts of the Event Horizon Telescope (EHT). The EHT is a global array of radio telescopes that operates at a wavelength of 1.3mm and possesses a theoretical diffraction-limit resolution of 25 $\mu$as, limited only by diffraction \cite{Doeleman:2009te}. In April 2019, the EHT collaboration made groundbreaking announcements regarding the detection of the shadow of M87*, the SMBH situated at the center of the elliptical galaxy Messier 87. These announcements were made through a series of influential papers \cite{EventHorizonTelescope:2019dse,EventHorizonTelescope:2019uob,EventHorizonTelescope:2019jan,EventHorizonTelescope:2019ths,EventHorizonTelescope:2019pgp,EventHorizonTelescope:2019ggy,EventHorizonTelescope:2021bee,EventHorizonTelescope:2021srq}. The rotating nature of the SMBH M87* has been investigated in \cite{Bambi:2019tjh}, while the deviations between the observational data for the SMBH M87* and the Schwarzschild model are clarified in \cite{Tian:2019yhn}. Additionally, in May 2022, the detection of the shadow of Sagittarius A* (Sgr A*), the SMBH located at the center of the Milky Way galaxy, was announced by the EHT collaboration. The analysed data was presented in a series of scientific articles \cite{EventHorizonTelescope:2022wkp,EventHorizonTelescope:2022apq,EventHorizonTelescope:2022wok,EventHorizonTelescope:2022exc,EventHorizonTelescope:2022urf,EventHorizonTelescope:2022xqj}.

BH shadows hold immense promise as a testing ground for exploring deviations from GR, including potential violations of the no-hair theorem \cite{Johannsen:2010ru,Broderick:2013rlq,Johannsen:2015hib,Johannsen:2015mdd,Psaltis:2018xkc,Yan:2019hxx,Khodadi:2020jij,Glampedakis:2023eek}. Some perturbative deviations from GR have been examined in \cite{Rummel:2019ads}. Following the EHT's groundbreaking detection, numerous studies have investigated the potential to extract valuable information from the shadows of M87* and Sgr A*. The observations of M87* are consistent with the predictions of the Schwarzschild-model of GR. However, the existence of systematic uncertainties in the EHT observations allows for the potential testing of alternative theories of gravity. This can be accomplished by comparing the observed BH image with the theoretical predictions for BH shadows derived from alternative gravity theories. In fact, according to the results of \cite{Narayan:2019imo}, where a simplified model of a spherically accreting BH has been investigated, the size of the observed shadow is a signature of the spacetime geometry and it is hardly influenced by accretion details. Thus, by comparing the shadows of M87* and Sgr A* with theoretical predictions, the authors of papers \cite{Khodadi:2020jij,Allahyari:2019jqz,Jusufi:2021fek,Shaikh:2021cvl,Bogush:2022hop,Chen:2022nbb,Chen:2022lct,Kuang:2022xjp,Uniyal:2022vdu,Vagnozzi:2022tba,Pantig:2022ely,Kuang:2022ojj,Rayimbaev:2022hca,Tang:2022hsu,Banerjee:2022iok,Das:2022tqi,Mustafa:2022xod,Mandal:2022oma,Li:2022eue,Lobos:2022jsz,Pantig:2022qak,Pantig:2022gih,Kumaran:2022soh,Atamurotov:2022nim,Sengo:2022jif,Ghosh:2022gka,Hu:2022lek,Atamurotov:2022knb,Sau:2022afl,Anjum:2023axh,Kalita:2023xlu,Ovgun:2023ego,Zubair:2023krl,Uniyal:2023inx,Xavier:2023exm,Ramadhan:2023ogm,Jana:2023sil,Ghorani:2023hkm,Yan:2023pxj,Meng:2023wgi,Gonzalez:2023rsd,Sahoo:2023czj,Nozari:2023flq,Zubair:2023cep,Heydari-Fard:2023ent} have established constraints on various modified GR theories. Particularly in \cite{Khodadi:2020jij}, the authors set constraints on the amount of scalar hair carried by two hairy BHs, while in \cite{Allahyari:2019jqz} constraints have been imposed on the parameters of the EH model and Bronnikov NLED model.

In this work we will investigate the properties of the geometry of a magnetically charged EH BH with scalar hair, first presented in \cite{Karakasis:2022xzm}, by studying geodesic motions in this geometry and the shadow of the BH. We will use the Hamilton-Jacobi (HJ) formalism, which is connected to the Euler-Lagrange formalism, in order to obtain the equations of motion. We shall first discuss the time-like geodesics for asymptotically flat, dS and AdS spacetimes and then the null geodesics for asymptotically flat spacetimes. We aim to determine the contribution of the scalar charge, the magnetic charge of the BH and the EH parameter to the spacetime curvature. Afterwards, we will discuss the motion of light in this geometry. NLED theories predict self-interacting properties for photons, and therefore, the light does not follow the null geodesics of these geometries. Nevertheless, in the papers \cite{Novello:1999pg,Obukhov:2002xa,Stuchlik:2019uvf}, it has been proven, following three different ways, that photons move along the null geodesics of an \emph{effective geometry}. This phenomenon will be discussed in detail later on in this article. Thus, we will determine the possible trajectories of light and the BH shadow. Finally, constraints will be imposed on the parameters of the BH by comparing our theoretical predictions for the BH shadow with the images of the shadows of M87* and Sgr A* released by EHT collaboration.

The work is organized as follows: In Section \ref{sec2}, we briefly describe the model of \cite{Karakasis:2022xzm}. The HJ formalism is presented in Section \ref{sec3}. In Section \ref{sec4}, we determine the particle equations of motion by utilizing the HJ formalism, numerically calculate the time-like and null geodesics, and comment on the contribution of the BH parameters to the particle motion. Calculations of the photon trajectories in the vicinity of the BH and the BH shadow occur in Section \ref{sec5}. In Section \ref{sec6}, the BH shadow is compared with the shadows of M87* and Sgr A* obtained by the EHT, in order for constraints to be derived for the parameters of the BH. Finally, we conclude our work and point out possible future work in Section \ref{sec7}.

\section{The Model} \label{sec2}
%\pagenumbering{arabic}

In this section, we will briefly discuss the BH solution of \cite{Karakasis:2022xzm}, which is a magnetically charged EH BH with scalar hair. This BH arises as a solution of the Einstein-Euler-Heisenberg theory minimally coupled to a scalar field. The action of this theory reads
\begin{equation}
    S=\int d^{4}x \sqrt{-g}\mathcal{L}=\int d^{4}x \sqrt{-g}\left(\frac{R}{16\pi G}-\frac{1}{2}\partial^{\mu}\varphi\partial_{\mu}\varphi-V(\varphi)+\frac{L(P,Q)}{16\pi}\right)~,
    \label{Action}
\end{equation}
where
\begin{equation}
    L(P,Q)=-P+\alpha P^{2}+\beta Q^{2}\label{L}~,
\end{equation}
and $\mathcal{L}$ denotes the Lagrangian density of the
theory, $g=det\left(g_{\mu\nu}\right)$, $R$ is the Ricci scalar for $g_{\mu\nu}$~\footnote{Our conventions and definitions throughout this work are: $(-,+,+,+)$ for the signature of the metric, the Riemann tensor is defined as
$R^\lambda_{\,\,\,\,\mu \nu \sigma} = \partial_\nu \, \Gamma^\lambda_{\,\,\mu\sigma} + \Gamma^\rho_{\,\, \mu\sigma} \, \Gamma^\lambda_{\,\, \rho\nu} - (\nu \leftrightarrow \sigma)$,
and the Ricci tensor and scalar are given by  $R_{\nu\alpha} = R^\lambda_{\,\,\,\,\nu \lambda \alpha}$ and $R= g^{\mu\nu}\, R_{\mu\nu}$ respectively.}, $P=F_{\mu\nu}F^{\mu\nu}$ and $Q=\epsilon_{\mu\nu\rho\sigma}F^{\mu\nu}F^{\rho\sigma}$. Additionally, $F_{\mu\nu}=\partial_{\mu}A_{\nu}-\partial_{\nu}A_{\mu}$ is the Faraday tensor (strength tensor), $\epsilon_{\mu\nu\rho\sigma}$ is the Levi-Civita tensor obeying $\epsilon_{\mu\nu\rho\sigma}\epsilon^{\mu\nu\rho\sigma}=-4!$ and $G$ is the Gravitational constant. We may define the actions that refer to the scalar field and electromagnetic (E/M) field as
\begin{equation}
    S_{\varphi}=\int d^{4}x \sqrt{-g}\left(-\frac{1}{2}\partial^{\mu}\varphi\partial_{\mu}\varphi-V(\varphi)\right)~,
    \label{ScalarAction}
\end{equation}
and
\begin{equation}
    S_{EM}=\int d^{4}x \sqrt{-g}\frac{L(P,Q)}{16\pi}~.
    \label{EMAction}
\end{equation}
The Einstein field equations can be determined by considering variations of the action (\ref{Action}) with respect to the metric tensor

\begin{equation}
    R_{\mu\nu}-\frac{1}{2}g_{\mu\nu}R=8\pi G T_{\mu\nu}~,
    \label{EinsteinEqs}
\end{equation}
where
\begin{equation}
    T_{\mu\nu}=T^{\varphi}_{\mu\nu}+T^{EM}_{\mu\nu}~,
\end{equation}
\begin{equation}
    T^{\varphi}_{\mu\nu}=-\frac{2}{\sqrt{-g}}\frac{\delta S_{\varphi}}{\delta g^{\mu\nu}}=\partial_{\mu}\varphi\partial_{\nu}\varphi-\frac{1}{2}g_{\mu\nu}g^{\rho\sigma}\partial_{\rho}\varphi\partial_{\sigma}\varphi-g_{\mu\nu}V(\varphi)
\end{equation}
and
\begin{equation}
    T^{EM}_{\mu\nu}=-\frac{2}{\sqrt{-g}}\frac{\delta S_{EM}}{\delta g^{\mu\nu}}=\frac{1}{16\pi}g_{\mu\nu}\left(-P+\alpha P^{2}+\beta Q^{2}\right)+\frac{1}{4\pi}F_{\mu\rho}F_{\nu}^{\;\rho}-\frac{1}{2\pi}\alpha P F_{\mu\rho}F_{\nu}^{\;\rho}-\frac{1}{\pi}\beta Q\epsilon_{\mu\rho\kappa\lambda}F_{\nu}^{\;\rho}F^{\kappa\lambda}~.\label{S-EtensorE/M}
\end{equation}
Upon considering variations of the action (\ref{Action}) with respect to the E/M field and the scalar field, we obtain the equations of motion of the E/M field and the scalar field respectively
\begin{align}
    \label{EMEOM}
    &\nabla_{\mu}\left(L_{P}F^{\mu\nu}+L_{Q}\epsilon^{\mu\nu\rho\sigma}F_{\rho\sigma}\right)=0~,\\
    \label{ScalarEOM}&\square\varphi=g^{\mu\nu}\partial_{\mu}\partial_{\nu}\varphi+\frac{1}{\sqrt{-g}}\partial_{\mu}\left(\sqrt{-g}g^{\mu\nu}\right)\partial_{\nu}\varphi=\frac{dV(\varphi)}{d\varphi}~,
\end{align}
where $L_{X}$ denotes the partial derivative of $L(X)$ with respect to $X$, $L_{XX}$ denotes the second partial derivative, $\square\varphi\equiv\nabla^{\mu}\nabla_{\mu}\varphi$ and $\nabla_{\mu}$ is the standard covariant derivative. We introduce a static and spherically symmetric ansatz for the metric
\begin{equation}
    ds^{2}=-b(r)dt^{2}+\frac{dr^{2}}{b(r)}+w^{2}(r)(d\theta^{2}+\sin^{2}\theta d\phi^{2})~,
    \label{Metric}
\end{equation}
and a gauge field of the form
\begin{equation}
    A_{\mu}=\left(\mathcal{A}(r),0,0,Q_{m}\cos \theta\right)~,
    \label{GaugeField}
\end{equation}
where $Q_{m}$ is the magnetic charge of the BH. The scalars $P$, $Q$ read
\begin{equation}
    P=\frac{2 Q_{m}^{2}}{w^{4}(r)}-2\mathcal{A}'^{2}(r)~,~Q=-\frac{8Q_{m}\mathcal{A}'(r)}{w^{2}(r)}~,\label{P}
\end{equation}
where the prime indicates radial derivative. Looking for magnetically charged solutions (not dyons, i.e. both electrically and magnetically charged particles), we impose $\mathcal{A}(r)=0$, which yields $Q=0$. By solving the Eqs. (\ref{EinsteinEqs}), (\ref{EMEOM}) and (\ref{ScalarEOM}), we determine the scalar field that supports the hairy structure and the scalar potential
\begin{equation}
    \varphi(r)=\frac{1}{\sqrt{16\pi G}}\ln\left(1+\frac{\nu}{r}\right)~,
\end{equation}
\begin{equation}
    \begin{split}
    V(\varphi)&=\frac{1}{8\pi G}\frac{1}{3\nu^{8}}\bigg[\nu^{8}\Lambda_{\text{eff}}\left(\cosh\left(\sqrt{16\pi G}\varphi\right)+2\right)-36GM\nu^{5}\left(\sqrt{16\pi G}\varphi\left(\cosh\left(\sqrt{16\pi G}\varphi\right)+2\right)-3\sinh\left(\sqrt{16\pi G}\varphi\right)\right)\\
    &+6\nu^{4}GQ_{m}^{2}\left(64\pi G\varphi^{2}+4(8\pi G\varphi^{2}+2)\cosh\left(\sqrt{16\pi G}\varphi\right)-12\sqrt{16\pi G}\varphi\sinh\left(\sqrt{16\pi G}\varphi\right)+\cosh\left(2\sqrt{16\pi G}\varphi\right)-9\right)\\
    &-4G\alpha Q_{m}^{4}\Big(288(8\pi G)\varphi^{2}+2\left(72(8\pi G)\varphi^{2}+71\right)\cosh\left(\sqrt{16\pi G}\varphi\right)-432\sqrt{16\pi G}\varphi\sinh\left(\sqrt{16\pi G}\varphi\right)\\
    &+100\cosh\left(2\sqrt{16\pi G}\varphi\right)-14\cosh\left(3\sqrt{16\pi G}\varphi\right)+\cosh\left(4\sqrt{16\pi G}\varphi\right)-229\Big)\bigg]~,
    \end{split} \label{ScalarPotential}
\end{equation}
while the metric functions $b(r)$ and $w(r)$ read
\begin{equation}
    \label{MetricFunction}
    \begin{split}
    b(r)=& c_{1}r(\nu+r)+\frac{\left(2r-c_{2}\right)(\nu+2r)-4GQ_{m}^{2}}{\nu^{2}}+\frac{8\alpha GQ_{m}^{4}\left(-\nu^{2}+12r^{2}+12\nu r\right)\left(\nu^{2}+3r^{2}+3\nu r\right)}{3\nu^{6}r^{2} (\nu+r)^{2}}+\frac{2}{\nu^{8}}\ln\left(\frac{r}{\nu+r}\right)\times\\
    &\left(-\nu^{5}r(c_{2}+\nu)(\nu+r)-2GQ_{m}^{2}r(\nu+r)\left(\nu^{4}-24\alpha Q_{m}^{2}\right)\ln\left(\frac{r}{\nu+r}\right)+48\alpha\nu GQ_{m}^{4}(\nu+2r)-2\nu^{5}GQ_{m}^{2}(\nu+2r)\right),
    \end{split}
\end{equation}
\begin{equation}
    w(r)=\sqrt{r^{2}+\nu r}~.
\end{equation}
This is the magnetically charged BH solution with scalar hair first presented in \cite{Karakasis:2022xzm}. One can verify that the extended Maxwell equations (\ref{EMEOM}) are identically satisfied upon considering a gauge field of the form (\ref{GaugeField}) with $\mathcal{A}(r)=0$. Setting 
\begin{equation}
    c_{1}=-\frac{4}{\nu^{2}}-\frac{\Lambda_{\text{eff}}}{3}~,~c_{2}=6GM-\nu~,
\end{equation}
the metric function $b(r)$ asymptotically reads
\begin{equation}
    b(r\to\infty)=-\frac{\Lambda_{\text{eff}}}{3}r^{2}-\frac{\Lambda_{\text{eff}}}{3}\nu r+1-\frac{2GM}{r}+G\frac{Qm^{2}+\nu M}{r^{2}}+\mathcal{O}\left(\left(\frac{1}{r}\right)^{3}\right)~,
\end{equation}
where $M$ is the mass of the BH, $Q_{m}$ is the magnetic charge introduced in Eq. (\ref{GaugeField}) and $\nu$ can be identified as the scalar charge of the scalar field, since it controls the $\mathcal{O}(r^{-1})$-term in the expansion of the scalar field
\begin{equation}
    \varphi(r\to\infty)=\frac{\nu}{\sqrt{16\pi G}r}+\mathcal{O}\left(\left(\frac{1}{r}\right)^{2}\right)~.
\end{equation}
Additionally, $\Lambda_{\text{eff}}$ plays the role of an effective cosmological constant. In fact, for vanishing scalar field, i.e. $\nu\to 0$, the scalar potential reads
\begin{equation}
    V(\varphi=0)=\frac{\Lambda_{\text{eff}}}{8\pi G}~,
\end{equation}
and the metric function is
\begin{equation}
    \label{EEHmetric}
    b(r)=-\frac{\Lambda_{\text{eff}}}{3}r^{2}+1-\frac{2GM}{r}+G\frac{Q_{m}^{2}}{r^{2}}-G\frac{2\alpha Q_{m}^{4}}{5r^{6}}~.
\end{equation}

The metric function (\ref{EEHmetric}), for $\Lambda_{\text{eff}}=0$, corresponds to the BH solution of the Einstein-Euler-Heisenberg theory, initially obtained in \cite{Yajima:2000kw} and also presented in \cite{Allahyari:2019jqz}. From now on we will use Planck units with $c=\hbar=G=1$. Additionally, later in the article, we shall work in units of mass by setting $M=1$, or equivalently by rescaling all dimensionful quantities by the appropriate power of $M$, such as $r/M\to r$, $Q_{m}/M\to Q_{m}$, $\nu/M\to\nu$, $\alpha/M^{2}\to\alpha$, $\Lambda_{\text{eff}}\cdot M^{2}\to\Lambda_{\text{eff}}$ and $L/M^{2}\to L$, where $L$ denotes the magnitude of the angular momentum of a test particle. The spacetime (\ref{MetricFunction}) is characterized by five constants, the BH mass ($M$), the scalar charge ($\nu$), the magnetic charge ($Q_{m}$), the EH parameter ($\alpha$) and the cosmological constant ($\Lambda_{\text{eff}}$). We aim to determine the effects of the parameters of the matter fields on both the geodesic motion of neutral particles and the BH shadow.

The vectors $K = \partial_{t}$ and $R = \partial_{\phi}$ constitute Killing vectors in our model
\begin{eqnarray}
&&K^{\mu} = (1,0,0,0) \; \text{and} \; K_{\mu} = (-b(r),0,0,0)~,\label{Kkilling}\\
&&R^{\mu} = (0,0,0,1) \; \text{and} \; R_{\mu} = (0,0,0,w^{2}(r)\sin^{2}\theta)~.
\end{eqnarray}
Without loss of generality, we consider motion in the plane $\theta = \pi/2$ and obtain $R_{\mu} = (0,0,0,w^{2}(r))$. The above Killing vectors are linked with the conservation of the energy and the magnitude of the total angular momentum of the test particle
\begin{eqnarray}
&&E=-K^{\mu}p_{\mu}=\text{constant}~, \label{E}\\
&&J = R^{\mu}p_{\mu}=\text{constant}~, \label{J}
\end{eqnarray}
where $p^{\mu}$ is the four-momentum of the test particle, which satisfies the geodesic condition. In \cite{Karakasis:2022xzm} it has been shown that the examined BH may have more than one horizons. We seek for geodesics and observables outside the event horizon, where the metric functions are regular. However, in Section \ref{sec5}, we will discuss the effect of the existence of three horizons on the BH shadow. Finally, it is important to mention that the scalar potential given in Eq. (\ref{ScalarPotential}) is well-defined. To be more precise, it seems to contain the integration constants, namely the BH mass ($M$), the scalar charge ($\nu$) and the BH magnetic charge ($Q_{m}$), however, by defining the constant parameters $\chi\equiv M/\nu^3$ and $\psi\equiv Q_m^2/\nu^4$, it becomes obvious that the potential does not explicitly depend on the integration constants. Instead, it depends on the constant ratios $\chi$ and $\psi$, as well as the cosmological constant and the EH parameter. In Section \ref{sec6}, constraints will be imposed on the BH magnetic charge and the scalar charge. Since mass units will be employed, where $M=1$, the aforementioned constraints will refer to the parameters $\chi$, $\psi$.

\section{Hamilton-Jacobi formalism }\label{sec3}

To obtain the equations of geodesic motion, we will apply the HJ formalism to our model. In this section we discuss how the HJ formalism is connected to the Euler-Lagrange and Hamilton formalisms. From the Lagrangian formulation one can construct the Hamiltonian and then, by performing a canonical transformation, the HJ equation can be obtained. In the Lagrangian formalism, for the determination of geodesics, we define a Lagrangian from the metric, and the corresponding action
\begin{equation}
    \frac{2}{m}\mathcal{L} = g_{\mu\nu}\dot{x}^{\mu}\dot{x}^{\nu}~,~\mathcal{S}(x^{\mu},\dot{x}^{\mu})=\int_{\tau_{i}}^{\tau_{f}}\mathcal{L}(x^{\mu},\dot{x}^{\mu},\tau)d\tau~,\label{ActionGeodesics}
\end{equation}
where the dot indicates derivative with respect to an affine parameter $\tau$, which will be the proper time for massive particles, and $m$ denotes the mass of the test particle. The Hamiltonian can be obtained by performing a Legendre transformation on Lagrangian
\begin{equation}
\mathcal{H}=p_{\mu} \dot{x}^{\mu}-\mathcal{L}=\frac{1}{2m} g^{\mu \nu} p_{\mu} p_{\nu}=-\frac{m}{2}~,
\label{Hamiltonian}
\end{equation}
where $\dot{x}^{\mu}$ was expressed in terms of the conjugate momentum $p_{\mu}$. In Eq. (\ref{Hamiltonian}), we have considered normalised momentum ($ g_{\mu\nu} p^{\mu} p^{\nu} + m^{2} = 0$). Equation (\ref{Hamiltonian}) is valid for the case of a neutral test particle. In the case of a test particle with charge $q$, we have to substitute the momentum $p^{\mu}$ by the expression $p^{\mu} - q A^{\mu}$ into Eq. (\ref{Hamiltonian}). Upon substituting the Lagrangian given in Eq. (\ref{Hamiltonian}) into Eq. (\ref{ActionGeodesics}), the action yields
\begin{equation}
    \mathcal{S}(x^{\mu},p_{\mu})=\int_{\tau_{i}}^{\tau_{f}}p_{\mu}(\tau)dx^{\mu}(\tau)-\int_{\tau_{i}}^{\tau_{f}}\mathcal{H}(x^{\mu},p_{\mu},\tau)d\tau~.
\end{equation}
A canonical transformation, according to which a total derivative ($dF(x^{\mu},p_{\mu},\tau)/d\tau$) is added to the Hamiltonian, does not influence the Hamiltonian dynamical evolution of the physical system. The coordinate transformation $X^{\mu}(x^{\mu},p_{\mu},\tau)$, $P_{\mu}(x^{\mu},p_{\mu},\tau)$ is a canonical transformation if it satisfies the equation:
\begin{equation}
    \label{CanonicalTransf}
    p_{\mu}dx^{\mu}-\mathcal{H}(x^{\mu},p_{\mu},\tau)d\tau-dF(x^{\mu},p_{\mu},\tau)=P_{\mu}dX^{\mu}-K(X^{\mu},P_{\mu},\tau)d\tau~,
\end{equation}
where $K(X^{\mu},P_{\mu},\tau)$ is the Hamiltonian in the new coordinate system. From Eq. (\ref{CanonicalTransf}) it becomes obvious that the canonical transformation implies
\begin{equation}
    \label{FCanTransf1}
    dF=p_{\mu}dx^{\mu}-P_{\mu}dX^{\mu}+(K-\mathcal{H})d\tau~.
\end{equation}
We can introduce a function $F_{1}(x^{\mu},X^{\mu},\tau)$ with
\begin{equation}
    \label{FCanTransf2}
    dF_{1}=\frac{\partial F_{1}}{\partial x^{\mu}}dx^{\mu}+\frac{\partial F_{1}}{\partial X^{\mu}}dX^{\mu}+\frac{\partial F_{1}}{\partial\tau}d\tau~,
\end{equation}
which defines the canonical transformation $p_{\mu}=p_{\mu}(x^{\mu},X^{\mu},\tau)$, $P_{\mu}=P_{\mu}(x^{\mu},X^{\mu},\tau)$. Through the comparison of Eqs. (\ref{FCanTransf1}) and (\ref{FCanTransf2}) we obtain
\begin{equation}
    \label{P2}
    p_{\mu}=\frac{\partial F_{1}}{\partial x^{\mu}}~,~P_{\mu}=-\frac{\partial F_{1}}{\partial X^{\mu}}~.
\end{equation}
The Hamiltonian in the new coordinates reads: $K=\mathcal{H}+\frac{\partial F_{1}}{\partial\tau}$. Interestingly, the Legendre transformation of the function $F_{1}(x^{\mu},X^{\mu},\tau)$ with respect to $X^{\mu}$
\begin{equation}
    \label{F2}
    F_{2}(x^{\mu},P_{\mu},\tau)=X^{\mu}(x^{\mu},P_{\mu},\tau)P_{\mu}+F_{1}(x^{\mu},X^{\mu}(x^{\mu},P_{\mu},\tau),\tau)
\end{equation}
also defines a canonical transformation. One can notice that the conjugate momentum of $X^{\mu}$ in Eq. (\ref{P2}) has the opposite sign from the usual one, and therefore, in the Legendre transformation (\ref{F2}), a sign opposite to the usual is considered. The equations that describe the corresponding canonical transformation can be obtained by integrating by parts the term $P_{\mu} dX^{\mu}$ in Eq. (\ref{FCanTransf1})
\begin{equation}
    dF_{2}(x^{\mu},P_{\mu},\tau)=d(F_{1}+X^{\mu}P_{\mu})=p_{\mu}dx^{\mu}+X^{\mu}dP_{\mu}+(K-\mathcal{H})d\tau~,
\end{equation}
and comparing the above equation with the differential of $F_{2}$
\begin{equation}
    dF_{2}(x^{\mu},P_{\mu},\tau)=\frac{\partial F_{2}}{\partial x^{\mu}}dx^{\mu}+\frac{\partial F_{2}}{\partial P_{\mu}}dP_{\mu}+\frac{\partial F_{2}}{\partial\tau}d\tau~.
\end{equation}
Thus, the canonical transformation $p_{\mu}=p_{\mu}(x^{\mu},P_{\mu},\tau)$, $X^{\mu}=X^{\mu}(x^{\mu},P_{\mu},\tau)$ is described by the relations
\begin{align}
    \label{p}
    p_{\mu}&=\frac{\partial F_{2}}{\partial x^{\mu}}~,\\
    \label{X}
    X^{\mu}&=\frac{\partial F_{2}}{\partial P_{\mu}}~,
\end{align}
and the Hamiltonian in the new coordinates is
\begin{equation}
    \label{KF2}
    K=\mathcal{H}(x^{\mu},p_{\sigma}, \tau)+\frac{\partial F_{2}}{\partial\tau}~.
\end{equation}
According to the HJ formalism, we consider a canonical transformation that implies a vanishing Hamiltonian $K(X^{\mu},P_{\mu},\tau)=0$. Consequently, the Hamilton equations for the new coordinates read
\begin{equation}
    \dot{X}^{\mu}=0~\text{and}~\dot{P}_{\mu}=0~,
\end{equation}
which imply
\begin{equation}
    X^{\mu}=\text{constant}~,~\text{and}~P_{\mu}=\text{constant}~,
\end{equation}
respectively. The above constants are arbitrary, and hence, we can choose the $X^{\mu}$ coordinates to be the initial values of $x^{\mu}$ in the initial time $T$. Additionally, due to the conservation of the energy (\ref{E}) and the magnitude of the total angular momentum (\ref{J}) of the test particle in our case, we can fix $P_{0}=-E$ and $P_{3}=J$. Upon substituting Eq. (\ref{p}) into Eq. (\ref{KF2}) and imposing $K=0$, we obtain
\begin{equation}
    \mathcal{H}\left(x^{\mu},\frac{\partial F_{2}}{\partial x^{\mu}},\tau\right)+\frac{\partial F_{2}}{\partial\tau}=0~,
\end{equation}
which is the HJ equation. Renaming the function $F_{2}(x^{\mu},P_{\mu},\tau)$ as $S(x^{\mu},P_{\mu},\tau)$ and considering the Hamiltonian (\ref{Hamiltonian}) and Eq. (\ref{p}), the HJ equation in our case implies
\begin{equation}
    \frac{1}{2m} g^{\mu \nu} \frac{\partial S}{\partial x^{\mu}} \frac{\partial S}{\partial x^{\nu}}+\frac{\partial S}{\partial \tau}=0~.
    \label{HJ}
\end{equation}

\section{Geodesic motion of uncharged particles}\label{sec4}

In this section, we aim to investigate the geodesic motion of uncharged particles, both massive and massless, in the geometry of a magnetically charged EH BH with scalar hair, which was initially obtained in \cite{Karakasis:2022xzm}. The corresponding metric is described by the Eq. (\ref{Metric}). First of all, the geodesic equations of motion, both for radial and angular motion, will be obtained by utilizing the HJ formalism presented in the previous section. Afterwards, the time-like and null geodesics will be explored in detail. To be more precise, we will determine every possible geodesic trajectory for the asymptotically (A)dS and flat cases by investigating an effective potential, which will be defined by the equations of motion. The impact of the cosmological constant $\Lambda_{\text{eff}}$ on the results will be explored. Then, the contribution of the BH parameters to the geodesics will be determined. The trajectories will be numerically calculated by exploiting the Scipy integrate.quad method using the Python programming language. This method, apart from the integral of the function, returns an estimate of the absolute error in the result, which corresponds to the integration method. The absolute errors in this paper are of the order of $10^{-8}$, and hence, do not affect the validity of the results.

\subsection{Determination of the equations of motion}

First, the geodesic equations will be specified by using the HJ formalism. The HJ Eq. (\ref{HJ}) adapted to our model reads
\begin{equation}
   \label{DiffEqS}
   -\dfrac{1}{b(r)}\left(\frac{\partial S}{\partial t}\right)^{2}+b(r)\left(\frac{\partial S}{\partial r}\right)^{2}+\frac{1}{w^{2}(r)}\left(\frac{\partial S}{\partial\theta}\right)^{2}+\frac{1}{w^{2}(r)\sin^{2}\theta}\left(\frac{\partial S}{\partial\phi}\right)^{2}+m^{2}= 0~.
\end{equation}
This equation refers to massive test particles with mass $m$. Later on in this article, we will explore the massless-particle case by imposing $m=0$ to the equations. The above partial differential equation can be solved by the method of separation of variables. It is convenient to use the following ansatz
\begin{equation}
   \label{AnsatzS}
   S(x^{\mu},P_{\mu},\tau)=-Et+S_{1}(r,P_{\mu})+S_{2}(\theta,P_{\mu})+J\phi+\frac{m}{2}\tau~,
\end{equation}
where $E,J$ can be identified as the conserved energy (\ref{E}) and angular momentum (\ref{J}) of the particle. For convenience, we substitute Eq. (\ref{p}) into Eqs. (\ref{E}) and (\ref{J}), and obtain
\begin{equation}
    E=-\frac{\partial S}{\partial t}~,~J=\frac{\partial S}{\partial \phi}~.
\end{equation}
Considering the ansatz (\ref{AnsatzS}), equation (\ref{DiffEqS}) reads
\begin{equation}
    \label{DiffEqS12}
    -E^{2}\frac{w^{2}(r)}{b(r)}+w^{2}(r)b(r)\left( \frac{\partial S_{1}}{\partial r}\right)^{2}+\left(\frac{\partial S_{2}}{\partial\theta}\right)^{2}+\frac{J^{2}}{\sin^{2}\theta}+w^{2}(r)m^{2}=0~.
\end{equation}
From Eq. (\ref{DiffEqS12}), we can identify a constant of motion
\begin{equation}
  L^{2}=\left(\frac{\partial S_{2}}{\partial\theta}\right)^{2}+\frac{J^{2}}{\sin^{2}\theta}~.
\end{equation}
In particular, without loss of generality, we may consider motion in the $\theta = \pi/2$ plane. Thus, $L$ is equal to the angular momentum $J$ and the Eq. (\ref{DiffEqS12}) yields
\begin{equation}
   -E^{2}\frac{w^{2}(r)}{b(r)}+w^{2}(r)b(r)\left( \frac{\partial S_{1}}{\partial r}\right)^{2}+L^{2}+w^{2}(r)m^{2}=0~.
\end{equation}
Later in the article, when we will investigate radial motion, we will set $L=0$ to the corresponding equations. We can also find a solution for the radial component of $S$
\begin{equation}
   S_{1}(r)=\pm\int\frac{dr}{b(r)}\sqrt{E^{2}-b(r)\left(m^{2}+\frac{L^{2}}{w^{2}(r)}\right)}~.
\end{equation}
The $t$-component of Eq. (\ref{X}) reads
\begin{equation}
   \dfrac{\partial S}{\partial E} = T~,
\end{equation}
where $T$ is the initial time. Consequently, upon taking the derivative of Eq. (\ref{AnsatzS}) with respect to the total energy of the test particle ($E$) and setting the initial time zero ($T=0$), we obtain
\begin{equation}
   t=\pm\int\frac{dr}{b(r)}E\left[E^{2}-b(r)\left(m^{2}+\frac{L^{2}}{w^{2}(r)}\right)\right]^{-1/2}~,
\end{equation}
which leads to the radial velocity in the coordinate-time framework
\begin{equation}
   \label{RadVelCTF}
   \dfrac{dr}{dt}=\pm\frac{b(r)}{E}\sqrt{E^{2}-b(r)\left(m^{2}+\frac{L^{2}}{w^{2}(r)}\right)}~.
\end{equation}
In the case of a massless particle (or a purely radial motion), in order to obtain the radial velocity in the coordinate-time framework, we impose $m=0$ (or $L=0)$, respectively, in the above equation. Considering the areas where $b(r) > 0$, we can define an effective potential
\begin{equation}
   \label{Veff}
   V_{\text{eff}}(r)=\sqrt{b(r)\left(m^{2}+\frac{L^{2}}{w^{2}(r)}\right)}~,
\end{equation}
which determines the geodesic motion of particles. The radial velocity (\ref{RadVelCTF}) is equal to zero at the horizons, where we have $b(r_{\text{H}})=0$. This does not mean that nothing can cross the event horizon, instead this implies that an observer outside the event horizon cannot see a particle passing through it. Upon working within the proper-time framework, we will be convinced for the analytic continuation of the space-time to the inner space of the event horizon. Additionally, considering the external region to the horizon, it is obvious that the radial velocity vanishes and changes direction when $V_{\text{eff}}=E$. Consequently, it is evident that $V_{\text{eff}}$ plays the role of the effective potential of the radial motion. In the subsequent subsections of the article, we will explicitly show that the effective potential determines the various possible motions that a massive or massless particle can execute. Until then, we will calculate the equation of motion for every possible case. In fact, from Eqs. (\ref{E}), (\ref{J}) and (\ref{RadVelCTF}), the equations of every possible geodesic motion follow.

The radial velocity of massive particles in the proper-time framework can be obtained from Eq. (\ref{RadVelCTF}) by considering the relation between the coordinate time and proper time imposed by Eq. (\ref{E}). The relation between the coordinate time and proper time in the case of massive particles reads
\begin{equation}
    \label{dtdtau}
    \frac{dt}{d\tau}=\frac{E}{mb(r)}~,
\end{equation}
and the radial velocity of massive particles in the proper-time framework is
\begin{align}
    \label{RadVelPTF}
    \dot{r}&=\frac{dr}{d\tau}=\frac{E}{mb(r)} \frac{dr}{dt}\Rightarrow\nonumber \\
    \dot{r}^{2}&=\frac{1}{m^{2}}\left(E^{2}-V_{\text{eff}}^{2}(r)\right)~.
\end{align}
Furthermore, upon considering the angular velocity of massive particles in the proper-time framework, which is yielded by Eq. (\ref{J})
\begin{equation}
    \label{dphidtau}
    \frac{d\phi}{d\tau}=\frac{L}{mw^{2}(r)}~,
\end{equation}
the trajectories of the angular motion, i.e. $\phi(r)$, can be determined as follows
\begin{eqnarray}
&&\dot{r}=\frac{dr}{d\phi}\dfrac{d\phi}{d\tau}=\frac{L}{mw^{2}(r)}\frac{dr}{d\phi}\Rightarrow\frac{dr}{d\phi}=\pm\frac{w^{2}(r)}{L}  \sqrt{E^{2} -  V_{\text{eff}}^{2}(r)}\Rightarrow\nonumber\\
&&\phi(r)=\pm L \int_{r_{0}}^{r}\frac{dr'}{w^{2}(r')\sqrt{E^{2} -  V_{\text{eff}}^{2}(r')}}~,\label{phi}
\end{eqnarray}
where we have considered a zero initial azimuthal angle. The circular orbits of test particles arise as special cases of the angular motion. Unstable circular orbits correspond to  maxima of the effective potential, while stable circular orbits correspond to minima of the effective potential. Therefore, the radius $r=r_{c}$ of the circular orbits nullifies the radial derivative of the effective potential
\begin{equation}
   \label{CircOrbCondit}
   \frac{dV_{\text{eff}}(r)}{dr}\bigg|_{r=r_{c}}=0\Rightarrow b'(r_{c})\left(m^{2}+\frac{L_{c}^{2}}{w^{2}(r_{c})}\right)-2w(r_{c})w'(r_{c})b(r_{c})\frac{L_{c}^{2}}{w^{4}(r_{c})}= 0~.
\end{equation}
In our case, due to the complexity of the metric functions, the radii of the circular trajectories cannot be exactly calculated. However, we can determine the exact periods of revolution for both stable and unstable circular orbits with respect to the proper time and coordinate time. First of all, we may solve the Eq. (\ref{CircOrbCondit}) with respect to the particle's angular momentum
\begin{equation}
   \label{Lc}
   L_{c}=mw^{2}(r_{c})\sqrt{\frac{b'(r_{c})}{2w(r_{c})w'(r_{c})b(r_{c})-b'(r_{c})w^{2}(r_{c})}}~.
\end{equation}
The Eq. (\ref{CircOrbCondit}) can be solved with respect to the angular momentum only in the case of massive particles. The periods of the circular orbits of massless particles will be calculated shortly. Equation (\ref{RadVelPTF}) implies that the condition for a vanishing radial velocity reads $E=V_{\text{eff}}(r_{c})$. Thus, the total energy of a massive test particle that executes a circular orbit reads
\begin{equation}
   \label{Ec}
   E_{c}=mb(r_{c})\sqrt{\frac{2w(r_{c})w'(r_{c})}{2w(r_{c})w'(r_{c})b(r_{c})-b'(r_{c})w^{2}(r_{c})}}~.
\end{equation}
The proper-time period of a circular orbit and the relation between the coordinate-time and proper-time periods are yielded by Eqs. (\ref{dphidtau}) and (\ref{dtdtau}) respectively
\begin{align}
    T_{\tau}&=\frac{2\pi mw^{2}(r_{c})}{L_{c}}~,\label{PeriodPTF}\\
    T_{t}&=T_{\tau}\frac{E_{c}}{mb(r_{c})}~.\label{PeriodCTF}
\end{align}
Upon substituting Eqs. (\ref{Lc}) and (\ref{Ec}) into the above equations, we obtain the periods of circular orbits in the proper-time and coordinate-time frameworks
\begin{align}
   T_{\tau}&=2\pi\sqrt{\frac{2w(r_{c})w'(r_{c})b(r_{c})-b'(r_{c})w^{2}(r_{c})}{b'(r_{c})}}~,\\
   T_{t}&=T_{\tau}\sqrt{\frac{2w(r_{c})w'(r_{c})}{2w(r_{c})w'(r_{c})b(r_{c})-b'(r_{c})w^{2}(r_{c})}}=2\pi\sqrt{\frac{2w(r_{c})w'(r_{c})}{b'(r_{c})}}~.
\end{align}
Finally, in the case of a purely radial motion of a massive particle, the equations of motion in coordinate-time and proper-time frameworks correspond to Eqs. (\ref{RadVelCTF}) and (\ref{RadVelPTF}) for $L=0$.

Considering a massless test particle, the relation between the coordinate time and an affine parameter $\tau$ is described by Eq. (\ref{E})
\begin{equation}
    \label{dtdtauML}
    \frac{dt}{d\tau}=\frac{E}{b(r)}~.
\end{equation}
By setting $m=0$ in Eq. (\ref{RadVelCTF}), the radial velocity of a massless test particle in the coordinate-time framework arises. The radial velocity in the framework of the affine parameter reads
\begin{equation}
    \label{RadVelPTFML}
    \left(\frac{dr}{d\tau}\right)^{2}=E^{2}-V_{\text{eff}}^{2}(r)~,
\end{equation}
where $V_{\text{eff}}$ is the effective potential (\ref{Veff}) calculated for $m=0$. Additionally, Eq. (\ref{J}) implies the form of the angular velocity of a massless particle
\begin{equation}
    \label{dphidtauML}
    \frac{d\phi}{d\tau}=\frac{L}{w^{2}(r)}~.
\end{equation}
The angular trajectories, in this case, are described by Eq. (\ref{phi}) by setting $m=0$. Interestingly, the radius of the circular orbits does not depend on the angular momentum of massless test particles, since the effective potential is proportional to $L$ in this case. Therefore, the periods of the circular motion of a massless test particle, in the frameworks of the coordinate-time and the affine parameter, are
\begin{align}
    T_{\tau}&=\frac{2\pi w^{2}(r_{c})}{L}~,\label{PeriodPTFmassless}\\
    T_{t}&=T_{\tau}\frac{E}{b(r_{c})}=2\pi\frac{w(r_{c})}{\sqrt{b(r_{c})}}~.\label{PeriodCTFmassless}
\end{align}
In the case of a purely radial motion of a massless particle, the equations of motion, in the coordinate-time framework and the framework of the affine parameter, correspond to Eqs. (\ref{RadVelCTF}) and (\ref{RadVelPTFML}) for $L=0$, $m=0$. We need to point out that the geodesic equations of motion for massless particles do not describe the motion of photons. The action (\ref{Action}) includes a self-interacting term for photons, which implies that light follows the null geodesics of an \emph{effective geometry}. The determination of the \emph{effective geometry} takes place in Section \ref{sec5}.

\subsection{Time-like geodesics}

The trajectories of uncharged massive test particles with mass $m$ correspond to the time-like geodesics of the geometry (\ref{Metric}). The shape of the effective potential (\ref{Veff}) implies the possible particle trajectories for the asymptotically (A)dS and flat cases. From now on, for the calculation of the time-like geodesics, we fix the values of the test particle's and BH parameters in order for the results to be straightforwardly compared. To be more precise, we consider $M=1$, $Q_{m}=0.2M$, $\alpha=0.002M^{2}$, $\nu=1M$, $m=0.001M$, while $L$ will be fixed as $L=0.0037M^{2}$ for angular motion and $L=0$ for radial motion. In the (A)dS case, we set $\Lambda_{\text{eff}}=(-)5\cdot 10^{-5}M^{-2}$, while in the asymptotically flat case we have $\Lambda_{\text{eff}}=0$. All of the following figures are plotted using these values unless stated otherwise. The aforementioned chosen set of parameters is reasonable, since the EH parameter $\alpha$ constitutes a small correction to our model and the scalar hair parameter $\nu=1M$ contributes to the results. The value $\nu=1M$ is reasonably large in order to have an impact on the observables, while the BH holds its properties. For extremely large values of the scalar charge the BH shrinks significantly. For this set of parameters, the BH in the asymptotically AdS spacetime possesses one event horizon at $r_{\text{H}}^{\text{AdS}}=1.5048M$. In the asymptotically flat case, the BH also has one event horizon located at $r_{\text{H}}^{\text{flat}}=1.5049M$, while in the asymptotically dS case the BH has one event horizon at $r_{\text{H}}^{\text{dS}}=1.5050M$ and one cosmological horizon at $r_{CH}^{\text{dS}}=243.4434M$. The cosmological horizon is the limit beyond which objects and regions are moving away from us faster than the speed of light due to the expansion of the universe caused by the positive cosmological constant $\Lambda_{\text{eff}}$. Consequently, any information or signals from these objects cannot reach us, and we are unable to observe or interact with them directly. As for what is to come, we will explore the shape of the effective potential (\ref{Veff}), determine all different trajectories, and depict the most representative ones. We will perform the above tasks for all three asymptotic cases, namely (A)dS and flat. Additionally, the contribution of the cosmological constant to the results will be commented on.

\subsubsection{Angular motion}

\begin{figure}
    \centering
    \includegraphics[width=0.3\textwidth]{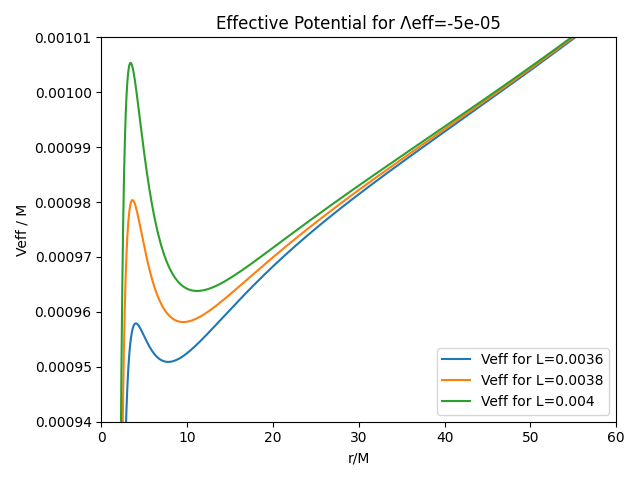}
    \includegraphics[width=0.3\textwidth]{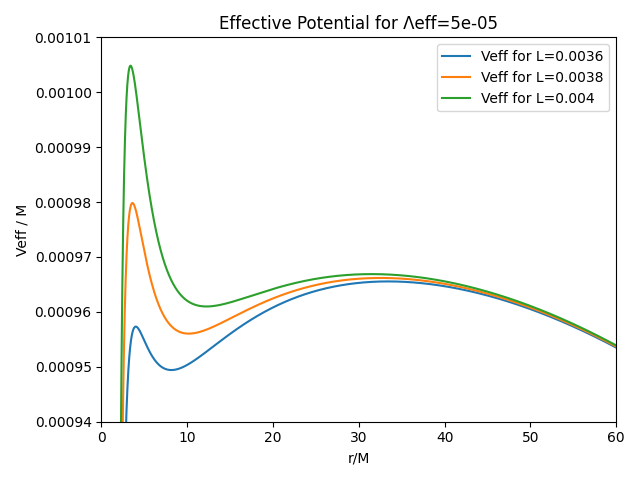}
    \includegraphics[width=0.3\textwidth]{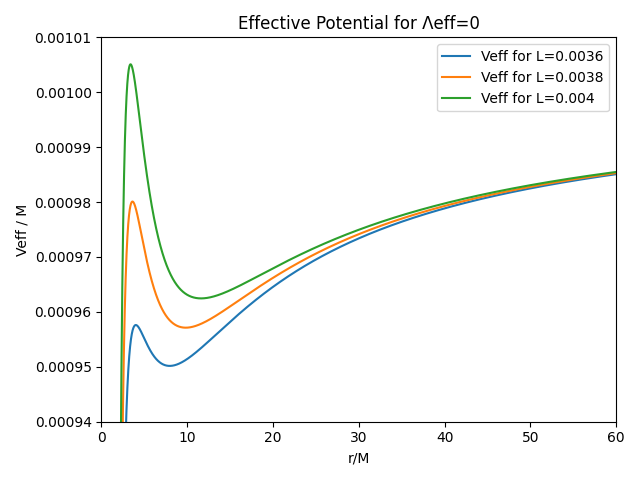}
    \caption{Effective potential $V_{\text{eff}}$ for different values of the angular momentum $L$ in the asymptotically AdS (left), dS (middle) and flat (right) cases.}
    \label{fig:1}
\end{figure}

In the case of particle angular motion ($L\neq 0$), we depict the effective potential (\ref{Veff}) in FIG. \ref{fig:1}, for various values of the particle angular momentum. From FIG. \ref{fig:1}, it becomes obvious that bound orbits exist, as well as stable and unstable circular orbits, for all three different asymptotic forms of the spacetime. Moreover, it is noticeable that the effective potential takes on larger values as the angular momentum increases. Therefore, a particle with a fixed total energy $E$ is more likely to fall into the BH when it possesses smaller angular momentum rather than larger. Consequently, for different values of particle's angular momentum and fixed total energy, we may observe different types of particle trajectories. However, all the possible trajectories can be obtained by fixing the value of the angular momentum and varying the total energy.

\begin{figure}
    \centering
    \includegraphics[width=0.3\textwidth]{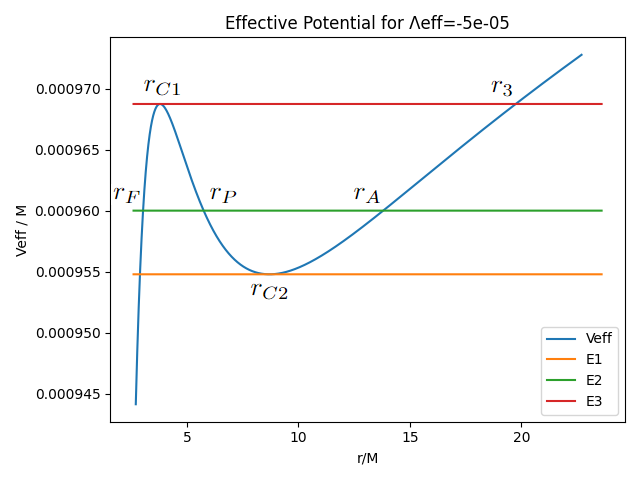}
    \includegraphics[width=0.3\textwidth]{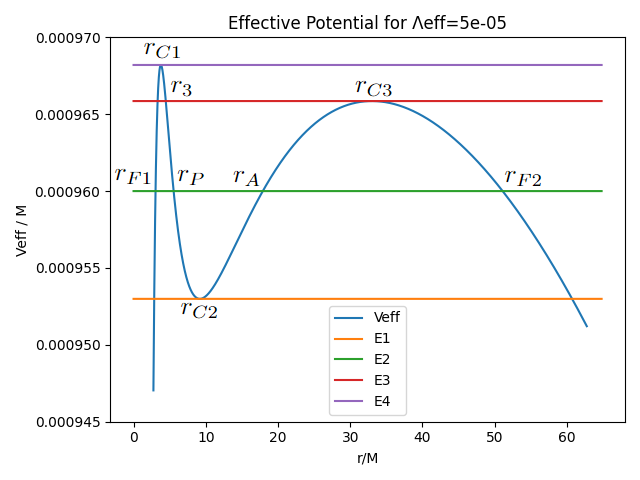}
    \includegraphics[width=0.3\textwidth]{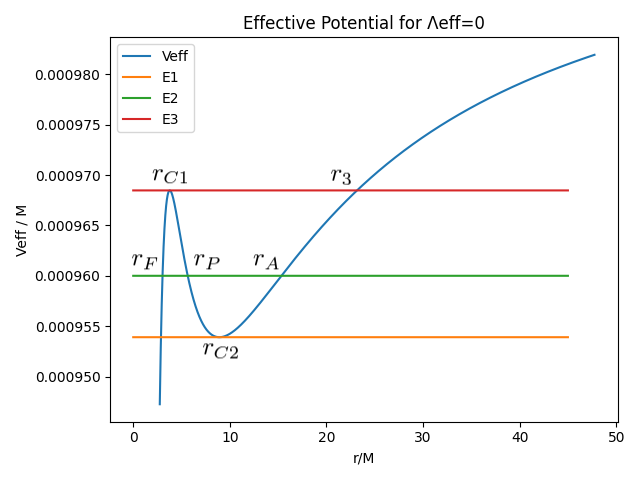}
    \caption{Effective potential for particle's angular motion. Left Panel: Plot for the AdS case with $E_{1}=9.55\cdot 10^{-4}M$, $E_{2}=9.60\cdot 10^{-4}M$, $E_{3}=9.69\cdot 10^{-4}M$. Middle Panel: Plot for the dS case with $E_{1}=9.53\cdot 10^{-4}M$, $E_{2}=9.60\cdot 10^{-4}M$, $E_{3}=9.66\cdot 10^{-4}M$, $E_{4}=9.68\cdot 10^{-4}M$. Right Panel: Plot for the asymptotically flat case with $E_{1}=9.54\cdot 10^{-4}M$, $E_{2}=9.60\cdot 10^{-4}M$, $E_{3}=9.68\cdot 10^{-4}M$.}
    \label{fig:2}
\end{figure}

\begin{figure}
    \centering
    \includegraphics[width=0.38\textwidth]{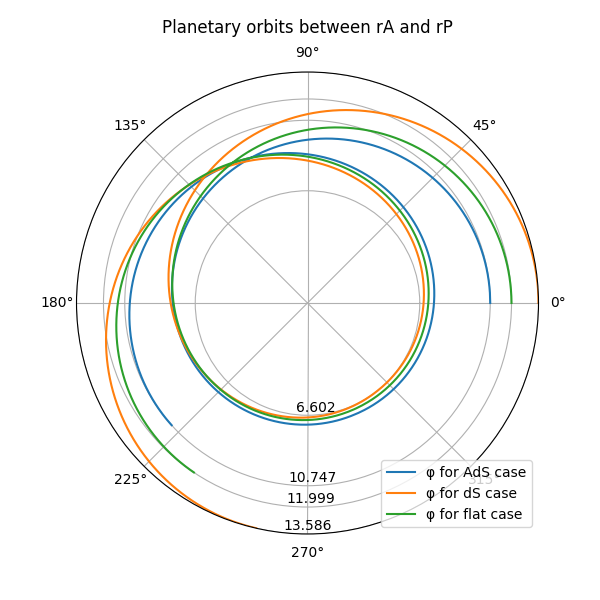}
    \caption{Planetary orbits with periastron precession for particle's total energy $E=9.56\cdot 10^{-4}M$. The massive test particle moves in an asymptotically AdS spacetime with $\Lambda_{\text{eff}}=-5\cdot10^{-5}M^{-2}$ (blue line), in an asymptotically flat spacetime with $\Lambda_{\text{eff}}=0$ (green line) and in an asymptotically dS spacetime with $\Lambda_{\text{eff}}=5\cdot10^{-5}M^{-2}$ (orange line). The particle in all three cases execute counterclockwise motion.}
    \label{fig:3}
\end{figure}

To extract more information about the possible particle trajectories in every asymptotic case, we present the effective potential given in Eq. (\ref{Veff}) for $L=0.0037M^{2}$, along with some noteworthy points in space, in FIG. \ref{fig:2}. There are different types of particle motion, which vary depending on both its total energy $E$ and the initial conditions of the motion. To be more precise, for all three asymptotic cases, there are planetary orbits for $E=E_{2}$, such that $E_{1}<E<E_{3}$, and $r$-coordinate varying in $[r_{P},r_{A}]$, where $r_{A}$ denotes the apastron and $r_{P}$ denotes the periastron of the planetary orbit. Using the Eq. (\ref{phi}), we calculate the corresponding trajectories, for $E=9.56\cdot 10^{-4}M$, and present them in FIG. \ref{fig:3}. The cosmological constant does not influence the shape of the planetary trajectories, instead it has an impact on the distance between the periastron and the apastron. More specifically, the repulsive forces of the dS spacetime induced by a positive cosmological constant increase the gap between the points $r_{P}$ and $r_{A}$ compared to the case of an asymptotically flat spacetime. On the contrary, the attractive forces caused by a negative cosmological constant, in an AdS spacetime, decrease the distance between the points $r_{P}$ and $r_{A}$ compared to the case of an asymptotically flat spacetime. The periastron and apastron in all three cases read $r_{P}^{\text{dS}}=6.602M$, $r_{P}^{\text{flat}}=6.824M$, $r_{P}^{\text{AdS}}=7.121M$ and $r_{A}^{\text{AdS}}=10.747M$, $r_{A}^{\text{flat}}=11.999M$, $r_{A}^{\text{dS}}=13.586M$. In the planetary orbits of FIG. \ref{fig:3}, a usual phenomenon in GR arises, the periastron  precession. This phenomenon is basically about the deviation of the planetary motion from a closed orbit. The periastron precession ($\Delta\phi$) can be calculated according to the formula
\begin{equation}
   \Delta\phi=2\phi(r_{P})-2\pi~,
\label{PerPrec}
\end{equation}
where $\phi(r_{P})$ is given by Eq. (\ref{phi}) by choosing the minus sign and the apastron as an initial position, i.e. $r_{0}=r_{A}$. Due to the different distances between the periastron and apastron, and the repulsive or attractive forces caused by the varying value of the cosmological constant, the periastron precession exhibits differences in each scenario. In particular, for the chosen set of parameters, the periastron precession, in the three cases, reads $\Delta\phi_{\text{AdS}}=3.963$, $\Delta\phi_{\text{flat}}=4.212$ and $\Delta\phi_{\text{dS}}=4.589$. These values can be ranked as follows
\begin{equation}
    \Delta\phi_{\text{AdS}}<\Delta\phi_{\text{flat}}<\Delta\phi_{\text{dS}}~.
\end{equation}

The stable circular orbits arise as special cases of the planetary orbits and correspond to the local minima of the effective potential. In all three cases, when the total energy of the test particle equals to $E_{1}$, see FIG. \ref{fig:2}, the particle moves in a stable circular orbit of radius $r_{C2}$. We can numerically calculate the radii of these orbits from Eq. (\ref{CircOrbCondit}), which are $r_{C2}^{\text{AdS}}=8.697M$, $r_{C2}^{\text{flat}}=8.919M$ and $r_{C2}^{\text{dS}}=9.181M$. Comparing these results, the circular orbit is closer to the BH in the AdS case compared to the case of an asymptotically flat spacetime, due to the attractive forces induced by the negative cosmological constant. Exactly the opposite effect is observed in the case of the dS spacetime.

\begin{figure}
    \centering
    \includegraphics[width=0.35\textwidth]{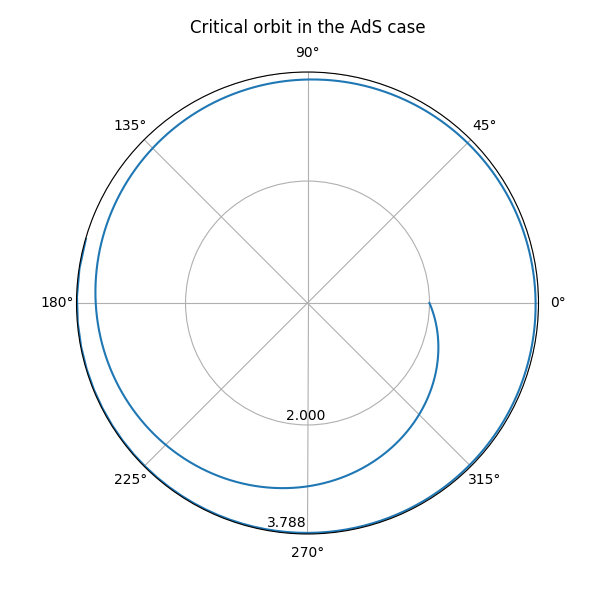}
    \includegraphics[width=0.35\textwidth]{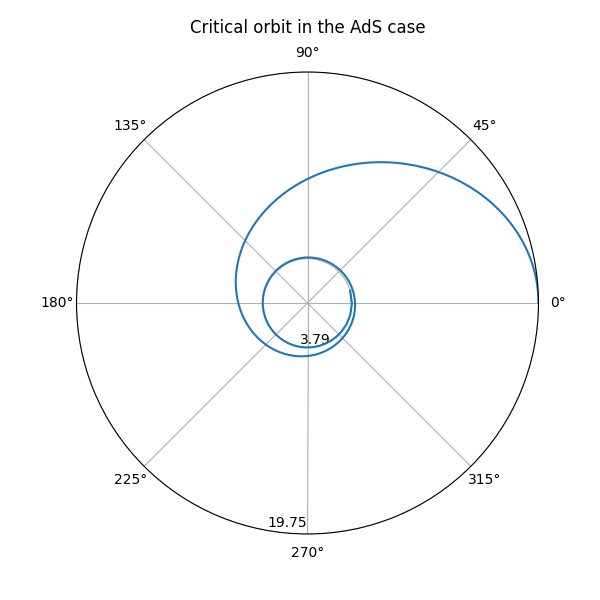}
    \caption{Left figure: Inner asymptotically unstable circular orbit of radius $r_{C1}=3.778M$. The particle's initial position is located at $r_{i}=2.000M$, such that $r_{\text{H}}^{\text{AdS}}<r_{i}<r_{C1}$. The particle executes clockwise motion. Right figure: Outer asymptotically unstable circular orbit of radius $r_{C1}=3.778M$. The particle's initial position is located at $r_{3}=19.745M$, such that $r_{3}>r_{C1}$. The particle executes counterclockwise motion. Both figures refer to the AdS case.}
    \label{fig:4}
\end{figure}

Moreover, there are asymptotically unstable circular orbits in all three cases. The unstable circular orbits correspond to the local maxima of the effective potential and their radii are calculated by the Eq. (\ref{CircOrbCondit}). Based on the information presented in FIG. \ref{fig:2}, a particle follows an unstable circular orbit with a radius of $r=r_{C1}$ when its total energy is equal to $E=E_{3}$ in both the AdS and flat cases. For the chosen set of parameters, the radii of these unstable circular orbits are $r_{C1}^{\text{AdS}}=3.788M$ and $r_{C1}^{\text{flat}}=3.784M$. Interestingly, in the dS case, we have two asymptotically unstable circular orbits of radii $r_{C1}^{\text{dS}}=3.779M$ and $r_{C3}=33.002M$. The inner one is obtained when the particle's total energy is equal to $E=E_{4}$, while the outer one when $E=E_{3}$. In fact, the outer unstable circular orbit is observed at the region where the repulsive forces induced by the positive cosmological constant eliminate the attractive gravitational forces of the BH. Particles can approach the unstable circular orbits from either smaller or larger radii. In FIG. \ref{fig:4}, we depict the asymptotically unstable circular trajectories of radius $r_{C1}^{\text{AdS}}$, in the AdS case. The initial particle's position $r_{i}$, in the left picture of FIG. \ref{fig:4}, satisfies the relation: $r_{\text{H}}^{\text{AdS}}<r_{i}=2.00M<r_{C1}^{\text{AdS}}$. In this case the particle approaches the unstable circular orbit from smaller radii. Additionally, the initial particle's position $r_{3}$, in the right picture of FIG. \ref{fig:4}, satisfies the relation: $r_{3}=19.745M>r_{C1}^{\text{AdS}}$. In this case the particle approaches the unstable circular orbit from larger radii. The asymptotic motions in the dS and flat cases are similar to those of FIG. \ref{fig:4}, and hence, are omitted. The periods of the circular orbits, both stable and unstable, can be calculated using the formulas (\ref{PeriodPTF}) and (\ref{PeriodCTF}).

\begin{figure}
    \centering
    \includegraphics[width=0.35\textwidth]{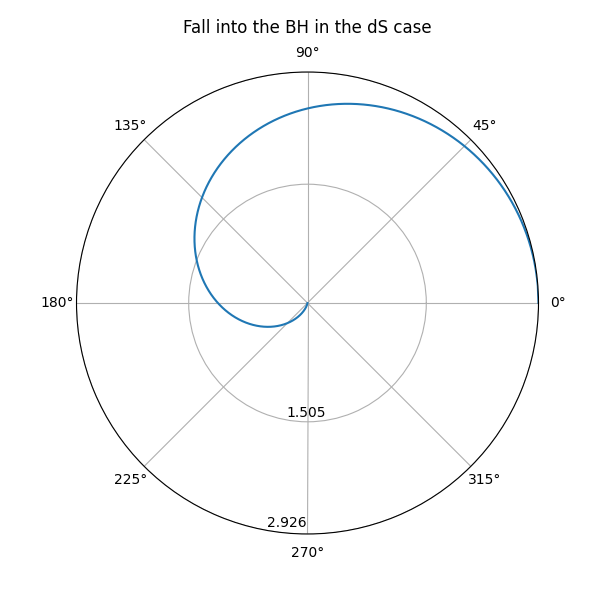}
    \includegraphics[width=0.35\textwidth]{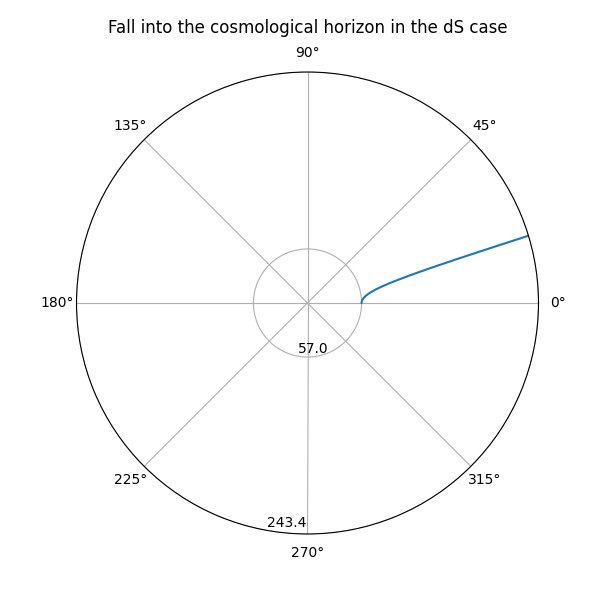}
    \caption{Left panel: Fall into the BH. The particle's initial position is located at $r_{F1}=2.926M$. The BH event horizon is positioned at $r_{\text{H}}^{\text{dS}}=1.5050M$. Right panel: Fall into the cosmological horizon. The particle's initial position is located at $r_{F2}=56.974M$, while the cosmological horizon is situated at $r_{CH}^{\text{dS}}=243.4434M$. In both graphs, the particle executes counterclockwise motion. Both graphs are plotted for particle's total energy $E=9.56\cdot10^{-4}M$ and refer to the dS case.}
    \label{fig:5}
\end{figure}

Furthermore, as it can be seen in FIG. \ref{fig:2}, in the asymptotically AdS and flat cases, particles with total energy $E=E_{2}<E_{3}$ and initial position $r_{i}$, such that $r_{i}\leq r_{F}$, are doomed to fall into the BH. This scenario also holds in the AdS case, when $E>E_{3}$, and in the flat case, when $E$ satisfies the inequality $E_{3}<E<m=0.001M$. The upper limit imposed on $E$ in the last inequality is the asymptotic value of the effective potential at the large-$r$ limit, $V_{\text{eff}}^{\text{flat}}(r\to+\infty)\to m$, in the flat case.  In the asymptotically flat case, for $E$ such that $E>m=0.001M$, the particle will cross the event horizon if its initial radial velocity is negative, otherwise it will go towards infinity. Additionally, we can explore the asymptotically dS case. In this case, the test particle will fall into the BH if its total energy $E$ satisfies the relation $E=E_{2}<E_{4}$ and simultaneously its initial position is $r_{i}$, such that $r_{i}\leq r_{F1}$. If we have $E>E_{4}$ and negative initial radial velocity the particle will also cross the event horizon, while if the initial radial velocity is positive, the particle will fall into the cosmological horizon. This will also be the case when the total energy $E$ satisfies the inequality $E_{3}<E<E_{4}$ or when we simultaneously have $E=E_{2}<E_{3}$ and $r_{F2}<r_{i}<r_{CH}^{\text{dS}}$, where $r_{i}$ is the initial position of the test particle. In FIG. \ref{fig:5}, we illustrate the trajectories in the dS case, for $E=9.56\cdot10^{-4}M$, where the particle falls into the BH (left panel) and into the cosmological horizon (right panel). The trajectories that fall into the event horizon, in the AdS and flat cases, are similar to the one depicted in the left panel of FIG. \ref{fig:5}, and therefore are omitted. It is noticeable that a particle can cross the event horizon and fall into the BH singularity, see the left picture of FIG. \ref{fig:5}. Consequently, the continuation of the spacetime inside the event horizon is evident. However, an observer in the outer region of the event horizon sees the particle approaching the event horizon, but never crossing it. This phenomenon will be illustrated later by exploring the radial motion of a massive test particle. A discussion on the geodesic continuation inside the event horizon in a case with three horizons is not included in this article and is postponed for a future work.

The chosen set of parameters reveals all the possible types of trajectories. The impact of the BH parameters on the effective potential will be discussed later in the article. There is a critical value of the test particle's angular momentum below which the innermost local maximum of the effective potential depicted in FIG. \ref{fig:1} is eliminated. This phenomenon holds in all three asymptotic cases. The critical value of the particle's angular momentum, for the chosen set of parameters, approximately reads $L_{\text{c}}=0.003M^{2}$. Thus, for $L<L_{\text{c}}$, there are no planetary or stable circular orbits. Instead, there are only orbits that fall into the BH or go towards the cosmological horizon or infinity depending on the value of the cosmological constant. In fact, the shape of the effective potential and the possible particle trajectories, for small values of particle's angular momentum, such that $L<L_{\text{c}}$, are similar to those in the case of purely radial motion where $L=0$. This case will be explored shortly.

In conclusion, we can have bound circular orbits regardless of the asymptotic form of spacetime. However, the inclusion of a positive (negative) cosmological constant will introduce a repulsive (attractive) force in the spacetime which will further influence the periastron precision as well as the radial distance from the BH in which circular orbits occur.

\subsubsection{Radial motion}

\begin{figure}
    \centering
    \includegraphics[width=0.3\textwidth]{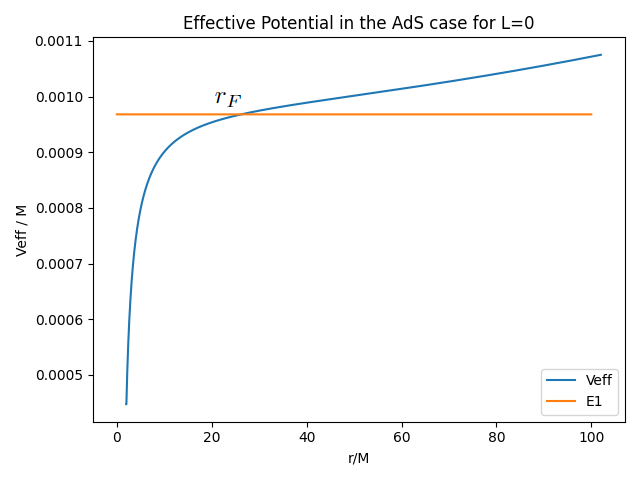}
    \includegraphics[width=0.3\textwidth]{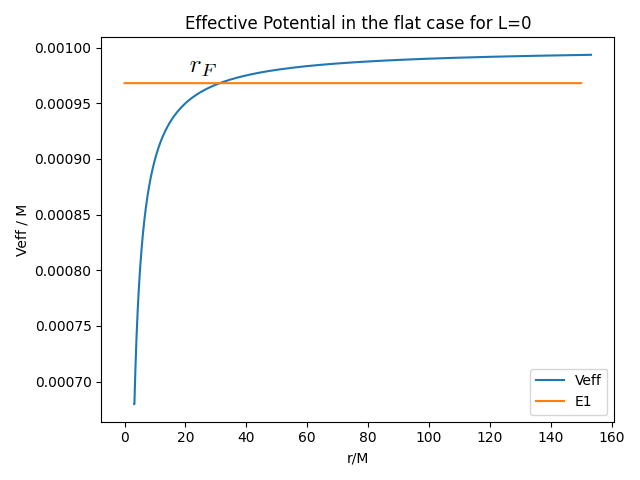}
    \includegraphics[width=0.3\textwidth]{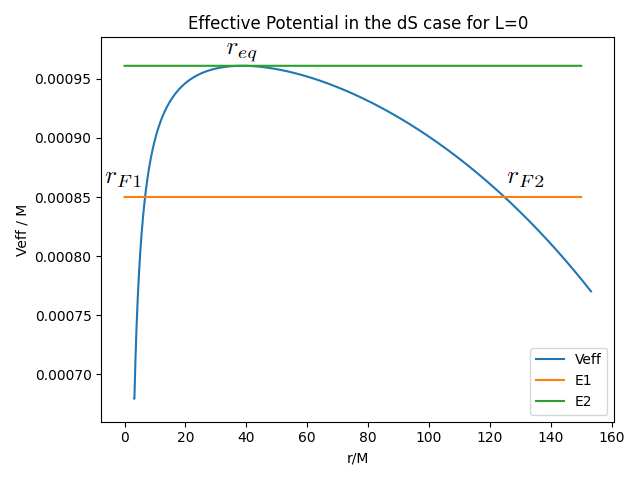}
    \caption{Effective potential for particle's radial motion. Left Panel: Plot for the AdS case with $E_{1}=9.68\cdot 10^{-4}M$. Middle Panel: Plot for the asymptotically flat case with $E_{1}=9.68\cdot 10^{-4}M$. Right Panel: Plot for the dS case with $E_{1}=8.50\cdot 10^{-4}M$ and $E_{2}=9.61\cdot 10^{-4}M$.}
    \label{fig:6}
\end{figure}

In the case of particle's radial motion ($L=0$), we depict the effective potential (\ref{Veff}) in FIG. \ref{fig:6}. In the AdS case, see the left panel of FIG. \ref{fig:6}, the massive test particle is doomed to fall into the BH for every value of its total energy $E$. The effective potential for an asymptotically flat spacetime is illustrated in the middle panel of FIG. \ref{fig:6}. A particle with energy $E$, such that $E<V_{\text{eff}}^{\text{flat}}(r\to+\infty)=m=0.001M$, in the asymptotically flat case, will fall into the BH. Additionally, for $E\geq m$ in the flat case, a particle with negative radial velocity is inevitably pulled into the BH, otherwise in the case of a positive velocity, it goes towards spatial infinity. The dS case, depicted in the right panel of FIG. \ref{fig:6}, is the most interesting. A particle with total energy $E=E_{1}<E_{2}$ and initial position $r_{i}$ will fall into either the BH event horizon or the cosmological horizon. The first case holds if the relation $r_{i}<r_{F1}$ is satisfied, while, if the condition $r_{i}>r_{F2}$ is met, the second case applies. Additionally, a trajectory of a particle with a total energy $E$, such that $E>E_{2}$, will cross the BH event horizon, in case of a negative radial particle's velocity, while it will traverse the cosmological horizon if the particle has a positive radial velocity. An interesting case arises when the particle's total energy $E$ equals the critical value $E_{2}$. In this case, the point $r_{eq}$ acts as an unstable equilibrium point between the attractive gravitational force of the BH and the repulsive nature of the expanding universe induced by the positive cosmological constant. A particle with $E=E_{2}$ located at this point will stay stationary until a slight perturbation in its energy occurs. Afterwards, it is doomed to fall into either the event horizon or the cosmological horizon.

\begin{figure}
    \centering
    \includegraphics[width=0.4\textwidth]{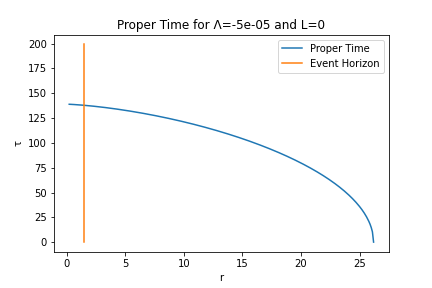}
 \includegraphics[width=0.4\textwidth]{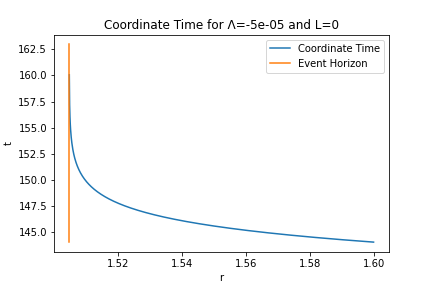}
    \caption{Radial motion towards the event horizon, in the proper-time (left) and coordinate-time (right) framework, for $E=9.68\cdot 10^{-4}M$, in the AdS case. The event horizon is located at $r_{\text{H}}^{\text{AdS}}=1.5048M$.}
    \label{fig:7}
\end{figure}

\begin{figure}
    \centering
    \includegraphics[width=0.4\textwidth]{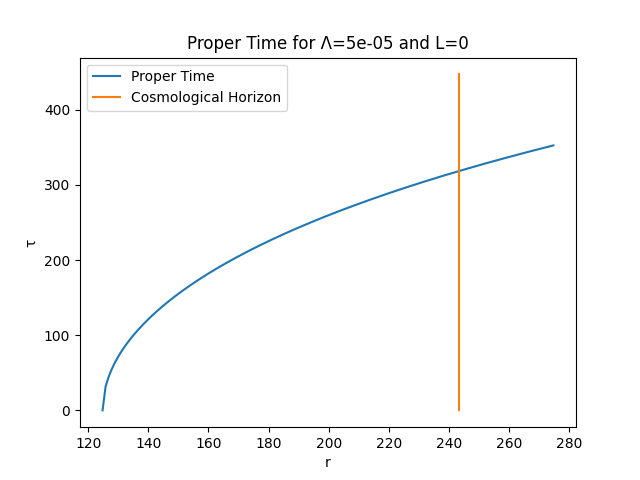}
 \includegraphics[width=0.38\textwidth]{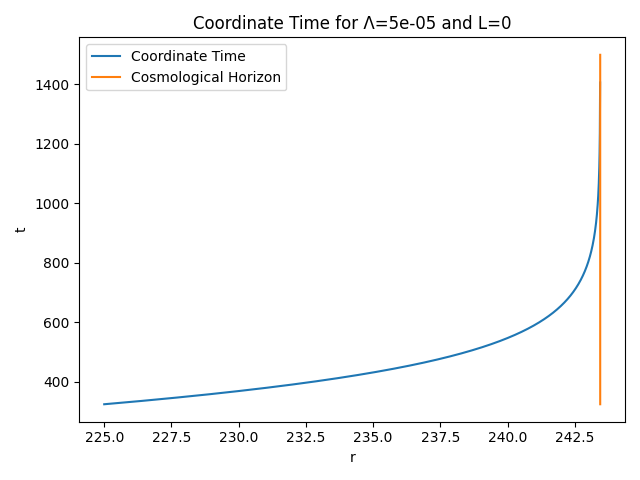}
    \caption{Radial motion towards the cosmological horizon, in the proper-time (left) and coordinate-time (right) framework, for $E=8.50\cdot10^{-4}M$, in the dS case. The cosmological horizon is located at $r_{CH}^{\text{AdS}}=243.4434M$.}
    \label{fig:8}
\end{figure}

The radial motion in the proper-time framework and the coordinate-time framework can be described by the functions $\tau(r)$ and $t(r)$ respectively. These functions are yielded by Eqs. (\ref{RadVelCTF}) and (\ref{RadVelPTF})
\begin{align}
    t(r)&=-E\int_{r_{F}}^{r}\frac{dr'}{b(r')\sqrt{E^{2}-V_{\text{eff}}^{2}(r')}}~,\label{t(r)}\\
    \tau(r)&=-m\int_{r_{F}}^{r}\frac{dr'}{\sqrt{E^{2}-V_{\text{eff}}^{2}(r')}}~.\label{tau(r)}
\end{align}
In FIG. \ref{fig:7}, we illustrate the particle's motion towards the BH event horizon described by the Eqs. (\ref{t(r)}) and (\ref{tau(r)}), in the AdS case. In the proper-time framework, see the left graph of FIG. \ref{fig:7}, the test particle crosses the BH event horizon and goes towards BH singularity. Consequently, the spacetime's continuation holds inside the event horizon. In the coordinate-time framework, see the right graph of FIG. \ref{fig:7}, the test particle approaches the BH event horizon but never reaches it. The interpretation of this phenomenon is that a regular observer in the universe cannot see a particle cross the BH event horizon. The trajectories that fall into the event horizon, in the asymptotically dS and flat cases, are similar to those presented in FIG. \ref{fig:7} and therefore are not presented. Additionally, in FIG. \ref{fig:8}, we plot the functions $\tau(r)$ and $t(r)$ in the case of a particle's motion towards the cosmological horizon. As it was expected, in the proper-time framework, the particle enters the cosmological horizon and the spacetime's continuation holds, while, in the coordinate-time framework, an observer in the universe testifies that the particle asymptotically reaches the cosmological horizon but never enters it.

\subsection{Null geodesics}

\begin{figure}
    \centering
    \includegraphics[width=0.4\textwidth]{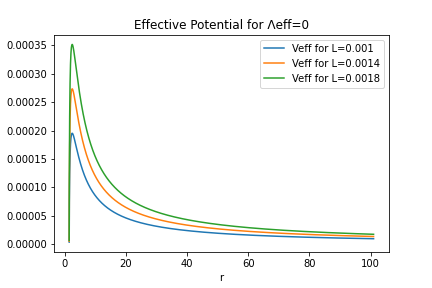}
    \includegraphics[width=0.4\textwidth]{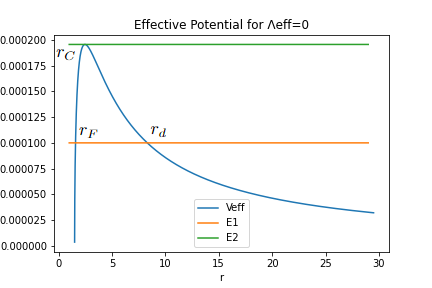}
    \caption{Effective potential of the angular motion of massless particles, in the asymptotically flat case, for different values of particle's angular momentum $L$ (left graph) and a particular case for $L=0.001M^{2}$, $E_{1}=1.00\cdot10^{-4}M$, $E_{2}=1.95\cdot10^{-4}M$ (right graph).}
    \label{fig:9}
\end{figure}

In this subsection, we will briefly investigate the motion of an uncharged massless particle and determine the null geodesics of the geometry under consideration. We use the set of parameters from the previous subsection with the only difference being in the mass of the test particle which is $m=0$. The null geodesics do not describe a photon motion, due to photon self-interactions induced by the EH NLED sector of the action (\ref{Action}). The photon propagation is considered in Section \ref{sec5}. The effective potential that influences the motion of massless particles is described by Eq. (\ref{Veff}), when the particle’s mass $m$ is set equal to zero. The three asymptotic cases, namely (A)dS and flat, show minimal differences, except for the dS case, which involves a cosmological horizon. Consequently, we will focus on the asymptotically flat case to explore the null geodesics in the vicinity of the BH. In FIG. \ref{fig:9}, we depict the effective potential of the null geodesics in the asymptotically flat case. To be more specific, the effective potential for various values of the massless particle's angular momentum $L$ is presented in the left panel of FIG. \ref{fig:9}. In the right panel of FIG. \ref{fig:9}, the effective potential for $L=0.001M^{2}$ is depicted along with some remarkable spatial points. As it can be seen in the left graph of FIG. \ref{fig:9}, an increase at the value of the angular momentum causes a rise of the values of the effective potential. However, the location of the local maximum is not affected by the changes of the value of the angular momentum. 

\begin{figure}
    \centering
    \includegraphics[width=0.33\textwidth]{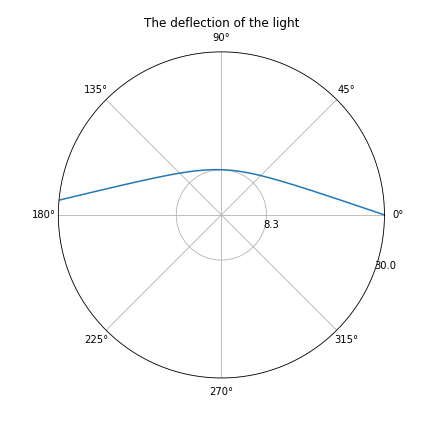}
    \caption{The deflection of massless particles on the BH, for $\Lambda_{eff} =0$, $E=1.00\cdot10^{-4}M$ and $L=0.001M^{2}$. Initial position $r=30M>r_{d}=8.3M$. Counterclockwise motion.}
    \label{fig:10}
\end{figure}

By investigating the plot of the effective potential in the right panel of FIG. \ref{fig:9}, we conclude that, for particle's total energy $E$ such that $E<E_{2}$, there are two different scenarios. Without loss of generality, we fix the value of the total energy at $E=E_{1}<E_{2}$. Thus, in the first scenario the initial position $r_{i}$ satisfies the inequality $r_{i}\leq r_{F}$ and the massless particle is fated to fall into the BH, while in the second scenario, we have $r_{i}\geq r_{d}$ and the particle is deflected on the effective potential. This phenomenon holds due to the curvature of the spacetime caused by the BH. The null geodesics that lead into the BH are similar to the time-like geodesics depicted in the left panel of FIG. \ref{fig:5}. Additionally, for $E>E_{2}$, a particle with a negative radial velocity will cross the BH event horizon, while a particle with a positive radial velocity will escape to infinity. Interestingly, there is an asymptotically unstable circular orbit of radius $r_{C}$, which corresponds to the local maximum of the effective potential. The asymptotically circular trajectories of massless particles are similar to those presented in FIG. \ref{fig:4} and therefore are omitted. The periods of revolution of the unstable circular orbits, in the proper-time framework and coordinate-time framework, can be calculated by Eqs. (\ref{PeriodPTFmassless}) and (\ref{PeriodCTFmassless}), respectively. In FIG. \ref{fig:10}, we plot the deflected trajectory, for $E=1.00\cdot10^{-4}M$, which can be calculated by Eq. (\ref{phi}) by setting $m=0$. In this case, the massless test particle moves form $r=30M>r_{d}=8.3M$ to $r=r_{d}$, where it changes direction (deflection of massless particles) and starts moving away from the BH. Finally, the radial motion of a massless particle, in the coordinate-time and proper-time framework, is described by Eqs. (\ref{RadVelCTF}) and (\ref{RadVelPTFML}), respectively, by setting $m=0$ and $L=0$. The corresponding trajectories in the asymptotically flat case are similar to those presented in FIG. \ref{fig:7} and hence are not presented.

\subsection{Contribution of black hole parameters to the geometry}

It is crucial to determine the contribution of the scalar hair and magnetic charge to the particle trajectories in the geometry under consideration. Such an analysis will reveal the effect of the BH parameters on the structure of spacetime. First of all, we will explore how the parameters $\nu$, $Q_{m}$ and $\alpha$ affect the effective potential that governs particle motion. Thus, we will plot the effective potential for various values of these parameters, in the asymptotically (A)dS and flat cases, and compare the results. It is also important to compare the results with the $\nu=0$ case, which corresponds to a magnetically charged EH BH without scalar hair \cite{Yajima:2000kw}, and the $Q_{m}=\alpha=0$ case, which corresponds to an asymptotically AdS BH with scalar hair \cite{Gonzalez:2013aca}. The BH of \cite{Gonzalez:2013aca} can also be asymptotically flat or dS, and it has been checked that a curvature singularity is present and shielded by an event horizon in each case. Moreover, we will discuss the contribution of the scalar hair to the periastron precession, and finally, some extremal scenarios and their contribution to the event horizon will be illustrated.

\begin{figure}
    \centering
    \includegraphics[width=0.3\textwidth]{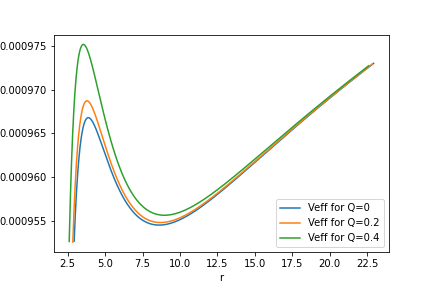}
    \includegraphics[width=0.3\textwidth]{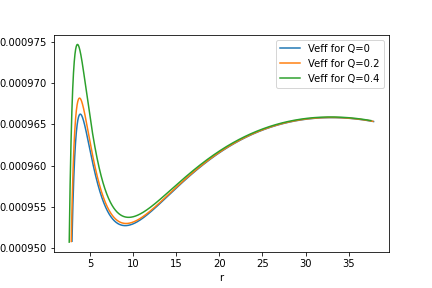}
    \includegraphics[width=0.3\textwidth]{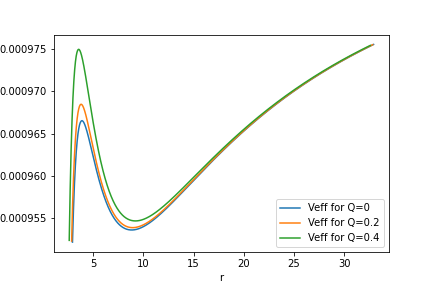}
    \caption{Effective potential, in the AdS (left), dS (middle) and flat (right) cases for angular particle's motion, $\alpha=0.002M^{2}$ and different values of the BH magnetic charge $Q_{m}$. The cosmological constant in the (A)dS and flat cases reads $\Lambda_{\text{eff}} = (-)5\cdot10^{-5}M^{-2}$ and $\Lambda_{\text{eff}} =0$, respectively.}
    \label{fig:11}
\end{figure}

To discuss the contribution of $Q_{m}$ and $\alpha$ to the effective potential, we fix the parameters: $m = 0.001M$, $\nu=1M$ and $L=0.0037M^{2}$. In FIG. \ref{fig:11}, we compare the effective potential in the vanishing $Q_{m}$ case with the one corresponding to a magnetically charged EH BH with scalar hair, in the (A)dS and flat cases.
In each case, the effective potential has similar shapes, and planetary orbits and stable and unstable circular orbits exist. However, the parameters of particle trajectories seem to differ in each case. For instance, the radii of stable and unstable circular orbits, and the periastron and apastron of the planetary orbits are some of the observable differences. Additionally, we observe that a larger magnetic charge results to a larger peak for the effective potential, and the repulsive gravitational nature of the magnetic charge will result to a BH spacetime in which the particles will need higher energy (keeping the angular momentum fixed) to fall into the BH, just like the Reissner-Nordstr\"om case.

\begin{figure}[h]
    \centering            \includegraphics[width=0.4\textwidth]{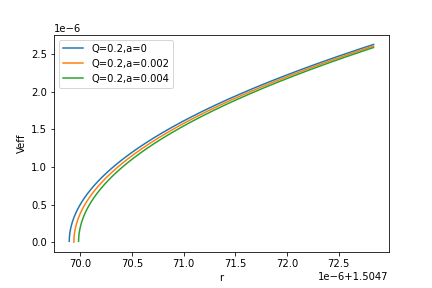}
    \caption{Effective potential, in the AdS case, for angular particle motion near the event horizon for different values of $\alpha$, and $Q_{m}=0.2M$, $m = 0.001M$, $\nu=1M$, $L=0.0037M^{2}$ and $\Lambda_{\text{eff}} = -5\cdot10^{-5}M^{-2}$. The radial coordinate $r$ is depicted for values of the form $1.5047+()\cdot10^{-6}M = r_H+ \Delta r_H$ where $\Delta r_H$ is of the order of $10^{-6}M$.}
    \label{fig:12}
\end{figure}

Moreover, we  mention that the EH correction does not affect the effective potential far away from the event horizon. In fact, at large distances the effect of the EH parameter appears as a $\mathcal{O}(r^{-6})$ term in the metric function, meaning that it is completely negligible
\begin{equation}
    b(r\to\infty) \sim -\frac{2\alpha Q_{m}^{4}}{5r^{6}}+\mathcal{O}\left(\frac{1}{r}\right)^{7}~.
\end{equation}
However, the EH parameter allows for the existence of more than one BH horizon, as we mentioned earlier, and contributes to the effective potential near the event horizon, as we depict in FIG. \ref{fig:12}. In particular, in FIG. \ref{fig:12}, we plot the effective potential for three different values of $\alpha$, in the AdS case. The plot starts from the event horizon $r_{\text{H}}$. As we increase the value of $\alpha$, the event horizon becomes slightly bigger. It is also obvious that, for $r>r_{\text{H}}+10^{-6}$, the three cases are indistinguishable. In conclusion, when the EH parameter ($\alpha$) remains small, i.e. approximately around $10^{-3}M^{2}$, it does not play a significant role in the trajectories of uncharged particles. However, it is expected that $\alpha$ plays a dramatic role near the singularity of the BH. In our model, $\alpha$ controls the behaviour of spacetime near the origin (Eqs. (18) and (26) in  \cite{Karakasis:2022xzm}).

\begin{figure}[h]
    \centering
    \includegraphics[width=0.4\textwidth]{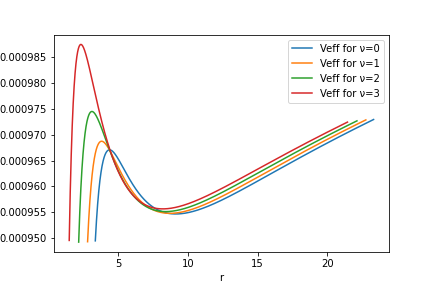}
    \includegraphics[width=0.36\textwidth]{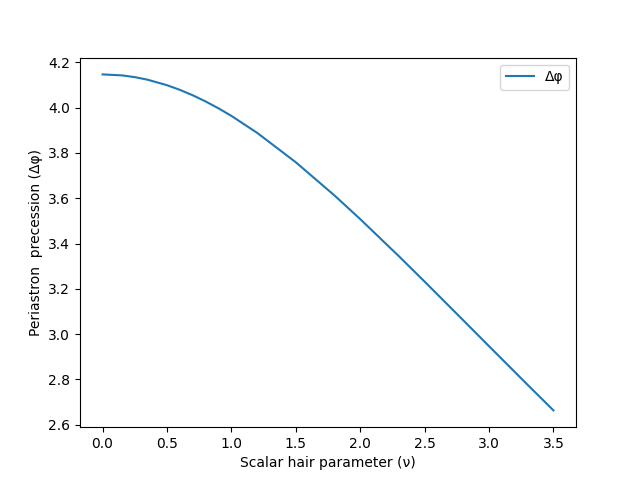}
    \caption{Left graph: Effective potential, in the AdS case for different values of $\nu$. Right graph: Periastron precession of particle's planetary orbits, in the AdS case, for different values of $\nu$. In both graphs, we have fixed $L=0.0037M^{2}$, $\Lambda_{\text{eff}} = -5\cdot10^{-5}M^{-2}$, $Q_{m}=0.2M$, $\alpha=0.002M^{2}$ and  $E=9.56\cdot10^{-4}M$.}
    \label{fig:13}
\end{figure}

To explore the contribution of the scalar charge to the geodesics, we fix the values of the parameters $Q_{m}=0.2M$ and $\alpha=0.002M^{2}$. For reasonable values of the parameters $Q_{m}$ and $\alpha$, such as $Q_{m}<3M$ and $\alpha<1M^{2}$, the scalar hair barely contributes to the effective potential when the scalar charge takes values such that $\nu<0.1M$. The corrections of the scalar hair become significant for $\nu\geq 1M$. That is why we calculated the time-like and null geodesics for $\nu=1M$ earlier. In the left graph of FIG. \ref{fig:13}, we present the effective potential for several values of the scalar charge including the hairless case where $\nu=0$. For the chosen non-zero values of $\nu$-parameter, the scalar hair plays an important role in our model by affecting the observables rather than the different types of geodesics. The effective potential is only plotted for the AdS case, since the conclusions regarding the impact of the scalar charge on the observables are the same compared to those in the other two cases, namely dS and flat. The potential is getting larger as we increase the scalar hair parameter, and hence, a larger total energy is needed for a particle to fall into the BH, keeping the particle's angular momentum fixed. The behaviour of the scalar hair parameter has the same effect as the magnetic charge. It can be easily verified that, in the Schwarzschild case, the effective potential takes smaller values as the BH mass increases. Consequently, both the parameters $Q_{m}$ and $\nu$ have the opposite gravitational effect when compared to the mass of the BH. A large gravitational mass gives a large BH, while large values of the charges $Q_{m}$, $\nu$ imply a small BH.
Additionally, in the right graph of FIG. \ref{fig:13}, we present the periastron precession given by Eq. (\ref{PerPrec}) with respect to the scalar charge ($\nu$). It is clear that the periastron precession is reduced and the orbits tend to be more ``closed", as the scalar charge increases. This phenomenon also demonstrates that the BH shrinks as the scalar charge increases.
\begin{figure}
    \centering
    \includegraphics[width=0.3\textwidth]{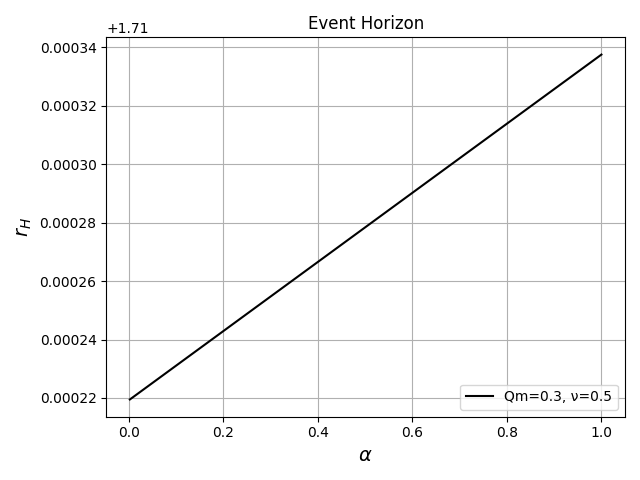}
    \includegraphics[width=0.3\textwidth]{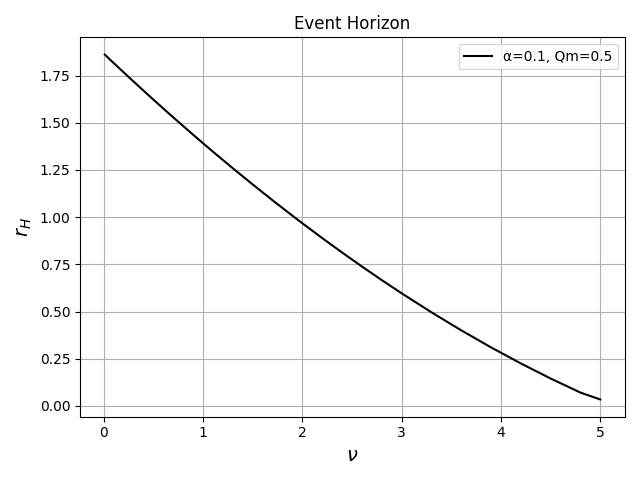}
    \includegraphics[width=0.3\textwidth]{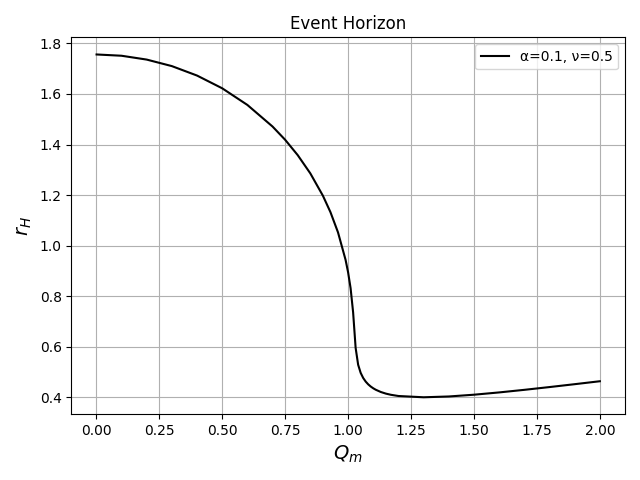}
    \caption{Radius of the BH event horizon as a function of the Euler-Heisenberg parameter (left graph), scalar charge (middle graph) and the magnetic charge (right graph), in the asymptotically flat case.}
    \label{fig:14}
\end{figure}

Furthermore, we can investigate the impact of the BH parameters, namely $\alpha,~\nu$ and $Q_{m}$, on the BH event horizon. For the following discussion, we consider values of the parameters that correspond to a BH with one horizon. Later in this subsection, we will explore in detail cases of BHs with more than one horizon. In the three graphs of FIG. \ref{fig:14}, the radius of the event horizon ($r_{\text{H}}$), which satisfies the condition $b(r_{\text{H}})=0$, is illustrated as a function of the three parameters. The plots are dedicated to the asymptotically flat case, and in each plot, two out of three parameter are fixed. As can be clearly observed from the plots, the BH grows as the $\alpha$-parameter increases, whereas it shrinks as the scalar charge rises. The impact of the $\alpha$-parameter on the event horizon is minor. The radius of the event horizon also follows downward trends as the BH magnetic charge increases, until it reaches a critical value, which is around $1.3M$ for the chosen set of parameters. When the magnetic charge surpasses its critical value, the BH's size increases while the magnetic charge takes larger values. This phenomenon is caused by the NLED terms. Let us consider the Einstein-Euler-Heisenberg metric of Eq. (\ref{EEHmetric}). For small values of the BH magnetic charge ($Q_{m}$), the Reissner-Nordstr\"om term $\left(Q_{m}^{2}/r^{2}\right)$ dominates over the NLED term $\left(-2\alpha Q_{m}^{4}/(5r^{6})\right)$, and therefore, the BH gets smaller as its magnetic charge increases. On the contrary, when the magnetic charge surpasses its critical value, the NLED term, which has the same sign as the mass-term of the metric, becomes more significant than the Reissner-Nordstr\"om term, and hence, the BH grows for large values of the magnetic charge. The same interpretation can be given in the hairy case. All the aforementioned results also hold in the (A)dS cases.

\begin{figure}[h]
    \centering
    \includegraphics[width=0.4\textwidth]{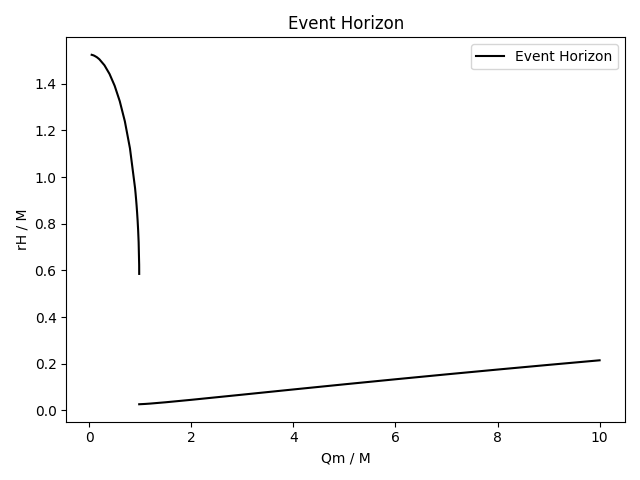}
    \includegraphics[width=0.4\textwidth]{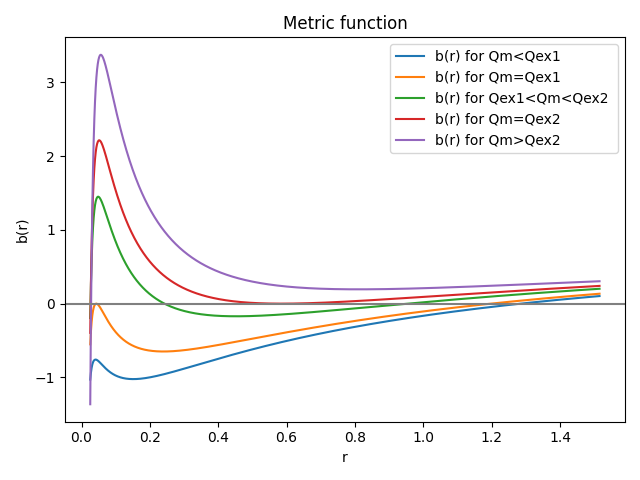}
    \caption{Left panel: Event horizon with respect to the BH magnetic charge around an extremal scenario. Right panel: Metric function $b(r)$ for various values of the BH magnetic charge. Both graphs are plotted for
    $\nu=1M$, $\alpha=0.002M^{2}$ and $\Lambda_{\text{eff}}=0$.}
    \label{fig:15}
\end{figure}

Interestingly, some extremal scenarios exist in which the radius of a horizon is a root of both the metric function ($b(r)$) and its first radial derivative. The conditions for these scenarios read
\begin{equation}
    b(r_{\text{ex}})=0=\frac{db}{dr}\Big|_{r=r_{\text{{ex}}}}~.\label{extremalConditions}
\end{equation}
In our case, these equations cannot be analytically solved due to the complexity of the metric function given by Eq. (\ref{MetricFunction}). For values of $\alpha$-parameter of the order of $10^{-1}M^{2}$ or larger, solutions to Eqs. (\ref{extremalConditions}) cannot be found. Thus, to investigate the consequences of the extremal scenarios, we consider a small value for the EH parameter ($\alpha=0.002M^{2}$) and an asymptotically flat spacetime ($\Lambda_{\text{eff}}=0$). Upon fixing one of the two charges, namely the scalar charge ($\nu$) or the magnetic charge ($Q_{m}$), we can determine the remaining charge and the BH horizon in an extremal scenario by utilizing Eq. (\ref{extremalConditions}). The graphs of FIG. \ref{fig:15} are dedicated to the case of a fixed scalar charge $\nu=1M$. To be more precise, the graph in the left panel of FIG. \ref{fig:15} demonstrates the behavior of the event horizon as the value of the magnetic charge $Q_{m}$ varies. For the chosen set of parameters, according to Eq. (\ref{extremalConditions}), we have two extremal scenarios that read $Q_{m}=Q_{\text{ex2}}=0.98098M$, $r_{\text{ex2}}=0.0266038M$ and $Q_{m}=Q_{\text{ex1}}=0.73928M$, $r_{\text{ex1}}=0.0433922M$. For small values of $Q_{m}$, such that $Q_{m}<Q_{\text{ex2}}$, the radius of the event horizon decreases as the $Q_{m}$ increases. At the second extremal scenario, where $Q_{m}=Q_{\text{ex2}}$, the graph becomes discontinuous. When we have $Q_{m}>Q_{\text{ex2}}$, the event horizon follows upward trends.

To give an interpretation to the discontinuity in the values of the event horizon, we depict the metric function $b(r)$ given by Eq. (\ref{MetricFunction}) for five values of the magnetic charge in the right panel of FIG. \ref{fig:15}. For small values of $Q_{m}$, such that $Q_{m}<Q_{\text{ex1}}$, the equation $b(r)=0$ has one positive solution, which corresponds to the radius of the event horizon of the BH. In the first extremal scenario, where $Q_{m}=Q_{\text{ex1}}$, a second innermost horizon arises corresponding to a positive double-root of equation $b(r)=0$. For values of the BH magnetic charge $Q_{m}$ such that $Q_{\text{ex1}}<Q_{m}<Q_{\text{ex2}}$, three positive solutions of the equation $b(r)=0$ exist corresponding to three BH horizons. The largest solution is identified as the BH event horizon, while the other two represent the radii of two inner horizons. Although the number of the BH horizons varies from one to three as the magnetic charge $Q_{m}$ increases in such a way that satisfies the relation $Q_{m}<Q_{\text{ex2}}$, the BH event horizon (outermost horizon) varies with a continuous way with respect to the magnetic charge. In the second extremal scenario, where $Q_{m}=Q_{\text{ex2}}$, the two larger solutions of the equation $b(r)=0$ coincide and form an event horizon. In this case, there is also an inner horizon as the right picture of FIG. \ref{fig:15} demonstrates. For $Q_{m}>Q_{\text{ex2}}$, the small positive solution of equation $b(r)=0$ becomes the only positive solution of this equation, and hence, the inner horizon becomes the BH event horizon. Consequently, when the system reaches the second extremal scenario, the radius of the event horizon undergoes a violent change, leading to the discontinuity in the left graph of FIG. \ref{fig:15}.

\begin{figure}[h]
    \centering
    \includegraphics[width=0.4\textwidth]{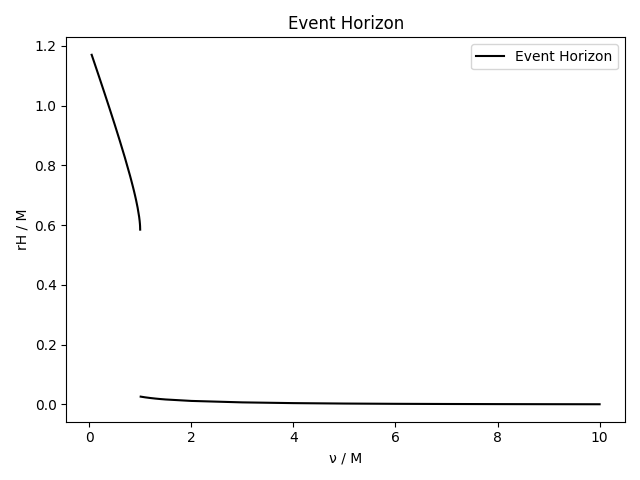}
    \includegraphics[width=0.4\textwidth]{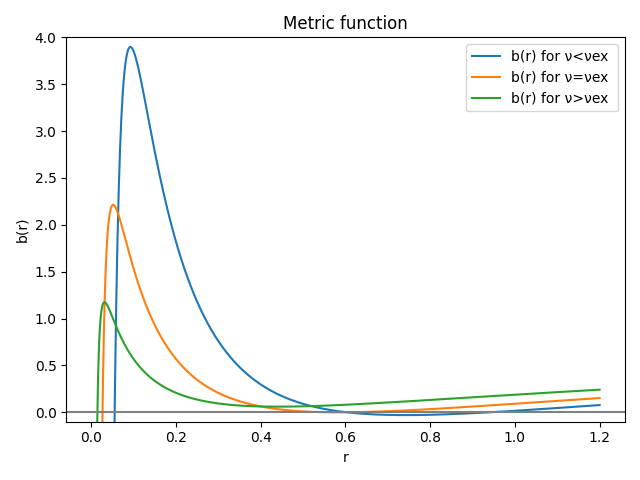}
    \caption{Left picture: Event horizon with respect to the scalar charge around an extremal scenario. Right picture: Metric function $b(r)$ for various values of the scalar charge. Both graphs are plotted for
    $Q_{m}=0.98098M$, $\alpha=0.002M^{2}$ and $\Lambda_{\text{eff}}=0$.}
    \label{fig:16}
\end{figure}

Similarly, if we fix the value of the magnetic charge $Q_{m}=0.98098M$, an extremal scenario arises that reads $r_{\text{ex}}=0.585054M$ and $\nu=\nu_{\text{ex}}=1M$. In the left picture of FIG. \ref{fig:16}, the BH event horizon is presented with respect to the scalar charge. In this case, we also have a point of discontinuity that corresponds to the extremal value of the scalar charge. In the right picture of FIG. \ref{fig:16}, the metric function given by Eq. (\ref{MetricFunction}) is depicted for three values of the scalar charge. For $\nu<\nu_{\text{ex}}$, the equation $b(r)=0$ has three positive roots with the largest of them to represent the BH event horizon. In the extremal case, where we have $\nu=\nu_{\text{ex}}$, the two larger solutions of equation $b(r)=0$ coincide. The double-root corresponds to the BH event horizon. In this case, there is also an inner horizon, as it can be clearly seen in the right panel of FIG. \ref{fig:16}. For $\nu>\nu_{\text{ex}}$, only the small horizon survives, which represents the BH event horizon. Therefore, similarly to the previous case, the extremal scenario induces the discontinuity in the left picture of FIG. \ref{fig:16}. Despite the presence of the discontinuity in the values of the radius of the BH event horizon, the BH seems to always shrink as the scalar charge increases, unlike the case of varying magnetic charge.

Finally, in case the EH parameter takes values around $\alpha=10^{-1}M^{2}$ or larger, there are no extremal scenarios, i.e. the conditions given in Eq. (\ref{extremalConditions}) cannot be satisfied. Additionally, for $\alpha<0.1M^{2}$, there are particular choices of the parameters $Q_{m}$ and $\nu$, for which there are no extremal scenarios. The sets of parameters that allow or not the existence of extremal scenarios cannot be explicitly specified, since the Eqs. (\ref{extremalConditions}) cannot be analytically solved. In the aforementioned cases, where there are no extremal scenarios, the BH has always one horizon. The horizon's radius continuously varies with respect to the BH parameters, as it can be observed from FIG. \ref{fig:14}. Last but not least, the discontinuity in the BH's size induced by the extremal scenarios allows for the existence of a stable circular orbit, an innermost unstable circular orbit and planetary orbits in the case of massless particles and photons. This phenomenon will be explored in the case of photons, which has observational interest, in Section \ref{sec5}.

\section{Photon propagation and black hole shadow}\label{sec5}

In this Section, we aim to determine the BH shadow in the asymptotically flat case, as the impact of the cosmological constant on the observables near the BH is limited. Since we work in a NLED context, photons do not follow the null geodesics of the geometry described by Eq. (\ref{Metric}). As a consequence of photon self-interactions implied by $L(P,Q)$, photons execute null-geodesic motion in an \emph{effective geometry}, initially explored in \cite{Novello:1999pg}. The authors of the articles \cite{Obukhov:2002xa,Stuchlik:2019uvf} come to the same conclusion as in \cite{Novello:1999pg} following different methods. In our work we will present the procedure of determining the \emph{effective geometry} as it is demonstrated in \cite{Novello:1999pg}. Then, we will determine the photon spheres of the BH and the BH shadow.

We assume that the wavefront surface is represented by the surface of discontinuity $\Sigma$.  According to the Hadamard method \cite{J. Hadamard}, the E/M fields are continuous at the surface $\Sigma$, while their derivative possesses a finite discontinuity there. We denote these conditions as follows
\begin{equation}
    \left[F_{\mu\nu}\right]_{\Sigma}=0~,~\left[\nabla_{\lambda}F_{\mu\nu}\right]_{\Sigma}=f_{\mu\nu}k_{\lambda}~,\label{Discontinuous}
\end{equation}
where the symbol
\begin{equation}
    \left[J\right]_{\Sigma}=\lim_{\delta\to0^{+}}\left(J|_{\Sigma+\delta}-J|_{\Sigma-\delta}\right)~,
\end{equation}
represents the discontinuity of the arbitrary function $J$ at the surface $\Sigma$. The tensor $f_{\mu\nu}$ denotes the discontinuity of the fields and the covector $k_{\mu}$ satisfies the equation
\begin{equation}
    k_{\mu}=\nabla_{\mu}\Sigma\label{PropVec}~,
\end{equation}
and is identified as the photon propagation vector. We will proceed further by considering a case in which $L=L(P)$. In other words, we assume that the scalar $Q$ is equal to zero, as it is in our case. Thus, by imposing the discontinuity condition of Eq. (\ref{Discontinuous}) on the generalized Maxwell equations given by Eq. (\ref{EMEOM}), we obtain
\begin{equation}
    L_{P}f^{\mu\nu}k_{\nu}+2L_{PP}f^{\alpha\beta}F_{\alpha\beta}F^{\mu\nu}k_{\nu}=0~.\label{Pr1}
\end{equation}
Moreover, by considering the discontinuity condition presented in Eq. (\ref{Discontinuous}), the Bianchi identity $\nabla_{\mu}F_{\nu\rho}+\nabla_{\rho}F_{\mu\nu}+\nabla_{\nu}F_{\rho\mu}=0$ yields
\begin{equation}
    k_{\mu}f_{\nu\rho}+k_{\rho}f_{\mu\nu}+k_{\nu}f_{\rho\mu}=0~.\label{Pr2}
\end{equation}
Scalar relations arise by contracting the Eq. (\ref{Pr1}) with $k_{\rho}F^{\rho}_{~\mu}$ and Eq. (\ref{Pr2}) with $k_{\alpha}g^{\alpha\rho}F^{\mu\nu}$
\begin{align}
    &k_{\mu}k_{\nu}g^{\mu\nu}+\frac{1}{f^{\alpha\beta}F_{\alpha\beta}}2F^{\mu\nu}f_{\nu}^{~\rho}k_{\rho}k_{\mu}=0~,\\
    &\frac{L_{P}}{f^{\alpha\beta}F_{\alpha\beta}}2F^{\mu\nu}f_{\nu}^{~\rho}k_{\rho}k_{\mu}+4L_{PP}F^{\mu\rho}F_{\rho}^{~\nu}k_{\mu}k_{\nu}=0~.
\end{align}
Upon substituting the one of the above equations into the other, we obtain
\begin{equation}
    \left(-L_{P}g^{\mu\nu}+4L_{PP}F^{\mu\rho}F_{\rho}^{~\nu}\right)k_{\mu}k_{\nu}=0~.\label{Null}
\end{equation}
The above analysis is valid for $f^{\alpha\beta}F_{\alpha\beta}\neq 0$. In the alternative scenario, where we have $f^{\alpha\beta}F_{\alpha\beta}=0$, it has been proven in \cite{Novello:1999pg} that photons follow the standard null geodesics of the original geometry. In such a case, photons would move only under the effect of gravity, and hence, this case does not include photon self-interactions induced by NLED. As Eq. (\ref{Null}) demonstrates, the cotangent vector $k_{\mu}$ can be treated as a null cotangent vector in the geometry described by
\begin{equation}
    \Tilde{g}^{\mu\nu}= -L_{P}g^{\mu\nu}+4L_{PP}F^{\mu\rho}F_{\rho}^{~\nu}~,\label{EffGeo}
\end{equation}
which is the \emph{effective geometry} that we are seeking for. In the Maxwell case, where we have $L(P)=-P$, the \emph{effective geometry} ($\Tilde{g}^{\mu\nu}$) degenerates into the standard geometry $g^{\mu\nu}$. We can illustrate the contribution of the stress-energy tensor of the E/M field to the \emph{effective geometry}. Generally, in the cases where the scalar $Q$ is zero, the stress-energy tensor of the E/M field can be written as $16\pi T^{\mu\nu}_{EM}=L(P)g^{\mu\nu}+4L_{P}F^{\mu\rho}F_{\rho}^{~\nu}$, which is in agreement with Eq. (\ref{S-EtensorE/M}). Thus, the \emph{effective geometry} reads
\begin{equation}
    \Tilde{g}^{\mu\nu}= -\left(L_{P}+\frac{LL_{PP}}{L_{P}}\right)g^{\mu\nu}+\frac{L_{PP}}{L_{P}}16\pi T^{\mu\nu}_{EM}~.\label{EffGeoSET}
\end{equation}
In order to prove that the propagation vector follows the null geodesics of the \emph{effective geometry}, we define the metric $\Tilde{g}_{\mu\nu}$ from the equation
\begin{equation}
    \Tilde{g}^{\mu\rho}\Tilde{g}_{\rho\nu}=\delta_{\nu}^{\mu}~,\label{Unit}
\end{equation}
and the tangent vector ($\Tilde{k}^{\mu}$)
\begin{equation}
    \Tilde{k}^{\mu}\doteq\Tilde{g}^{\mu\nu}k_{\nu}=\Tilde{g}^{\mu\nu}g_{\nu\rho}k^{\rho}~,
    \label{PropVectors}
\end{equation}
which is defined in the \emph{effective geometry}. Eq. (\ref{PropVectors}) also illustrates the relation between the propagation vectors in the effective ($\Tilde{k}^{\mu}$) and original ($k^{\mu}$) geometries. It is also necessary to establish an underlying Riemannian structure for the manifold corresponding to the \emph{effective geometry}. This involves determining a set of Levi-Civita connection coefficients $\Tilde{\Gamma}^{\alpha}_{~\mu\nu}=\Tilde{\Gamma}^{\alpha}_{~\nu\mu}$, which are necessary for the existence of a covariant derivative $\Tilde{\nabla}_{\mu}$ such that
\begin{equation}
    \Tilde{\nabla}_{\alpha}\Tilde{g}^{\mu\nu}\equiv\partial_{\alpha}\Tilde{g}^{\mu\nu}+\Tilde{\Gamma}^{\mu}_{~\alpha\rho}\Tilde{g}^{\rho\nu}+\Tilde{\Gamma}^{\nu}_{~\alpha\rho}\Tilde{g}^{\mu\rho}=0~.\label{metricity condition}
\end{equation}
Thus, the effective connection coefficients are completely determined by the usual Christoffel formula. By differentiating Eq. (\ref{Null}), we obtain
\begin{equation}
    2\Tilde{g}^{\mu\nu}k_{\mu}\partial_{\alpha}k_{\nu}+k_{\mu}k_{\nu}\partial_{\alpha
    }\Tilde{g}^{\mu\nu}=0~.
\end{equation}
By substituting $\partial_{\alpha}\Tilde{g}^{\mu\nu}$ from Eq. (\ref{metricity condition}) into the above equation, we have
\begin{equation}
    \Tilde{k}^{\mu}\Tilde{\nabla}_{\alpha}k_{\mu}=0~.
\end{equation}
The cotangent vector $k_{\mu}$ is defined as the derivative of a scalar, as given in Eq. (\ref{PropVec}). Therefore, we have $\Tilde{\nabla}_{\alpha}k_{\mu}=\Tilde{\nabla}_{\mu}k_{\alpha}$, and hence, the propagation vector satisfies the geodesic condition
\begin{equation}
    \Tilde{k}^{\mu}\Tilde{\nabla}_{\mu}k_{\nu}=0~.
\end{equation}
The vectors $K = \partial_{t}$ and $R = \partial_{\phi}$ constitute Killing vectors of the \emph{effective geometry} and correspond to the conserved total energy $E$ and angular momentum $L$ of photons. Thus, the Eqs. (\ref{E}) and (\ref{J}) imply
\begin{equation}
    \Tilde{k}^{t}=-\frac{E}{\Tilde{g}_{tt}}~,~\Tilde{k}^{\phi}=\frac{L}{\Tilde{g}_{\phi\phi}}~.
\end{equation}
Taking into account the Eq. (\ref{P}) for $\mathcal{A}(r)=0$, the scalars $L$, $L_{P}$ and $L_{PP}$ in our case read
\begin{equation}
    L=-\frac{2 Q_{m}^{2}}{w^{4}(r)} +\frac{4\alpha Q_{m}^{4}}{w^{8}(r)}~,~L_{P}=-1+\frac{4\alpha Q_{m}^{2}}{w^{4}(r)}~,~L_{PP}=2\alpha~.
\end{equation}
Additionally, the Eq. (\ref{EffGeoSET}) yields the inverse effective metric tensor
\begin{equation}
    \Tilde{g}^{\mu\nu}=\text{diag}\left(-\frac{g(r)}{b(r)},~g(r)b(r),~\frac{h(r)}{w^{2}(r)},~\frac{h(r)}{w^{2}(r)\sin^{2}\theta}\right)~,
\end{equation}
while the line element of the \emph{effective geometry} is implied by taking the inverse of $\Tilde{g}^{\mu\nu}$
\begin{equation}
    d\Tilde{s}^{2}=-\frac{b(r)}{g(r)}dt^{2}+\frac{1}{g(r)b(r)}dr^{2}+\frac{w^{2}(r)}{h(r)}d\theta^{2}+\frac{w^{2}(r)\sin^{2}\theta}{h(r)}d\phi^{2}~,\label{EffLineElement}
\end{equation}
where
\begin{equation}
    g(r)=1-\frac{4\alpha Q_{m}^{2}}{w^{4}(r)}~,~h(r)=1-\frac{12\alpha Q_{m}^{2}}{w^{4}(r)}~.
\end{equation}
The effective metric is static and spherically symmetric like the original one. Eventually, photon equations of motion, which are identified as the null geodesic equations of the \emph{effective geometry}, arise from the above discussion and read
\begin{align}
    \Tilde{k}^{t}&=\frac{g(r)}{b(r)}E~,\\
    \Tilde{k}^{\phi}&=\frac{h(r)}{w^{2}(r)}L~,\\
    \Tilde{k}^{r}&=\pm g(r)E\sqrt{1-b(r)\frac{h(r)\ell^{2}}{g(r)w^{2}(r)}}~,\label{RadVelPhoton}
    %\\
    %\frac{dr}{d\phi}&=\pm w^{2}(r)\sqrt{\frac{g^{2}(r)}{h^{2}(r)\ell^{2}}-\frac{g(r)b(r)}{h(r)w^{2}(r)}}~,\label{drdphiPhoton}
\end{align}
where the photon motion is developed at the equatorial plane $\theta=\pi/2$ and $\ell=L/E$ denotes the impact parameter. The above equations of motion are similar to those of \cite{Stuchlik:2019uvf}, where the authors work on a regular Bardeen BH. However, our \emph{effective geometry} (\ref{EffLineElement}) differs from the one obtained in \cite{Allahyari:2019jqz}, where a magnetically charged EH BH without scalar hair is discussed. The difference between the results does not exclusively lie in the existence of the scalar hair in our case, instead the effective line element of Eq. (2.25) of \cite{Allahyari:2019jqz} does not correspond to the inverse effective metric of Eq. (2.21) of \cite{Allahyari:2019jqz}, and therefore, photons do not follow the null geodesics of the effective metric (2.25) of \cite{Allahyari:2019jqz} as it is claimed. In the analysis that follows, we will explore the BH shadows in both the cases of hairy and hairless BHs, and thus, the properties of the magnetically charged EH BH with and without scalar hair will arise.

In order to explore the photon trajectories and the BH shadow, let us define the effective potential that governs the photon motion
\begin{equation}
    V_{\text{eff}}(r)=b(r)\frac{h(r)\ell^{2}}{g(r)w^{2}(r)}~,\label{VeffPhoton}
\end{equation}
which is present in the expression of the photon radial velocity given by Eq. (\ref{RadVelPhoton}). The function $g(r)$ that multiplies the square root in Eq. (\ref{RadVelPhoton}) is a strictly increasing function and hence does not influence the types of photon trajectories, such as the circular orbits. Interestingly, we can obtain the propagation vector in the original geometry ($k^{\mu}$) by utilizing Eq. (\ref{PropVectors})
\begin{align}
    k^{t}&=\frac{E}{b(r)}~,\\
    k^{\phi}&=\frac{L}{w^{2}(r)}~,\\
    k^{r}&=\pm E\sqrt{1-b(r)\frac{h(r)\ell^{2}}{g(r)w^{2}(r)}}~.\label{RadVelPhotonOri}
    %\\
    %\frac{dr}{d\phi}&=\pm w^{2}(r)\sqrt{\frac{g^{2}(r)}{h^{2}(r)\ell^{2}}-\frac{g(r)b(r)}{h(r)w^{2}(r)}}~,\label{drdphiPhoton}
\end{align}
Also in this case the photon radial velocity (\ref{RadVelPhotonOri}) is governed by the effective potential provided by Eq. (\ref{VeffPhoton}).

Let us start our analysis by considering certain values for the parameters, such as $\alpha\geq 10^{-1}M^{2}$, for which there are no extremal scenarios, i.e. the conditions described by Eq. (\ref{extremalConditions}) cannot be satisfied. For these sets of parameters, the photon effective potential given by Eq. (\ref{VeffPhoton}) is similar to the one depicted in the left panel of FIG. \ref{fig:9}. Therefore, there are three different kinds of trajectories, as in the case of massless particles explored in Section \ref{sec4}. A photon has the potential to be deflected by the BH, to fall into it, or to adopt asymptotically unstable circular orbits around it. The unstable circular orbits correspond to the maximum of the effective potential. The radius of the unstable circular orbit can be identified as the radius of the BH photon sphere $r_{\text{ph}}$. Additionally, in order for a photon to follow an asymptotically unstable circular orbit, the photon's radial velocity must vanish at $r=r_{\text{ph}}$. Thus, the conditions that describe an unstable circular orbit are
\begin{equation}
    V_{\text{eff}}(r_{\text{ph}})=1~,~\frac{dV_{\text{eff}}}{dr}\Big|_{r=r_{\text{ph}}}=0~.\label{ConditionsPh}
\end{equation}
The condition $V_{\text{eff}}(r_{\text{ph}})=1$ is implied by the nullification of the photon radial velocity given by Eq. (\ref{RadVelPhoton}). This holds, since the function $g(r)$, which multiplies the square root in Eq. (\ref{RadVelPhoton}), always reaches the value of zero at a radius smaller than $r_{\text{ph}}$. The radius of the photon sphere ($r_{\text{ph}}$) is obtained by applying the condition in the right of Eq. (\ref{ConditionsPh}). By utilizing the condition $V_{\text{eff}}(r_{\text{ph}})=1$, we can determine the specific value of the impact parameter $\ell_{\text{ph}}$ associated with the unstable circular orbits. Moreover, as we aim to determine the BH shadow, we are interested in the region exterior to the event horizon. The event horizon, in the original geometry given by Eq. (\ref{Metric}), arises as a solution of the equation $b(r_{\text{H}})=0$.  Additionally, the photon sphere, in the \emph{effective geometry} described by Eq. (\ref{EffLineElement}), must be located in the region where the function $h(r)$ is positive, in order for the \emph{effective geometry} not to flip its signature during the photon motion. Fortunately, the radius of the photon sphere $r_{\text{ph}}$ consistently remains bigger than the radii $r_{\text{H}}$ and $r_{\text{h}}$, which satisfy the equations $b(r_{\text{H}})=0$ and $h(r_{\text{h}})=0$, respectively. If the photon sphere was hidden behind either the spherical surface of radius $r_{\text{H}}$ or $r_{\text{h}}$, then all the possible photon trajectories with negative initial radial velocity would fall into the BH. This would lead to a divergent size of the BH shadow for certain choices of BH parameters. In addition to the unstable circular orbits, there is also interest in the photon trajectories that fall into the BH. In this case, a photon, after crossing the horizons of radii $r_{\text{H}}$ and $r_{\text{h}}$, is always deflected before it reaches the radius $r_{\text{g}}$, which satisfies the equation $g(r_{\text{g}})=0$. This may lead to interesting phenomena regarding the geodesic completeness of the spacetime. However, such an analysis would be out of the scope of the current paper.

As it is demonstrated in the review paper \cite{Perlick:2021aok}, the radius of a BH shadow ($R_{\text{sh}}$) observed by a very distant observer, in a spherically symmetric geometry, coincides with the impact parameter $\ell_{\text{ph}}$ associated with the unstable circular orbits, or in other words, with the photon sphere. In this case, the impact parameter satisfies the relation on the left of  Eq. (\ref{ConditionsPh}). Therefore, the radius of a BH shadow reads
\begin{equation}
    R_{\text{sh}}=\ell_{\text{ph}}\equiv\sqrt{\frac{g(r_{\text{ph}})}{h(r_{\text{ph}})b(r_{\text{ph}})}}w(r_{\text{ph}})~. \label{Shadow}
\end{equation}
Additionally, the angular size of a BH shadow measured by a very distant observer is
\begin{equation}
    \phi_{0}\approx \frac{\ell_{\text{ph}}}{r_{0}}~,\label{AngularMagnitude}
\end{equation}
where $r_{0}$ is the radial distance between the BH and the observer.

\begin{figure}
    \centering
    \includegraphics[width=0.4\textwidth]{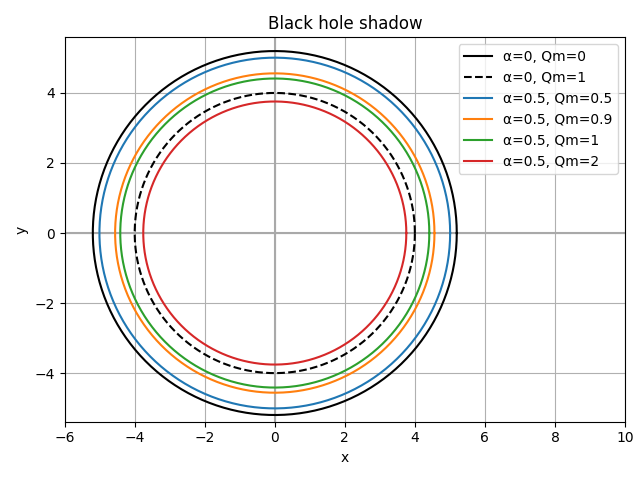}
    \includegraphics[width=0.4\textwidth]{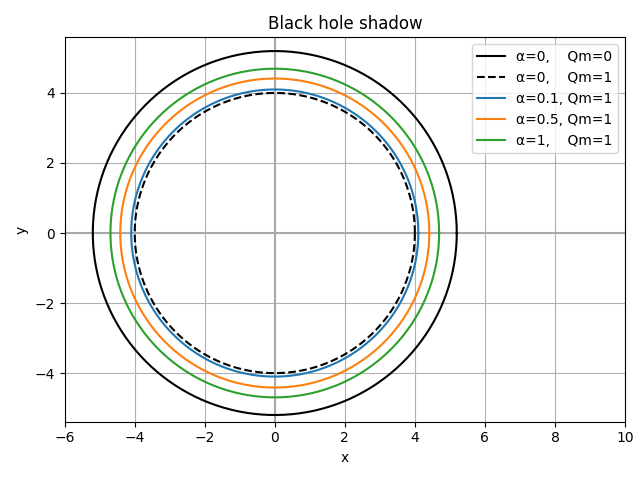}
    \caption{BH shadow, as observed by a distant observer, in the Schwarzschild ($\alpha=0$, $Q_{m}=0$), extremal Reissner-Nordstr\"om ($\alpha=0$, $Q_{m}=1M$) and EH without scalar hair ($\alpha\neq0$, $Q_{m}\neq0$, $\nu=0$) cases. BH shadows are plotted for various values of the BH magnetic charge (left figure) and the EH parameter (right figure). Both figures are plotted in mass units, where $M=1$.}
    \label{fig:17}
\end{figure}

In Table \ref{Table:1}, we present numerical calculations of the radii of the \emph{effective geometry}'s horizons described by the equations $b(r_{\text{H}})=0$ and $h(r_{\text{h}})=0$, as well as the radii of the photon sphere ($r_{\text{ph}}$) and the BH shadow ($\ell_{\text{ph}}$), in the case without scalar hair ($\nu=0$). This case is described by the metric function (\ref{EEHmetric}) for $\Lambda_{\text{eff}}=0$, which corresponds to a magnetically charged EH BH. We present the values of radii $r_{\text{H}}$ and $r_{\text{h}}$ in order to verify that they are always smaller than the radius of the photon sphere. We do not examine the behavior of the function $g(r)$, as it consistently takes smaller values than the function $h(r)$ for every radius $r$. Table \ref{Table:1} is focused on cases with a large EH parameter, in which extremal scenarios are not present. In other words, the conditions specified in Eq. (\ref{extremalConditions}) cannot be satisfied, and the corresponding BHs possess only one horizon - the event horizon. As we highlighted in Section \ref{sec4}, the BH grows as the EH parameter increases, while it shrinks as the magnetic charge increases up to a critical value. Table \ref{Table:1} demonstrates that the shadow of the BH behaves similarly to its size as the EH parameter and BH magnetic charge vary. In FIG. \ref{fig:17}, we depict the BH shadow for a wide range of the BH parameters. The magnetically charged BHs seem to form a smaller shadow than the Schwarzschild BH. In the left picture of FIG. \ref{fig:17}, the BH shadow seems to shrink as the BH magnetic charge increases. On the contrary, the right picture of FIG. \ref{fig:17} illustrates that the BH shadow rises as the EH parameter increases.
\begin{table}[]
    \centering
    \begin{center}
    \begin{tabular}{|| p{0.6cm} | p{0.6cm} || p{1.5cm} | p{1.5cm} | p{1.5cm} | p{1.5cm} || p{0.6cm} | p{0.6cm} || p{1.5cm} | p{1.5cm} | p{1.5cm} | c ||}
    \hline
         \centering$\alpha$ & \centering$Q_{m}$ & \centering$\ell_{\text{ph}}$ & \centering$r_{\text{ph}}$ & \centering$r_{\text{H}}$ & \centering$r_{\text{h}}$ & \centering$\alpha$ & \centering$Q_{m}$ & \centering$\ell_{\text{ph}}$ & \centering$r_{\text{ph}}$ & \centering$r_{\text{H}}$ & $r_{\text{h}}$ \\
         \hline
         \centering$0$ & \centering$0$ & \centering$3\sqrt{3}$ & \centering$3$ & \centering$2$ & \centering$-$ & \centering$0.5$ & \centering$0.5$ & \centering$5.00711$ & \centering$2.85435$ & \centering$1.86662$ & $1.10668$ \\
         \hline
         \centering$0$ & \centering$0.5$ & \centering$4.96791$ & \centering$2.82288$ & \centering$1.86603$ & \centering$-$ & \centering$0.5$ & \centering$0.9$ & \centering$4.5594$ & \centering$2.52251$ & \centering$1.46724$ & $1.48477$ \\
         \hline
         \centering$0$ & \centering$0.9$ & \centering$4.31923$ & \centering$2.29373$ & \centering$1.43589$ & \centering$-$ &  \centering$0.5$ & \centering$1$ & \centering$4.41229$ & \centering$2.43326$ & \centering$1.27507$ & $1.56508$ \\
         \hline
         \centering$0$ & \centering$1$ & \centering$4$ & \centering$2$ & \centering$1$ & \centering$-$ & \centering$0.5$ & \centering$2$ & \centering$3.75569$ & \centering$2.73007$ & \centering$1.01624$ & $2.21336$ \\
         \hline
         \centering$0.1$ & \centering$0.5$ & \centering$4.97577$ & \centering$2.82908$ & \centering$1.86614$ & \centering$0.740083$ & \centering$1$ & \centering$0.5$ & \centering$5.04599$ & \centering$2.88674$ & \centering$1.86721$ & $1.31607$ \\
         \hline
         \centering$0.1$ & \centering$0.9$ & \centering$4.37047$ & \centering$2.34022$ & \centering$1.44278$ & \centering$0.992925$ & \centering$1$ & \centering$0.9$ & \centering$4.75995$ & \centering$2.72215$ & \centering$1.4928$ & $1.7657$ \\
         \hline
         \centering$0.1$ & \centering$1$ & \centering$4.09949$ & \centering$2.1051$ & \centering$1.15097$ & \centering$1.04664$ & \centering$1$ & \centering$1$ & \centering$4.69251$ & \centering$2.71505$ & \centering$1.34804$ & $1.86121$ \\
         \hline
         \centering$0.1$ & \centering$2$ & \centering$2.13463$ & \centering$1.76277$ & \centering$0.673719$ & \centering$1.48017$ & \centering$1$ & \centering$2$ & \centering$4.57445$ & \centering$3.29223$ & \centering$1.20438$ & $2.63215$ \\
         \hline
    \end{tabular}
    \end{center}
    \caption{Numerical solutions of Eq. (\ref{ConditionsPh}) and $b(r_{\text{H}})=0=h(r_{\text{h}})$, in the hairless case, for $\nu=0$ and a wide range of values of the EH parameter ($\alpha$) and the BH magnetic charge ($Q_{m}$). The $r_{\text{ph}}$ denotes the radius of the photon sphere, while the $\ell_{\text{ph}}$ represents the radius of the BH shadow. The results are given in mass units where $M=1$.}
    \label{Table:1}
\end{table}

\begin{figure}
    \centering
    \includegraphics[width=0.325\textwidth]{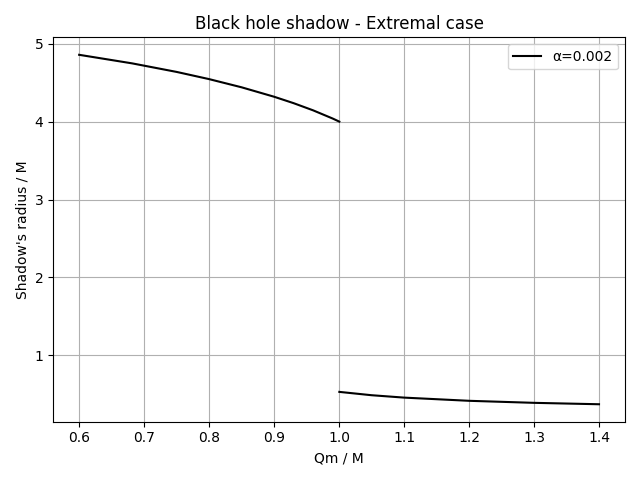}
    \includegraphics[width=0.325\textwidth]{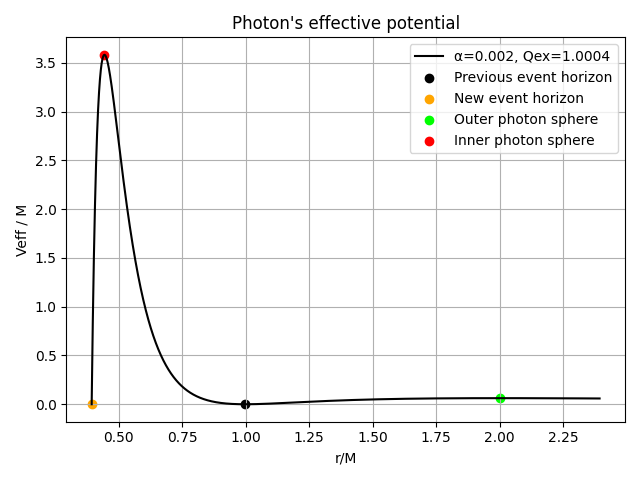}
    \includegraphics[width=0.325\textwidth]{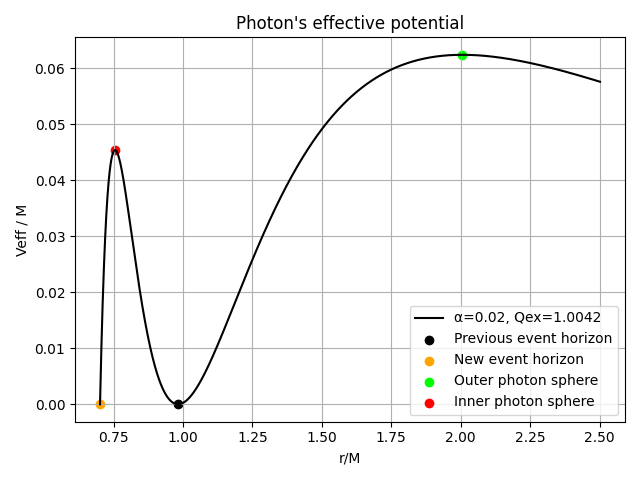}
    \caption{Left graph: Radius of the BH shadow, as observed by a distant observer, in the EH with no scalar hair case, varying with respect to the BH magnetic charge. This figure is plotted for $\alpha=0.002M^{2}$. In this case the extremal scenario is obtained for $Q_{\text{ex}}=1.00040M$. Middle and Right graphs: photon effective potential in extremal scenarios for $\alpha=0.002M^{2}$ (middle graph) and $\alpha=0.02M^{2}$ (right graph). These graphs are plotted in the region outside the horizons of the \emph{effective geometry}, where the effective metric does not change its sign. All the graphs are plotted in mass units, where $M=1$.}
    \label{fig:18}
\end{figure}

In Table \ref{Table:2}, some extremal scenarios of the hairless case are explored. We focus on extremal scenarios around which the BH event horizon suddenly relocates. In the left picture of FIG. \ref{fig:18}, we fix $\alpha=0.002M^{2}$ and plot the BH shadow with respect to the BH magnetic charge, which varies around its extremal value presented in Table \ref{Table:2}. As it can be seen in this picture, the extremal scenario also implies a point of discontinuity in the values of the radius of the BH shadow. Additionally, extremal scenarios permit the presence of a stable circular orbit, an innermost unstable circular orbit, and planetary orbits. To reveal the origins of these phenomena, we depict in the middle and right pictures of FIG. \ref{fig:18} the photon effective potential for two extremal scenarios in which the EH parameter reads $\alpha=0.002M^{2}$ and $\alpha=0.02M^{2}$, respectively. FIG. \ref{fig:18} illustrates that due to the sudden decrease in the event horizon's radius, as discussed in Section \ref{sec4}, a second local maximum of the effective potential (red point) emerges in the region where the metric does not change its sign. This occurrence implies the presence of an inner unstable circular orbit, which in the case of photons is identified with a second photon sphere. A local minimum of the photon effective potential is also revealed corresponding to a stable circular orbit of photons. In the case where the EH parameter takes values such that $\alpha\approx10^{-3}M^{2}$ or smaller, the inner photon sphere corresponds to a larger local maximum of the photon effective potential compared to the outer one, see the middle panel of FIG. \ref{fig:18}, and therefore, the inner photon sphere is correlated with the BH shadow. Thus, the sudden displacement of the BH event horizon caused by the extremal scenario, along with the emergence of a second innermost photon sphere associated with the BH shadow, indicates a point of discontinuity in the values of the radius of the BH shadow as they vary with respect to the BH magnetic charge, see the left picture of FIG. \ref{fig:18}. This phenomenon is induced by the properties of the original spacetime and also holds in the case where the scalar hair is present.

For an EH parameter of the order of $10^{-2}M^{2}$, the effective potential's local maximum corresponding to the outer photon sphere (green point) is larger than that of the inner photon sphere (red point), as the right graph of FIG. \ref{fig:18} demonstrates. In this case, one needs to consider where the observer and the light sources that cause the BH shadow are located, in order to specify the BH shadow. For instance, considering that the BH shadow is formed by light sources, which are located in a great distance from the BH, the BH shadow observed by a distant observer is correlated with the photon sphere that corresponds to the largest local maximum of the photon effective potential. The aforementioned phenomena fade out if the EH parameter takes a value around $\alpha=0.1M^{2}$ or larger, since these cases do not involve an extremal scenario.
\begin{table}[]
    \centering
    \begin{tabular}{|| p{0.8cm} | p{1.3cm} || p{1.5cm} | p{1.5cm}| p{1.5cm} | c ||}
    \hline
         \centering$\alpha$ & \centering$Q_{\text{ex}}$ & \centering$\ell_{\text{ph}}$ & \centering$r_{\text{ph}}$ &\centering$r_{\text{ex}}\equiv r_{\text{H}}$ & $r_{\text{h}}$ \\
         \hline
         \centering$0.001$ & \centering$1.00020$ &\centering$4.00025$ & \centering$2.0003$ & \centering$0.999196$ & $0.331008$ \\
         \hline
         \centering$0.002$ & \centering$1.00040$ &\centering$4.0005$ & \centering$2.0006$ & \centering$0.998384$ & $0.393677$ \\
         \hline
         \centering$0.003$ & \centering$1.00060$ &\centering$4.00074$ & \centering$2.0009$ & \centering$0.997565$ & $0.435719$ \\
         \hline
         \centering$0.004$ & \centering$1.00081$ &\centering$4.00098$ & \centering$2.0012$ & \centering$0.996737$ & $0.468258$ \\
         \hline
    \end{tabular}
    \caption{Numerical solutions of Eq. (\ref{ConditionsPh}) and $b(r_{\text{H}})=0=h(r_{\text{h}})$ in the extremal cases given by Eq. (\ref{extremalConditions}), for $\nu=0$ and a wide range of values of the EH parameter ($\alpha$). The $r_{\text{ph}}$ denotes the radius of the photon sphere, while the $\ell_{\text{ph}}$ represents the radius of the BH shadow. The results are given in mass units where $M=1$. }
    \label{Table:2}
\end{table}

\begin{figure}
    \centering
    \includegraphics[width=0.325\textwidth]{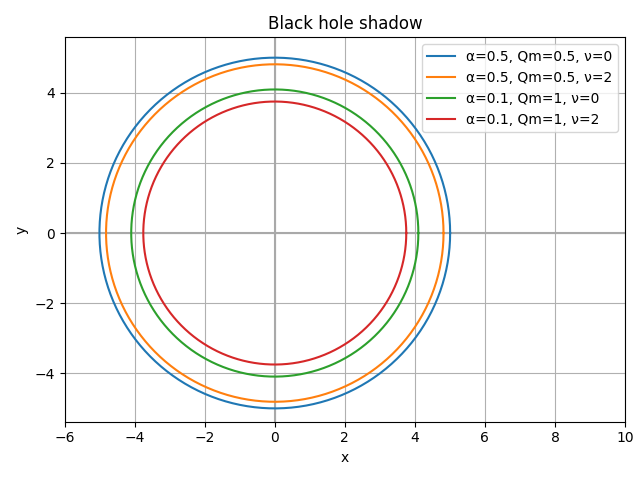}
    \includegraphics[width=0.325\textwidth]{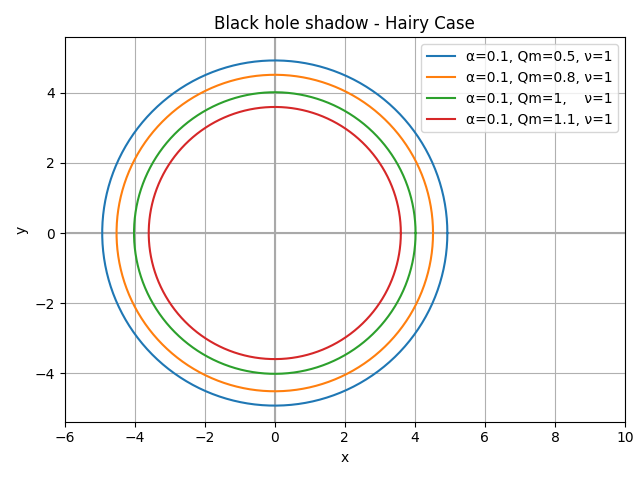}
    \includegraphics[width=0.325\textwidth]{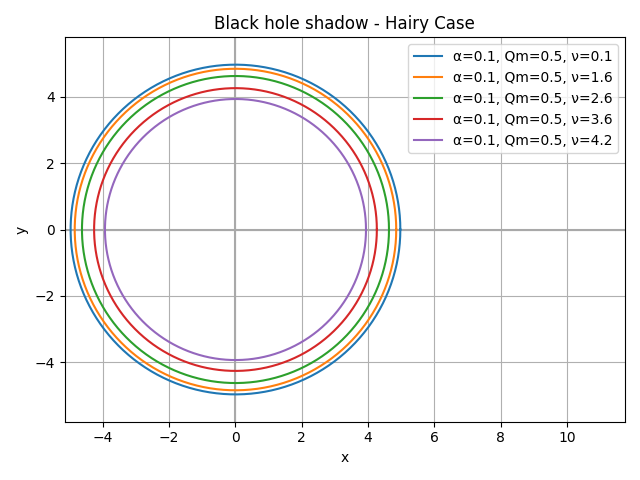}
    \caption{Shadows of a magnetically charged EH BH with scalar hair, as observed by a distant observer. The hairy EH BH shadow is compared with the shadow of an EH BH without scalar hair in the left graph, whereas the impact of the magnetic charge and the scalar charge on the hairy EH BH shadow is illustrated in the middle and right graphs respectively. All graphs are plotted in mass units, where $M=1$.}
    \label{fig:19}
\end{figure}

Let us proceed further with the case of a hairy BH. In Table \ref{Table:3}, we display the values for the radii of the horizons within the \emph{effective geometry}, characterized by the equations $b(r_{\text{H}}) = 0$ and $h(r_{\text{h}}) = 0$. Additionally, we provide the radii of the photon sphere ($r_{\text{ph}}$) and the shadow of the BH ($\ell_{\text{ph}}$). In every case the horizons of radii $r_{\text{H}}$ and $r_{\text{h}}$ are located inside the photon sphere of radius $r_{\text{ph}}$. Moreover, Table \ref{Table:3} illustrates that, as the scalar charge ($\nu$) increases, the BH shadow shrinks. The cases presented in Table \ref{Table:3} do not hold an extremal scenario. In FIG. \ref{fig:19}, we depict the shadow of the BH across a broad spectrum of its parameters. As the left graph of FIG. \ref{fig:19} demonstrates, the shadows of the hairy BHs seem to be smaller compared to the shadows in the hairless scenarios. Additionally, in the middle and right graph of FIG. \ref{fig:19}, the BH shadow gets smaller as the magnetic and scalar charges increase. Consequently, the impact of the magnetic charge and scalar hair on the BH's size, as discussed in Section \ref{sec4},  mirrors its effect on the BH shadow.
\begin{table}[]
    \centering
    \begin{center}
    \begin{tabular}{|| p{0.8cm} | p{0.6cm} | p{0.6cm} || p{1.3cm} | p{1.3cm} | p{1.3cm} | p{1.3cm} || p{0.8cm} | p{0.6cm} | p{0.6cm} || p{1.3cm} | p{1.3cm} | p{1.3cm} | c ||}
    \hline
         \centering$\alpha$ & \centering$Q_{m}$ & \centering$\nu$ & \centering$\ell_{\text{ph}}$ & \centering$r_{\text{ph}}$ & \centering$r_{\text{H}}$ & \centering$r_{\text{h}}$ & \centering$\alpha$ & \centering$Q_{m}$ & \centering$\nu$ & \centering$\ell_{\text{ph}}$ & \centering$r_{\text{ph}}$ & \centering$r_{\text{H}}$ & $r_{\text{h}}$ \\
         \hline
         \centering$0.002$ & \centering$0.2$ & \centering$0.5$ & \centering$5.15025$ & \centering$2.72304$ & \centering$1.73606$ & \centering$0.05575$ & \centering$0.1$ & \centering$1$ & \centering$0.5$ & \centering$4.07993$ & \centering$1.8501$ & \centering$0.894846$ & $0.826079$ \\
         \hline
         \centering$0.002$ & \centering$0.2$ & \centering$1$ & \centering$5.11697$ & \centering$2.47285$ & \centering$1.50489$ & \centering$0.03008$ & \centering$0.1$ & \centering$1$ & \centering$1$ & \centering$4.01978$ & \centering$1.5846$ & \centering$0.610411$ & $0.659933$ \\
         \hline
         \centering$0.002$ & \centering$0.2$ & \centering$2$ & \centering$4.97979$ & \centering$1.97201$ & \centering$1.08075$ & \centering$0.01537$ &  \centering$0.1$ & \centering$1$ & \centering$2$ & \centering$3.75459$ & \centering$1.02141$ & \centering$0.174554$ & $0.447565$ \\
         \hline
         \centering$0.002$ & \centering$0.5$ & \centering$0.5$ & \centering$4.95606$ & \centering$2.5725$ & \centering$1.62233$ & \centering$0.12411$ & \centering$0.5$ & \centering$0.5$ & \centering$0.5$ & \centering$4.99564$ & \centering$2.60428$ & \centering$1.62293$ & $0.884568$ \\
         \hline
         \centering$0.002$ & \centering$0.5$ & \centering$1$ & \centering$4.91969$ & \centering$2.32099$ & \centering$1.39125$ & \centering$0.07224$ & \centering$0.5$ & \centering$0.5$ & \centering$1$ & \centering$4.96099$ & \centering$2.35411$ & \centering$1.39189$ & $0.714391$ \\
         \hline
         \centering$0.002$ & \centering$0.5$ & \centering$2$ & \centering$4.76869$ & \centering$1.81439$ & \centering$0.967458$ & \centering$0.03801$ & \centering$0.5$ & \centering$0.5$ & \centering$2$ & \centering$4.81821$ & \centering$1.85393$ & \centering$0.968269$ & $0.491558$ \\
         \hline
         \centering$0.1$ & \centering$0.5$ & \centering$0.5$ & \centering$4.96387$ & \centering$2.57867$ & \centering$1.62244$ & \centering$0.53117$ & \centering$0.5$ & \centering$1$ & \centering$0.5$ & \centering$4.40072$ & \centering$2.18687$ & \centering$1.0315$ & $1.33493$ \\
         \hline
         \centering$0.1$ & \centering$0.5$ & \centering$1$ & \centering$4.92784$ & \centering$2.32741$ & \centering$1.39138$ & \centering$0.39315$ & \centering$0.5$ & \centering$1$ & \centering$1$ & \centering$4.36627$ & \centering$1.94862$ & \centering$0.802246$ & $1.14301$ \\
         \hline
         \centering$0.1$ & \centering$0.5$ & \centering$2$ & \centering$4.77849$ & \centering$1.82206$ & \centering$0.967618$ & \centering$0.24408$ & \centering$0.5$ & \centering$1$ & \centering$2$ & \centering$4.2336$ & \centering$1.50992$ & \centering$0.417515$ & $0.85728$ \\
         \hline
    \end{tabular}
    \end{center}
    \caption{Numerical solutions of Eq. (\ref{ConditionsPh}) and $b(r_{\text{H}})=0=h(r_{\text{h}})$ for a wide range of values of the scalar charge ($\nu$), the EH parameter ($\alpha$) and the BH magnetic charge ($Q_{m}$). The $r_{\text{ph}}$ denotes the radius of the photon sphere, while the $\ell_{\text{ph}}$ represents the radius of the BH shadow. The results are given in mass units where $M=1$.}
    \label{Table:3}
\end{table}

\begin{figure}
    \centering
    \includegraphics[width=0.3\textwidth]{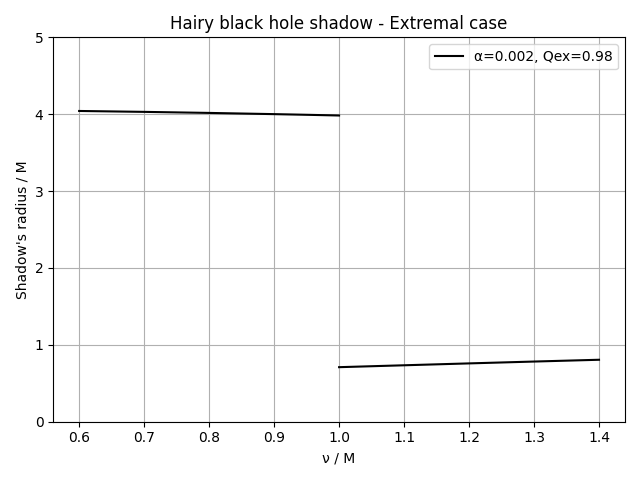}
    \includegraphics[width=0.3\textwidth]{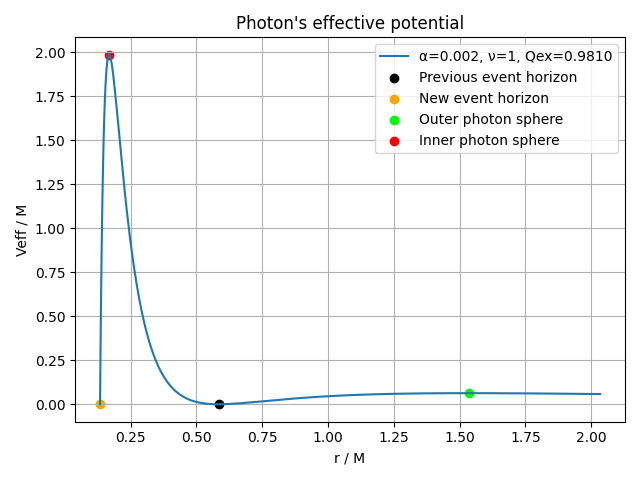}
    \includegraphics[width=0.3\textwidth]{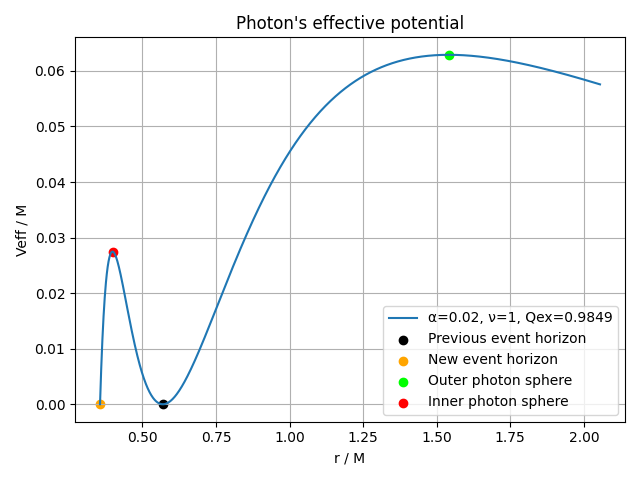}
    \caption{Left panel: Hairy BH shadow varying with respect to the scalar charge around the extremal scenario for $\alpha=0.002M^{2}$ and $Q_{ex}=0.9810M$. Middle panel: photon effective potential in the extremal scenario for $\alpha=0.002M^{2}$ and $Q_{ex}=0.9810M$. Right panel: photon effective potential in the extremal scenario for $\alpha=0.02M^{2}$ and $Q_{ex}=0.9849M$. The middle and right graphs are plotted in the region outside the horizons of the \emph{effective geometry}, where the effective metric does not change its sign. All graphs are plotted in mass units where $M=1$.}
    \label{fig:20}
\end{figure}

Furthermore, we investigate some extremal scenarios of the hairy case in Table \ref{Table:4}. In the left panel of FIG. \ref{fig:20}, we fix $\alpha=0.002M^{2}$ and $Q_{ex}=0.98M$, and plot the BH shadow with respect to the scalar charge, which varies around its extremal value demonstrated in Table \ref{Table:4}. The discontinuity in this graph is due to the extremal scenario. Similarly to the hairless case, the shift of the BH event horizon reveals a second innermost photon sphere. In the extremal scenario where $\alpha=0.002M^{2}$ and $Q_{ex}=0.98M$, the photon effective potential, which possesses an inner photon sphere (red point), is presented in the middle panel of FIG. \ref{fig:20}. In the extremal scenarios where the EH parameter and the BH magnetic charge are of the orders of $10^{-3}M^{2}$ and $1M$, respectively, the inner photon sphere corresponds to the largest local maximum of the photon effective potential, see the middle panel of FIG. \ref{fig:20}. Hence, the BH shadow is correlated with this photon sphere. In the extremal scenarios where $\alpha\sim10^{-2}M^{2}$ and $Q_{ex}\sim1$, the outer photon sphere (green point) corresponds to the largest local maximum of the photon effective potential, as it is observed in the right panel of FIG. \ref{fig:20}. In these cases, we need to decide where the light sources and the observer are located in order for the BH shadow to be determined.
\begin{table}[]
    \centering
    \begin{tabular}{|| p{0.8cm} | p{1.3cm} | p{0.6cm} || p{1.5cm} | p{1.5cm}| p{1.5cm} | c ||}
    \hline
         \centering$\alpha$ & \centering$Q_{\text{ex}}$ & \centering$\nu_{\text{ex}}$ & \centering$\ell_{\text{ph}}$ & \centering$r_{\text{ph}}$ &\centering$r_{\text{ex}}\equiv r_{\text{H}}$ & $r_{\text{h}}$ \\
         \hline
         \centering$0.002$ & \centering$0.99554$ & \centering$0.5$ &\centering$3.99629$ & \centering$1.75959$ & \centering$0.770474$ & $0.215542$ \\
         \hline
         \centering$0.002$ & \centering$0.98098$ & \centering$1$ &\centering$3.9832$ & \centering$1.53618$ & \centering$0.585054$ & $0.134013$ \\
         \hline
         \centering$0.002$ & \centering$0.92265$ & \centering$2$ &\centering$3.92428$ & \centering$1.13738$ & \centering$0.322297$ & $0.069082$ \\
         \hline
         \centering$0.004$ & \centering$0.99595$ & \centering$0.5$ &\centering$3.99677$ & \centering$1.76019$ & \centering$0.768842$ & $0.279813$ \\
         \hline
         \centering$0.004$ & \centering$0.981395$ & \centering$1$ &\centering$3.98367$ & \centering$1.53677$ & \centering$0.583464$ & $0.181919$ \\
         \hline
         \centering$0.004$ & \centering$0.92311$ & \centering$2$ &\centering$3.9247$ & \centering$1.13796$ & \centering$0.320828$ & $0.096469$ \\
         \hline
         \centering$0.007$ & \centering$0.99664$ & \centering$0.5$ &\centering$3.99755$ & \centering$1.76117$ & \centering$0.766075$ & $0.347882$ \\
         \hline
         \centering$0.007$ & \centering$0.98209$ & \centering$1$ &\centering$3.98443$ & \centering$1.53774$ & \centering$0.58076$ & $0.235295$ \\
         \hline
         \centering$0.007$ & \centering$0.92388$ & \centering$2$ &\centering$3.92536$ & \centering$1.13891$ & \centering$0.31831$ & $0.12846$ \\
         \hline
    \end{tabular}
    \caption{Numerical solutions of Eq. (\ref{ConditionsPh}) and $b(r_{\text{H}})=0=h(r_{\text{h}})$ in the extremal cases given by Eq. (\ref{extremalConditions}), for a wide range of values of the scalar charge ($\nu$) and the EH parameter ($\alpha$). The $r_{\text{ph}}$ denotes the radius of the photon sphere, while the $\ell_{\text{ph}}$ represents the radius of the BH shadow. The results are given in mass units where $M=1$.}
    \label{Table:4}
\end{table}

In conclusion, the BH shadow as observed by a distant observer behaves in the same way as the BH event horizon when the BH magnetic charge and scalar charge vary. Additionally, extremal scenarios lead to discontinuities not only in the event horizon radius of the BH but also in the radius of its shadow.

\section{Comparison of the black hole shadow with the shadows of M87* and Sgr A* obtained by the Event Horizon Telescope}\label{sec6}

In this Section, we compare the shadow of magnetically charged EH BHs with and without scalar hair with the shadows of the SMBHs M87* and Sgr A* obtained by the EHT collaboration \cite{EventHorizonTelescope:2019dse,EventHorizonTelescope:2022wkp}. As it was discussed in Section \ref{sec5}, the BH parameters have an impact on the BH shadow. Thus, by comparing the theoretically predicted BH shadow with the observational data, we aim to constrain the values of the BH parameters.

From \cite{EventHorizonTelescope:2019dse} is derived that the angular size of M87*, as observed from Earth, is $\theta_{\text{M87*}}=(42\pm3)\mu as$, its mass is $M_{\text{M87*}}= \left(6.5\pm0.2|_{\text{stat}} \pm 0.7 |_{\text{sys}} \right) \cdot10^{9}M_{\odot}$, while our distance to M87* is considered to be $D_{\text{M87*}}=16.8^{+0.8}_{-0.7} Mpc$. Using Planck units and applying error propagation, Eq. (\ref{AngularMagnitude}) implies that the diameter of the shadow of M87* in mass units reads
\begin{equation}
    d_{\text{M87*}}\approx\frac{\theta_{\text{M87*}}D_{\text{M87*}}}{M_{\text{M87*}}}=11\pm1.5~.\label{M87}
\end{equation}
Similarly, from \cite{EventHorizonTelescope:2022wkp} is obtained that the angular size of Sgr A*, as observed from Earth, is $\theta_{\text{Sgr A*}}=(48.7\pm7)\mu as$, its mass is $M_{\text{Sgr A*}}= \left(4.297 \pm 0.013\right) \cdot10^{6}M_{\odot}$, while our distance to the center of Milky Way where the SMBH Sgr A* is located is $D_{\text{Sgr A*}}=\left(8277\pm33\right) pc$. Thus, the diameter of the shadow of Sgr A* in mass units reads
\begin{equation}
    d_{\text{Sgr A*}}\approx\frac{\theta_{\text{Sgr A*}}D_{\text{Sgr A*}}}{M_{\text{Sgr A*}}}=9.5\pm1.4~.\label{Sgr}
\end{equation}

\begin{figure}
    \centering
    \includegraphics[width=0.38\textwidth]{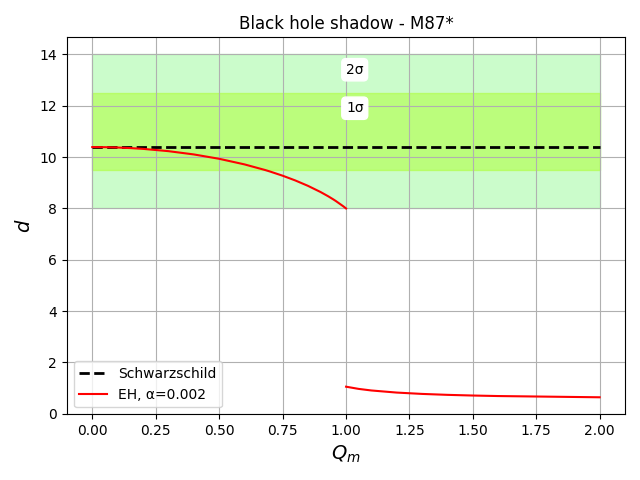}
    \includegraphics[width=0.38\textwidth]{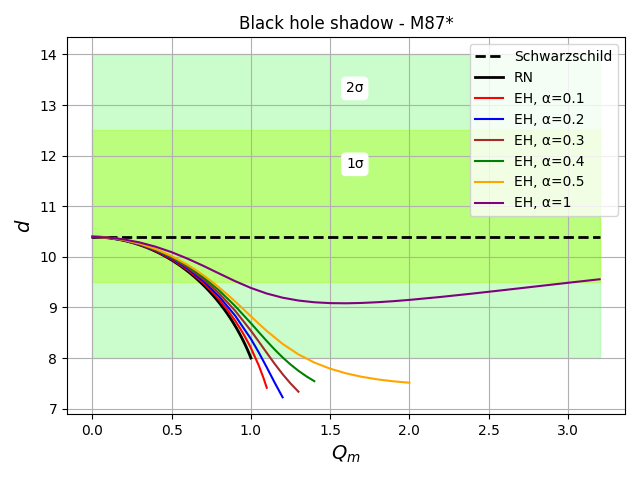}
    \includegraphics[width=0.38\textwidth]{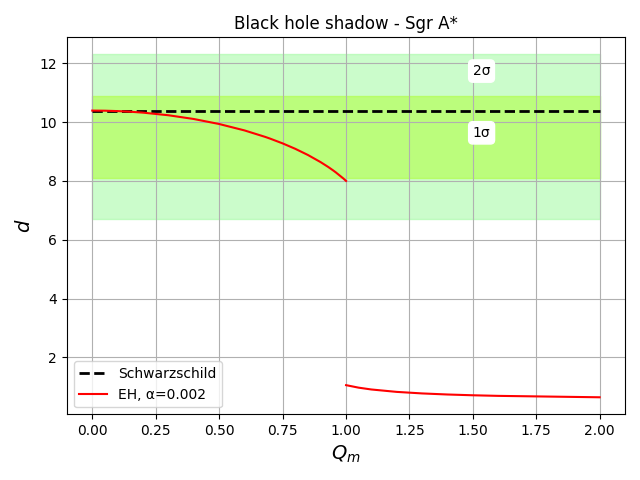}
    \includegraphics[width=0.38\textwidth]{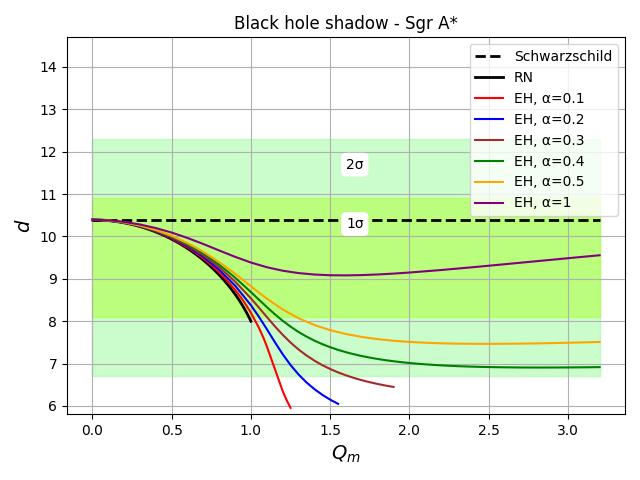}
    \caption{Shadow diameter ($d$) of an EH BH as a function of the BH magnetic charge for a small value (left graphs) and large values (right graphs) of the EH parameter. The shaded regions indicate the values of the shadow diameter consistent with the shadow of the SMBH M87* (upper row) and SMBH Sgr A* (lower row) detected by the EHT, with details in Eqs. (\ref{M87}) and (\ref{Sgr}) respectively. The narrow shaded regions give the 1$\sigma$ confidence regions, whereas the wider shaded regions give the 2$\sigma$ confidence regions. All graphs are in mass units ($M=1$).}
    \label{fig:21}
\end{figure}

To proceed further, we will consider the hairless case. In the graphs of FIG. \ref{fig:21}, we plot the shadow diameter of an EH BH with respect to the BH magnetic charge for various values of the $\alpha$-parameter. Additionally, we have shaded two regions on the graphs to indicate the range of shadow diameters consistent with the shadows of the SMBHs M87* (upper row) and Sgr A* (lower row), which were detected by the EHT, at 1$\sigma$ and 2$\sigma$ confidence levels. From these graphs, we can impose upper and lower bounds on the magnetic charge of the EH BH regarding the points where the lines cross the shaded regions.

In the left graphs of FIG. \ref{fig:21}, we plot the shadow diameter with respect to the BH magnetic charge for $\alpha=0.002M^{2}$. For values of the EH parameter of the order of $10^{-2}M^{2}$ or smaller, the plots of the shadow diameter are not distinguishable and are similar to the one in the Reissner-Nordstr\"om case, where we have $\alpha=0$. This holds for values of the BH magnetic charge up to the value of $1M$, where the extremal scenario of the Reissner-Nordstr\"om case is obtained. Thus, we only present the plots for $\alpha=0.002M^{2}$ as a representative case of small values of the EH parameter. Considering the small-$\alpha$ cases, the system reaches an extremal scenario, around the value of $Q_{m}=1M$. When we have $Q_{m}>1$, the shadow diameter takes very small values, as the left graphs of FIG. \ref{fig:21} demonstrate. The discontinuity in the values of the shadow diameter is induced by the extremal scenario, as we extensively discuss in Section \ref{sec5}.

In the right graphs of FIG. \ref{fig:21}, the shadow diameter is illustrated for various values of the EH parameter. In these graphs, we consider cases associated with values of the EH parameter of the order of $10^{-1}M^{2}$ or larger, where extremal scenarios are absent. As it can be clearly observed from these graphs, the shadow diameter takes larger values as the EH parameter increases. This phenomenon has been discussed in detail in Section \ref{sec5}. It is a reasonable result, since the BH's size also grows as the EH parameter rises, see the discussion in Section \ref{sec4}. Additionally, as we mention in Section \ref{sec4}, in the cases where the extremal scenarios are absent, when the value of the BH magnetic charge surpasses a critical value, which is around $1.3M$ when the EH parameter takes values around $1M^{2}$, the BH gets larger as the magnetic charge increases. This phenomenon mirrors the behavior of the BH shadow presented in the right graphs of FIG. \ref{fig:21}. Nevertheless, for extremely large values of the BH magnetic charge, the shadow diameter increases with a significantly low pace. Consequently, using the value of $Q_{m}=1000M$ as an extreme example, the shadow diameter barely reaches the values: $d=4.014$ for $\alpha=0.1$, $d=8.97$ for $\alpha=0.5$ and $d=12.68$ for $\alpha=1$. Interestingly, for values of $\alpha$ around $1$, the shadow diameter remains in the shaded regions, in both M87* and Sgr A* cases, for practically every value of the magnetic charge. However, this discussion holds for large values of the $\alpha$-parameter and large values of the BH magnetic charge.

In Table \ref{Table:5}, we present the bounds of the values of the BH magnetic charge for various values of the EH parameter, at 1$\sigma$ and 2$\sigma$ confidence levels, considering the constraints imposed by the observational data for the shadow of the SMBH M87*. Significantly, BHs in the NLED scenario are allowed to hold a greater amount of magnetic charge when compared to the Reissner-Nordstr\"om case ($\alpha=0$). This is because the presence of the EH parameter implies larger BH shadows. It is also noteworthy that the upper bounds of the BH magnetic charge, at 2$\sigma$ confidence level, in cases with extremal scenarios, are identical to the extremal values of the BH magnetic charge associated with extremal scenarios where the number of BH horizons decreases from three to one. For large values of the $\alpha$-parameter, such as $\alpha=1M^{2}$, there are no constraints in the values of the BH magnetic charge at 2$\sigma$ confidence level. Instead, at 1$\sigma$ confidence level, both upper and lower bounds exist, because the line for $\alpha=1M^{2}$ crosses twice the 1$\sigma$ confidence region. Therefore, for $\alpha=1M^{2}$, the allowed values of the BH magnetic charge ($Q_{m}$) are $|Q_{m}|\in[0,0.9139]\cup[3.0374,+\infty)$. We use the infinity symbol ($+\infty$) to refer to enormously large values of the BH magnetic charge when compared to the mass, such as $Q_{m}>1000M$.
\begin{table}
    \centering
    \begin{tabular}{|| p{0.8cm} || p{2.6cm} | p{2.6cm} | p{2.6cm} | c ||}
    \hline
    \multicolumn{5}{||c||}{Constraints on the BH magnetic charge - M87*} \\
    \hline
         \centering$\alpha$ & \centering$\text{Upper bound}~(1\sigma)$ & \centering$\text{Lower bound}~(1\sigma)$ & \centering$\text{Upper bound}~(2\sigma)$ & $\text{Lower bound}~(2\sigma)$  \\
         \hline
         \centering$0$ & \centering$0.6806$ & \centering$-$ &\centering$1$ & $-$ \\
         \hline
         \centering$0.002$ & \centering$0.6809$ & \centering$-$ &\centering$1.0004$ & $-$ \\
         \hline
         \centering$0.007$ & \centering$0.6815$ & \centering$-$ &\centering$1.0015$ & $-$ \\
         \hline
         \centering$0.01$ & \centering$0.6818$ & \centering$-$ &\centering$1.0020$ & $-$ \\
         \hline
         \centering$0.1$ & \centering$0.6933$ & \centering$-$ &\centering$1.0297$ & $-$ \\
         \hline
         \centering$0.2$ & \centering$0.7071$ & \centering$-$ &\centering$1.0689$ & $-$ \\
         \hline
         \centering$0.3$ & \centering$0.7222$ & \centering$-$ &\centering$1.1232$ & $-$ \\
         \hline
         \centering$0.4$ & \centering$0.7389$ & \centering$-$ &\centering$1.2041$ & $-$ \\
         \hline
         \centering$0.5$ & \centering$0.7576$ & \centering$-$ &\centering$1.3412$ & $-$ \\
         \hline
         \centering$1$ & \centering$0.9139$ & \centering$3.0374$ &\centering$-$ & $-$ \\
         \hline
    \end{tabular}
    \caption{Constraints on the amount of magnetic charge carried by the BH for various values of the EH parameter. The constraints are imposed by comparing the theoretically calculated diameter of the BH shadow with the observational data for the SMBH M87* given in \cite{EventHorizonTelescope:2019dse}. The results are given in mass units where $M=1$.}
    \label{Table:5}
\end{table}

In Table \ref{Table:6}, we present the bounds of the values of the BH magnetic charge derived by the comparison between the theoretically calculated BH shadow and the shadow of the SMBH Sgr A* obtained by the EHT. In the case of the SMBH Sgr A*, larger values of the magnetic charge are allowed in comparison with the M87* case. For values of the EH parameters of the order of $10^{-3}M^{2}$ or less, at 1$\sigma$ confidence level, the allowed values of the magnetic charge vary from zero to slightly below its extremal value, which determines the extremal scenario according to which the number of BH horizons decreases from three to one. For values of the EH parameter, such that $\alpha>0.4M^{2}$, for which extremal scenarios do not exist, there are no constraints for the values of the BH magnetic charge at $2\sigma$ confidence level. Additionally, for $\alpha=1M^{2}$, every value of the magnetic charge is practically allowed, at 1$\sigma$ confidence level.

To conclude the hairless case, based on the data in Tables \ref{Table:5} and \ref{Table:6}, NLED-induced interactions enable an EH BH to carry more magnetic charge compared to a Reissner-Nordstr\"om BH.
\begin{table}[]
    \centering
    \begin{tabular}{|| p{0.8cm} || p{2.6cm} | p{2.6cm} | p{2.6cm} | c ||}
    \hline
    \multicolumn{5}{||c||}{Constraints on the BH magnetic charge - Sgr A*} \\
    \hline
         \centering$\alpha$ & \centering$\text{Upper bound}~(1\sigma)$ & \centering$\text{Lower bound}~(1\sigma)$ & \centering$\text{Upper bound}~(2\sigma)$ & $\text{Lower bound}~(2\sigma)$  \\
         \hline
         \centering$0$ & \centering$0.6806$ & \centering$-$ &\centering$1$ & $-$ \\
         \hline
         \centering$0.002$ & \centering$1.0004$ & \centering$-$ &\centering$1.0004$ & $-$ \\
         \hline
         \centering$0.007$ & \centering$1.0015$ & \centering$-$ &\centering$1.0015$ & $-$ \\
         \hline
         \centering$0.01$ & \centering$1.0020$ & \centering$-$ &\centering$1.0020$ & $-$ \\
         \hline
         \centering$0.1$ & \centering$1.0152$ & \centering$-$ &\centering$1.1655$ & $-$ \\
         \hline
         \centering$0.2$ & \centering$1.0514$ & \centering$-$ &\centering$1.3112$ & $-$ \\
         \hline
         \centering$0.3$ & \centering$1.1005$ & \centering$-$ &\centering$1.6174$ & $-$ \\
         \hline
         \centering$0.4$ & \centering$1.1711$ & \centering$-$ &\centering$-$ & $-$ \\
         \hline
         \centering$0.5$ & \centering$1.2846$ & \centering$10$ &\centering$-$ & $-$ \\
         \hline
         \centering$1$ & \centering$10$ & \centering$-$ &\centering $-$ & $-$ \\
         \hline
    \end{tabular}
    \caption{Constraints on the amount of magnetic charge carried by the BH for various values of the EH parameter. The constraints imposed by comparing the theoretically calculated diameter of the BH shadow with the observational data for the SMBH Sgr A* given in \cite{EventHorizonTelescope:2022wkp}. The results are given in mass units where $M=1$.}
    \label{Table:6}
\end{table}

\begin{figure}
    \centering
    \includegraphics[width=0.38\textwidth]{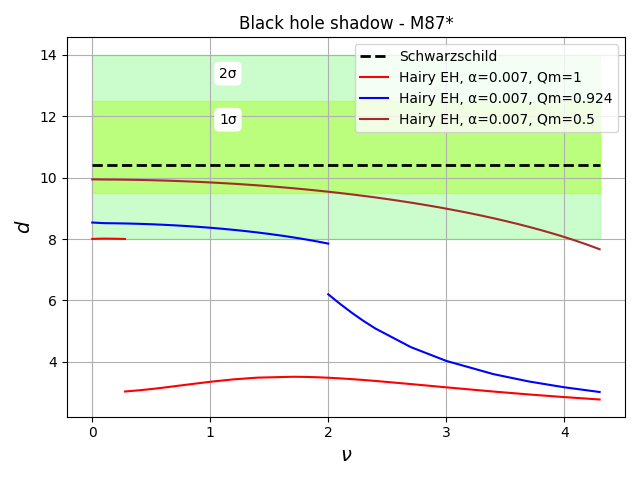}
    \includegraphics[width=0.38\textwidth]{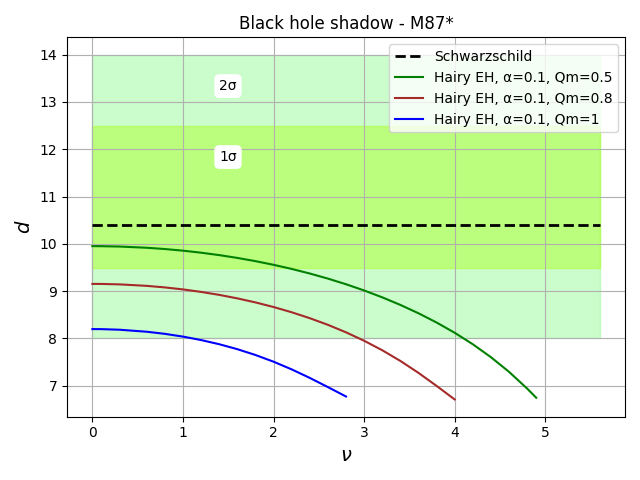}
    \includegraphics[width=0.38\textwidth]{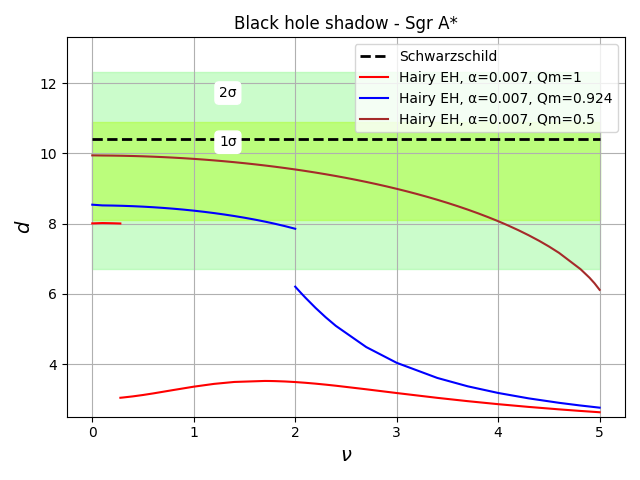}
    \includegraphics[width=0.38\textwidth]{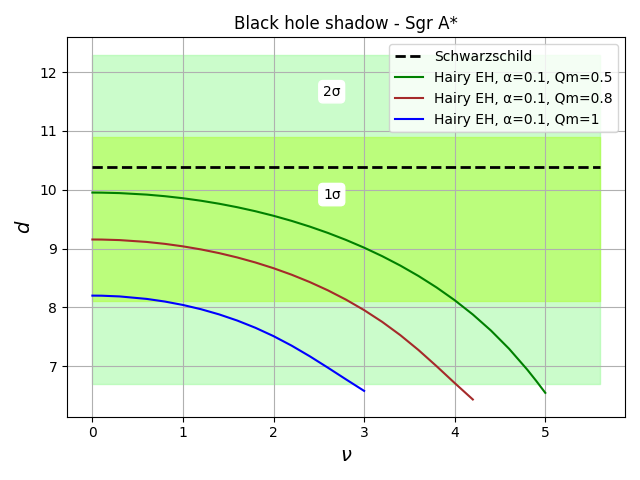}
    \caption{Shadow diameter ($d$) of a hairy EH BH as a function of the scalar charge for various values of the BH magnetic charge, and a small value (left graphs) and large values (right graphs) of the EH parameter. The shaded regions indicate the values of the shadow diameter consistent with the shadow of the SMBHs M87* (upper row) and Sgr A* (lower row) detected by the EHT, with details provided in Eqs. (\ref{M87}) and (\ref{Sgr}). The narrow shaded regions represents the 1$\sigma$ confidence regions, whereas the wider shaded regions give the 2$\sigma$ confidence regions. All graphs are in mass units ($M=1$).}
    \label{fig:22}
\end{figure}

Proceeding further with the hairy case, in FIG. \ref{fig:22}, we present the shadow diameter as a function of the scalar charge for various values of the BH magnetic charge and the EH parameter. In the left graphs of FIG. \ref{fig:22}, we utilize a small value for the EH parameter ($\alpha=0.007M^{2}$). For values of the EH parameter of the order of $10^{-2}M^{2}$ or less, whether an extremal scenario occurs or not depends on the chosen magnetic charge value. For small  values of the EH parameter, such as $\alpha=0.007M^{2}$, extremal scenarios exist if the magnetic charge takes values around $1M$. Thus, in the left graphs of FIG. \ref{fig:22}, we present plots with and without extremal scenarios. In the right graphs of FIG. \ref{fig:22}, we plot the shadow diameter for a large value of the $\alpha$-parameter ($\alpha=10^{-1}M^{2}$), for which extremal scenarios are absent. From the plots of FIG. \ref{fig:22}, bounds can be imposed on the scalar charge, which correspond to the points where the lines cross the shaded regions. The shaded regions in FIG. \ref{fig:22} represent the range of shadow diameters consistent with the shadows of the SMBHs M87* (upper row) and Sgr A* (lower row) at $1\sigma$ and $2\sigma$ confidence levels.

In Tables \ref{Table:7} and \ref{Table:8}, we illustrate the bounds of the scalar charge for various values of the EH parameter and magnetic charge at $1\sigma$ and $2\sigma$ confidence levels. Since the shadow diameter constantly decreases as the scalar charge increases, only upper bounds can be imposed on the scalar charge. Moreover, for some values for the $\alpha$-parameter and the magnetic charge, there are no values for the scalar charge in order for the shadow diameter to be consistent with the shadows of M87* or Sgr A*. These scenarios are denoted as ``rejected" in Tables \ref{Table:7} and \ref{Table:8}. Interestingly, the hairy EH BH model aligns more closely with the observations of the SMBH Sgr A* rather than those of M87*, because more ``rejected" scenarios exist in the case of M87* compared to the case of Sgr A*.
\begin{table}[]
    \centering
    \begin{tabular}{|| p{0.8cm} | p{0.8cm} || p{2.6cm} | p{2.6cm} | p{2.6cm} | c ||}
    \hline
    \multicolumn{6}{||c||}{Constraints on the scalar charge - M87*} \\
    \hline
         \centering$\alpha$ & \centering$Q_{m}$ & \centering$\text{Upper bound}~(1\sigma)$ & \centering$\text{Lower bound}~(1\sigma)$ & \centering$\text{Upper bound}~(2\sigma)$ & $\text{Lower bound}~(2\sigma)$  \\
         \hline
         \centering$0.007$ & \centering$0.5$ & \centering$2.090$ & \centering$-$ &\centering$4.054$ & $-$ \\
         \hline
         \centering$0.007$ & \centering$0.924$ & \centering$\text{rejected}$ & \centering$\text{rejected}$ &\centering$1.789$ & $-$ \\
         \hline
         \centering$0.007$ & \centering$1$ & \centering$\text{rejected}$ & \centering$\text{rejected}$ &\centering$0.277$ & $-$ \\
         \hline
         \centering$0.1$ & \centering$0.5$ & \centering$2.134$ & \centering$-$ &\centering$4.102$ & $-$ \\
         \hline
         \centering$0.1$ & \centering$0.8$ & \centering$\text{rejected}$ & \centering$\text{rejected}$ &\centering$2.948$ & $-$ \\
         \hline
         \centering$0.1$ & \centering$1$ & \centering$\text{rejected}$ & \centering$\text{rejected}$ &\centering$1.114$ & $-$ \\
         \hline
         \centering$0.2$ & \centering$0.5$ & \centering$2.180$ & \centering$-$ &\centering$4.157$ & $-$ \\
         \hline
         \centering$0.2$ & \centering$0.8$ & \centering$\text{rejected}$ & \centering$\text{rejected}$ &\centering$3.128$ & $-$ \\
         \hline
         \centering$0.2$ & \centering$1$ & \centering$\text{rejected}$ & \centering$\text{rejected}$ &\centering$1.672$ & $-$ \\
         \hline
         \centering$0.5$ & \centering$0.5$ & \centering$2.320$ & \centering$-$ &\centering$4.337$ & $-$ \\
         \hline
         \centering$0.5$ & \centering$0.8$ & \centering$\text{rejected}$ & \centering$\text{rejected}$ &\centering$3.766$ & $-$ \\
         \hline
         \centering$0.5$ & \centering$1$ & \centering$\text{rejected}$ & \centering$\text{rejected}$ &\centering$3.178$ & $-$ \\
         \hline
    \end{tabular}
    \caption{Constraints on the scalar charge carried by the BH for various values of the EH parameter and the BH magnetic charge. The constraints imposed by comparing the theoretically calculated diameter of the BH shadow with the observational data for the SMBH M87* given in \cite{EventHorizonTelescope:2019dse}. The results are given in mass units where $M=1$.}
    \label{Table:7}
\end{table}

\begin{table}[]
    \centering
    \begin{tabular}{|| p{0.8cm} | p{0.8cm} || p{2.6cm} | p{2.6cm} | p{2.6cm} | c ||}
    \hline
    \multicolumn{6}{||c||}{Constraints on the scalar charge - Sgr A*} \\
    \hline
         \centering$\alpha$ & \centering$Q_{m}$ & \centering$\text{Upper bound}~(1\sigma)$ & \centering$\text{Lower bound}~(1\sigma)$ & \centering$\text{Upper bound}~(2\sigma)$ & $\text{Lower bound}~(2\sigma)$  \\
         \hline
         \centering$0.007$ & \centering$0.5$ & \centering$3.971$ & \centering$-$ &\centering$4.812$ & $-$ \\
         \hline
         \centering$0.007$ & \centering$0.924$ & \centering$1.622$ & \centering$-$ &\centering$2$ & $-$ \\
         \hline
         \centering$0.007$ & \centering$1$ & \centering$\text{rejected}$ & \centering$\text{rejected}$ &\centering$0.277$ & $-$ \\
         \hline
         \centering$0.1$ & \centering$0.5$ & \centering$4.018$ & \centering$-$ &\centering$4.924$ & $-$ \\
         \hline
         \centering$0.1$ & \centering$0.8$ & \centering$2.835$ & \centering$-$ &\centering$4.008$ & $-$ \\
         \hline
         \centering$0.1$ & \centering$1$ & \centering$0.792$ & \centering$-$ &\centering$2.874$ & $-$ \\
         \hline
         \centering$0.2$ & \centering$0.5$ & \centering$4.070$ & \centering$-$ &\centering$5.061$ & $-$ \\
         \hline
         \centering$0.2$ & \centering$0.8$ & \centering$3.009$ & \centering$-$ &\centering$4.457$ & $-$ \\
         \hline
         \centering$0.2$ & \centering$1$ & \centering$1.437$ & \centering$-$ &\centering$3.763$ & $-$ \\
         \hline
         \centering$0.5$ & \centering$0.5$ & \centering$4.241$ & \centering$-$ &\centering$5.545$ & $-$ \\
         \hline
         \centering$0.5$ & \centering$0.8$ & \centering$3.618$ & \centering$-$ &\centering$6.077$ & $-$ \\
         \hline
         \centering$0.5$ & \centering$1$ & \centering$2.944$ & \centering$-$ &\centering$6.868$ & $-$ \\
         \hline
    \end{tabular}
    \caption{Constraints on the scalar charge carried by the BH for various values of the EH parameter and the BH magnetic charge. The constraints imposed by comparing the theoretically calculated diameter of the BH shadow with the observational data for the SMBH Sgr A* given in \cite{EventHorizonTelescope:2022wkp}. The results are given in mass units where $M=1$.}
    \label{Table:8}
\end{table}

\begin{figure}
    \centering
    \includegraphics[width=0.38\textwidth]{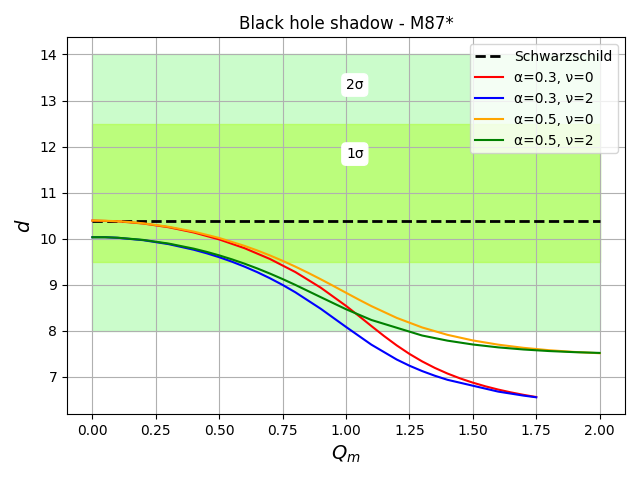}
    \includegraphics[width=0.38\textwidth]{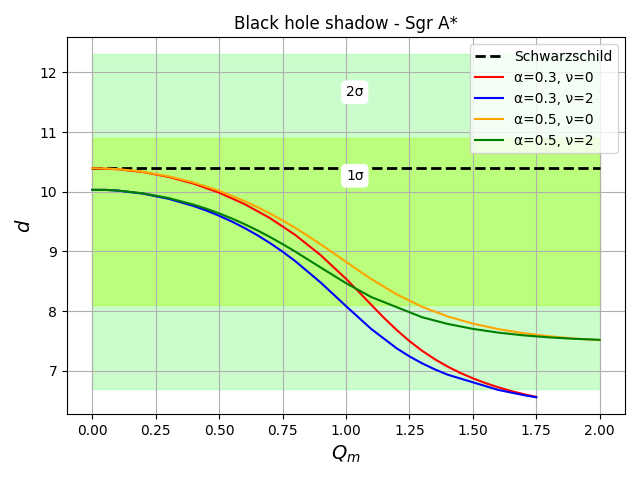}
    \caption{Shadow diameter ($d$) of hairy ($\nu=2M$) and hairless ($\nu=0$) EH BHs as a function of the BH magnetic charge. The shaded regions indicate the values of the shadow diameter consistent with the shadow of the SMBHs M87* (left graph) and Sgr A* (right graph) detected by the EHT, with details provided in Eqs. (\ref{M87}) and (\ref{Sgr}). The narrow shaded regions represents the 1$\sigma$ confidence regions, whereas the wider shaded regions give the 2$\sigma$ confidence regions. All graphs are in mass units ($M=1$).}
    \label{fig:23}
\end{figure}

Finally, in FIG. \ref{fig:23}, we compare the BH shadow in the hairy and hairless cases by illustrating the shadow diameter as a function of BH magnetic charge for zero and non-zero values of the scalar charge. As can be clearly observed from these graphs, hairless BHs can possess a greater amount of magnetic charge compared to that carried by hairy BHs in order for the results to be consistent with the observations of EHT. Interestingly, when the magnetic charge takes large values, its contribution to the size of the BH shadow is more significant compared to the contribution of the scalar charge. Consequently, for large values of the magnetic charge, the shadows of the hairy and hairless BHs are indistinguishable.   

\section{Conclusions}

\label{sec7}

In this work, we examined the trajectories of uncharged particles in a magnetically charged BH spacetime dressed with a scalar hair within the framework of EH electrodynamics. We considered both time-like and null cases. Using numerical integration of the equations of motion, we represented the different types of motion in the asymptotically AdS, dS, and flat spacetime cases. Our findings indicate that the periastron shift decreases as the scalar charge becomes stronger. Additionally, in the dS case, there is a larger periastron shift due to the repulsive nature of a positive cosmological constant.

Furthermore, we found that the magnetic and scalar charges of the BH have a repulsive effect on particle motion, which is opposite to the impact of BH mass. Thus, as these charges increase, the BH shrinks. In the case of the BH magnetic charge, this behavior holds up to a critical value. To be more precise, when the magnetic charge surpasses its critical value, which varies based on other BH parameters, its impact on spacetime mirrors that of the BH mass. On the contrary, the EH parameter introduced by modified electrodynamics has minimal effects on particle motion, and the BH appears to grow as the EH parameter increases. We have also concluded that the EH parameter influences the causal structure of spacetime. Due to the effects of NLED on the BH, it can have up to three horizons. The number of horizons depends on BH parameters and significantly contributes to the size of both the event horizon and shadow of the BH. For instance, some extremal scenarios exist according which the number of BH horizons decreases from 3 to 1. These scenarios imply some points of discontinuity in the values of radii of both the event horizon and BH shadow.

Moreover, we investigated the photon propagation in NLED and determined the \emph{effective geometry} in which photons follow null geodesics. We also defined the effective potential that governs photon motion and calculated the photon spheres and BH shadows. BH shadows have similar behaviour to that of the event horizon as BH parameters vary. Finally, we compared the theoretically calculated BH shadows in both hairy and hairless cases with the shadows of the SMBHs M87* and Sgr A* observed by the EHT. This analysis allowed us to impose bounds on both magnetic and scalar charges at $1\sigma$ and $2\sigma$ confidence levels. Both hairy and hairless models were found to be more consistent with the observations for the SMBH Sgr A*. Moreover, the NLED interactions enable BHs to carry a larger amount of magnetic charge compared to that carried by a Reissner-Nordstr\"om BH. Interestingly, for large values for the $\alpha$-parameter, such as $\alpha=1M^{2}$, every value of the magnetic charge is consistent with the EHT observations. Additionally, for smaller values of the magnetic charge, larger values of the scalar charge are permitted, and vice versa.

In future work, it would be interesting to consider the motion of charged particles in this spacetime in order to understand the effect of the EH parameter in this context. It might also be worthwhile to explore the geodesic motion of particles in spacetimes with a magnetic monopole induced from a global monopole, which has a well-defined ADM mass, such as the one described in \cite{Chatzifotis:2022ubq}. In this case, the spacetime includes a deficit angle that will contribute to the results.

\section*{Acknowledgments}
The research project was supported by the Hellenic Foundation for Research and Innovation (H.F.R.I.) under the “3rd Call for H.F.R.I. Research Projects to support Post-Doctoral Researchers” (Project Number: 7212).
The work of D.P.T. is supported by a National Technical University of Athens Master-programme ({\it Physics and Technological Applications}) award of scientific excellence.

\end{document}